\documentclass[twocolumn]{aastex63}
\usepackage{xspace} 
\usepackage{graphicx} 
\usepackage{natbib}
\usepackage{amsmath}
\usepackage{afterpage}
\newcommand{\nraoblurb}{The National Radio Astronomy Observatory is
a facility of the National Science Foundation operated under cooperative
agreement by Associated Universities, Inc.}
\newcommand{\y}[1]{\ensuremath{y_{#1}}\xspace}

\newcommand{\kms}{\ensuremath{\,{\rm km\,s^{-1}}}\xspace}

\newcommand{\m}{\ensuremath{\,{\rm m}}\xspace}

\newcommand{\kpc}{\ensuremath{\,{\rm kpc}}\xspace}
\newcommand{\K}{\ensuremath{\,{\rm K}}\xspace}

\newcommand{\khz}{\ensuremath{\,{\rm kHz}}\xspace}
\newcommand{\mhz}{\ensuremath{\,{\rm MHz}}\xspace}
\newcommand{\ghz}{\ensuremath{\,{\rm GHz}}\xspace}

\newcommand{\h}[1]{\ensuremath{^{#1}{\rm H}}\xspace}

\newcommand{\hii}{H\,{\sc ii}}
\newcommand{\cii}{C\,{\sc ii}}

\revised{\today, accepted for ApJ publication -- tvw}

\shorttitle{Metallicity Structure}
\shortauthors{Wenger et al.}

\begin{document}

\title{Metallicity Structure in the Milky Way Disk Revealed by Galactic \hii\ Regions}

\author{Trey V. Wenger} 
\affiliation{Dominion Radio Astrophysical Observatory, Herzberg Astronomy and
  Astrophysics Research Centre, National Research Council, P.O. Box 248,
  Penticton, BC V2A 6J9, Canada.}
\affiliation{Astronomy Department, University of Virginia, P.O. Box
  400325, Charlottesville, VA 22904-4325, USA.}
\affiliation{National Radio Astronomy Observatory, 520 Edgemont Road,
  Charlottesville, VA 22903, USA.}
\email{Trey.Wenger@nrc-cnrc.gc.ca}

\author{Dana S. Balser}
\affiliation{National Radio Astronomy Observatory, 520 Edgemont Road,
  Charlottesville, VA 22903, USA.}

\author{L. D. Anderson}
\affiliation{Department of Physics and Astronomy, West Virginia
  University, Morgantown, WV 26505, USA.}
\affiliation{Center for Gravitational Waves and Cosmology, West
Virginia University, Morgantown, Chestnut Ridge Research Building, 
Morgantown, WV 26505, USA.}
\affiliation{Adjunct Astronomer at the Green Bank Observatory, 
P.O. Box 2, Green Bank, WV 24944, USA.}

\author{T. M. Bania}
\affiliation{Institute for Astrophysical Research, Astronomy 
  Department, Boston University, 725 Commonwealth Ave., Boston, MA 
  02215, USA.}

\begin{abstract}
  The metallicity structure of the Milky Way disk stems from the
  chemodynamical evolutionary history of the Galaxy. We use the
  National Radio Astronomy Observatory Karl G. Jansky Very Large Array
  to observe \({\sim}8-10\ghz\) hydrogen radio recombination line and
  radio continuum emission toward 82 Galactic \hii\ regions. We use
  these data to derive the electron temperatures and metallicities for
  these nebulae.  Since collisionally excited lines from metals (e.g.,
  oxygen, nitrogen) are the dominant cooling mechanism in
  \hii\ regions, the nebular metallicity can be inferred from the
  electron temperature. Including previous single dish studies, there
  are now 167 nebulae with radio-determined electron temperature and
  either parallax or kinematic distance determinations.  The
  interferometric electron temperatures are systematically 10\% larger
  than those found in previous single dish studies, likely due to
  incorrect data analysis strategies, optical depth effects, and/or
  the observation of different gas by the interferometer.  By
  combining the interferometer and single dish samples, we find an
  oxygen abundance gradient across the Milky Way disk with a slope of
  \(-0.052\pm0.004\,\text{dex kpc}^{-1}\). We also find significant
  azimuthal structure in the metallicity distribution.  The slope of
  the oxygen gradient varies by a factor of \({\sim}2\) when
  Galactocentric azimuths near \({\sim}30^\circ\) are compared with
  those near \({\sim}100^\circ\). This azimuthal structure is
  consistent with simulations of Galactic chemodynamical evolution
  influenced by spiral arms.
\end{abstract}


\section{Introduction}

The present day chemical structure of the Milky Way disk is an
important constraint on models of Galactic chemodynamical evolution
\citep[e.g.,][]{chiappini2003,minchev2014,snaith2015,minchev2018}.
Radial metallicity gradients, for example, are found in both the Milky
Way and other spiral galaxies in studies using collisionally excited
lines in ionized star forming regions
\citep[e.g.,][]{searle1971,shaver1983} and stellar abundances
\citep[e.g.,][]{hayden2014,bovy2014}. These gradients reveal the
history of star formation, stellar migration, and chemical enrichment
by stars across galactic disks \citep{minchev2018}. Stellar and
gaseous tracers provide complementary information about the
chemodynamical history of the Galaxy. The chemical abundances of stars
represent the enrichment of the interstellar medium (ISM) when the
stars were born, whereas the abundances of gaseous tracers represent
the end product of billions of years of stellar evolution and ISM
enrichment.

Evidence for azimuthal variations in galactic radial metallicity
gradients is observed in both the Milky Way \citep[e.g.,][hereafter,
  B15]{balser2015} and other galaxies
\citep[e.g.,][]{sanchez-menguiano2016,sanchez-menguiano2017,ho2017,ho2018}.
Azimuthal abundance variations in the Milky Way are identified in
multiple elements and tracers, such as the oxygen abundances of
\hii\ regions (e.g, B15) and the iron abundances of Cepheids
\citep[e.g.,][]{luck2006,pedicelli2009}. Such variations are not
expected in an old and well-mixed galaxy \citep{balser2011}, and
chemodynamical models of galaxies typically assume axisymmetric
metallicity gradients \citep[e.g.,][]{chiappini2003}. Azimuthal
variations may be caused by streaming motions and radial migration
induced by galactic bars \citep{dimatteo2013}, spiral arms
\citep{grand2016,ho2017,spitoni2019,molla2019b}, and/or perturbations
from minor galaxy interactions \citep{bird2012}.

Here we expand the Galactic \hii\ region metallicity surveys of
\citet{quireza2006b}, \citet{balser2011}, and B15 to create a more
complete map of metallicity structure in the Milky Way disk and to
search for evidence of azimuthal variations in the Galactic radial
metallicity gradient. \hii\ regions are the sites of recent high-mass
star formation. These nebulae are an ideal tracer of Galactic
metallicity structure because (1) they live for \({\lesssim}10\) Myr,
and they therefore reveal the current enrichment of the ISM; (2) their
distances can be derived accurately using maser parallax measurements
\citep[e.g.,][]{reid2014} or kinematic techniques
\citep[e.g.,][]{wenger2018b}; and (3) their metallicities are easily
derived using optical and infrared collisionally excited lines or
inferred from the nebular electron temperatures. The radio
recombination line (RRL) and radio continuum emission from
\hii\ regions are an extinction-free diagnostic of the nebular
electron temperature \citep{mezger1967a}, which is empirically related
to the \hii\ region metallicity \citep{shaver1983}. Radio wavelength
observations of \hii\ regions can reveal metallicity structure across
the Milky Way disk due to the lack of dust extinction.

The local thermodynamic equilibrium (LTE) electron temperature of an
ionized gas can be derived from the RRL-to-continuum brightness ratio
when the nebula is optically thin (B15). The electron temperature
surveys of Galactic \hii\ regions by B15, \citet{balser2011}, and
\citet{quireza2006b} used single dish telescopes. Although these
instruments are extremely sensitive to faint RRL emission, they are
not ideal for measuring accurate RRL-to-continuum brightness ratios
because of the uncertainties in the continuum brightnesses. The single
dish continuum brightness of an \hii\ region is measured by scanning
the telescope across the source in multiple directions. Then, a
baseline fit to the diffuse background continuum emission is
removed. The accuracy of the radio continuum brightness is limited by
the ability to accurately remove this diffuse component.

An interferometer is the ideal tool for measuring the RRL-to-continuum
brightness ratio of Galactic \hii\ regions. By their nature,
interferometers are not sensitive to large scale, diffuse emission,
such as the non-thermal radio continuum emission that permeates the
Galactic plane.  We measure the total continuum flux density of
nebulae more accurately with an interferometer than with a single dish
telescope if the angular size of the source is smaller than the
largest angular scale of the telescope. Too, interferometer data can
be constructed as a high angular resolution image or data cube. These
images and cubes reduce source confusion and can provide maps of
electron temperature variations across a resolved nebula. Finally,
interferometers like the National Radio Astronomy Observatory (NRAO)
Karl G. Jansky Very Large Array (VLA) simultaneously measure both
radio continuum and RRL emission. Any systematic calibration or
weather issues affecting the data will be removed in the
RRL-to-continuum flux ratio.

We use the VLA to derive the nebular electron temperatures and
metallicities of Galactic \hii\ regions across the Milky Way disk.  A
subset of these nebulae overlap with previous single dish surveys,
which allows us to compare the interferometer-derived electron
temperatures with those derived from single dish observations.

\section{Target Sample}

Recent RRL surveys have more than doubled the number of known Galactic
\hii\ regions
\citep{bania2010,bania2012,anderson2014,anderson2015a,anderson2015b,anderson2018,wenger2019}. The
\textit{Widefield Infrared Survey Explorer (WISE)} Catalog of Galactic
\hii\ Regions (hereafter, \textit{WISE} Catalog) contains the infrared
and radio properties of more than 2000 known nebulae
\citep{anderson2014}. To derive accurate electron temperatures, we
require the subset of \textit{WISE} Catalog nebulae observable by the
VLA. Our selection criteria are nebulae with 1) a single RRL velocity
component, 2) a maser parallax measurement or an accurate kinematic
distance, and 3) a predicted RRL flux density \(>1.7\,\text{mJy
  beam\(^{-1}\)}\).

\begin{figure}
  \centering
  \includegraphics[width=\linewidth]{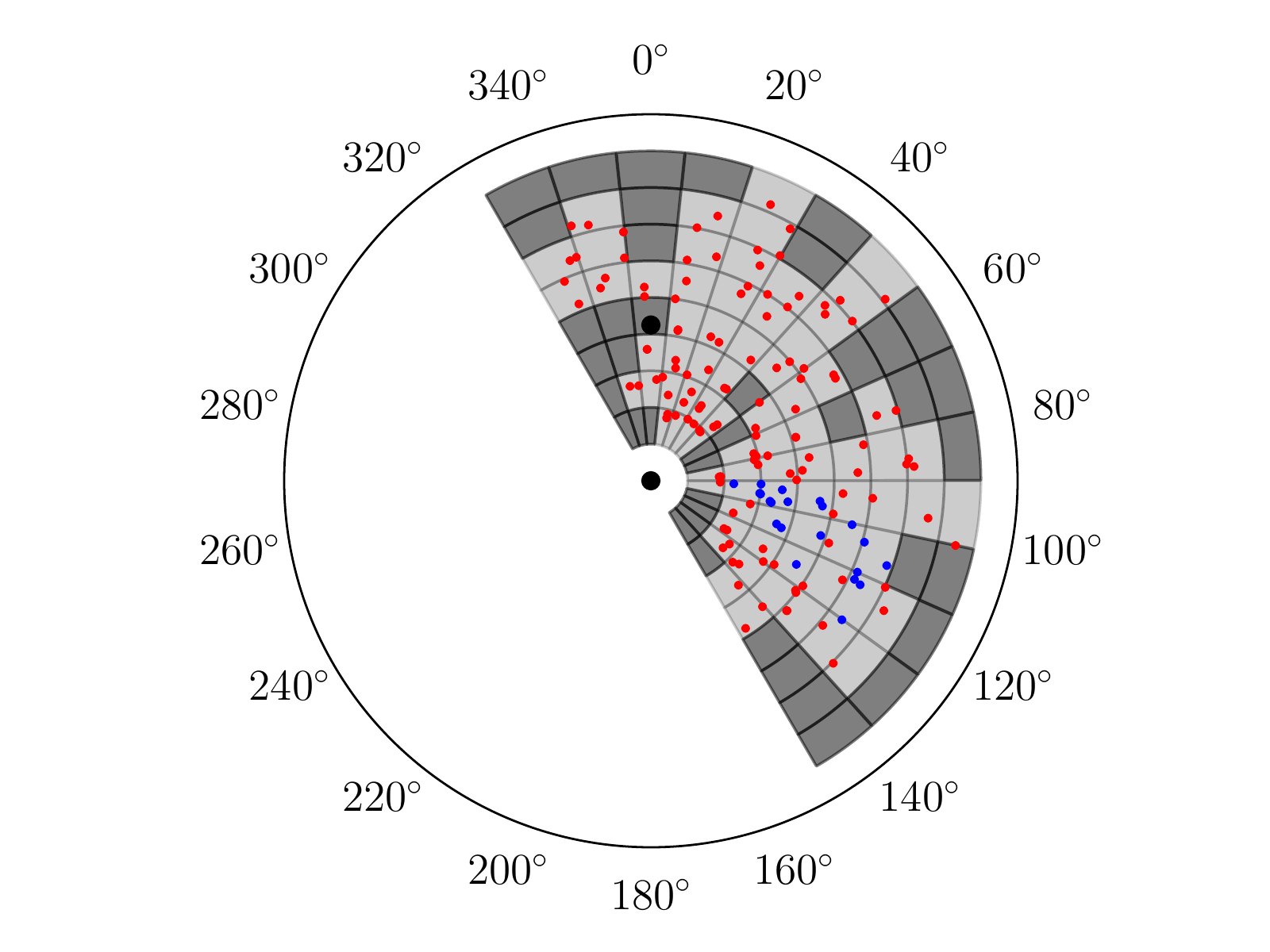}
  \caption{Galactocentric positions and Milky Way disk coverage of the
    VLA survey \hii\ regions. The Galactic Center is the black point
    at the origin and the Sun is the black point 8.34\kpc\ in the
    direction \(\theta = 0^\circ\). The colored points are the
    \hii\ regions in the pilot survey (blue) and main survey
    (red). The Galactic disk is divided into 120 bins of size
    \(12^\circ\) in Galactocentric azimuth, over the azimuth range
    \(-30^\circ\) to \(150^\circ\), and \(2\kpc\) in Galactocentric
    radius, up to \(18\kpc\). Bins that contain at least one nebulae
    are colored light gray, whereas empty bins are dark gray.}
  \label{fig:candidates}
\end{figure}

When this survey began, the \textit{WISE} Catalog contained RRL
measurements of \({\sim}1200\) unique Galactic \hii\ regions. Many of
these nebulae are clustered in \hii\ region groups or complexes, and a
single dish observation will see the combined emission from multiple
discrete sources. These star forming complexes are the source of
ionizing photons, which may leak out into and ionize the diffuse
ISM. In these cases, the RRL spectrum of the \hii\ region will show
multiple velocity components from either multiple discrete
\hii\ regions or a mix of \hii\ regions and diffuse ionized gas. The
presence of spectrally confused, or blended, RRL components will limit
our ability to derive the nebular RRL flux density
accurately. Therefore, we remove \({\sim}100\) nebulae with multiple
velocity component RRLs in the \textit{WISE} Catalog.

In order to study Galactic metallicity structure, accurate distances
to tracers are needed. Therefore, we further limit the \textit{WISE}
Catalog sample to those nebulae with published maser parallax
measurements and/or accurate kinematic distances. We adopt the
kinematic distance uncertainty model of \citet{anderson2012} to
estimate the accuracy of kinematic distances in the \textit{WISE}
Catalog. Because we aim to generate a Galactocentric map of the Milky
Way metallicity structure, we require kinematic distance accuracies
such that the uncertainty in the Galactocentric radius is \(\sigma_R <
2\kpc\) and the uncertainty in Galactocentric azimuth is
\(\sigma_\theta < 20^\circ\). Out of our sample of \({\sim}1100\)
single velocity RRL component nebulae, 107 have an associated maser
parallax measurement and 364 have a kinematic distance meeting these
accuracy thresholds. This brings our total sample of \hii\ regions to
471 nebulae.

Finally, we identify the subset of this sample with previously
measured RRL flux densities bright enough to be detected by the VLA in
a 10 minute observation. The point source sensitivity of the VLA with
this integration time is \({\sim}2\,\text{mJy beam\(^{-1}\)}\) per
\(31.25\,\khz\) channel at \({\sim}9\,\ghz\). By smoothing the spectra
to \(5\,\kms\) resolution and averaging 7 hydrogen RRL transitions, we
estimate a spectral rms noise of \({\sim}0.3\,\text{mJy
  beam\(^{-1}\)}\) per channel. We thus require our sample of
\hii\ regions to have a predicted \(9\,\ghz\) RRL flux density greater
than 5 times this sensitivity limit, \({\sim}1.7\,\text{mJy
  beam\(^{-1}\)}\).

All previously measured RRL flux densities of northern sky
\hii\ regions in the \textit{WISE} catalog were made with single dish
telescopes around \({\sim}9\,\ghz\). We first scale the observed RRL
brightness temperatures to exactly \(9\,\ghz\) assuming the RRL
brightness temperature is proportional to the RRL frequency (B15). We
convert these scaled RRL brightness temperatures to point source flux
densities assuming telescope gains of \({\sim}2\,\text{K
  Jy\(^{-1}\)}\) for the Green Bank Observatory (GBO) Green Bank
Telescope \citep[GBT;][]{balser2011}, \({\sim}0.27\,\text{K
  Jy\(^{-1}\)}\) for the NRAO 140 Foot Telescope \citep[hereafter, 140
  Foot;][]{balser2016}, and \({\sim}5\,\text{K Jy\(^{-1}\)}\) for the
Arecibo Observatory \citep{bania2012}. Any source with a predicted
\(9\,\ghz\) RRL flux density \(S_{L, 9\ghz} > 1.7\,\text{mJy
  beam\(^{-1}\)}\) fulfills our sensitivity criterion. This threshold
removes only 10 nebulae from our sample, bringing the total number of
observable \hii\ regions to 461.

The VLA is not sensitive to emission on scales larger than
\({\sim}145\,\text{arcsec}\) in the D (most compact) configuration at
\({\sim}9\,\ghz\). If we assume that the radio size of an \hii\ region
is approximately half of the infrared size \citep[e.g.,][]{bihr2016},
then \(30\%\) of the \hii\ regions in our sample have radio diameters
greater than this largest angular scale. Our observations will not be
sensitive to these angularly large nebulae if their emission is
uniform on such large spatial scales. We expect to detect clumpy
emission within these large \hii\ regions, however, so we do not use
any size restriction when defining our sample.

\afterpage{
  \begin{longrotatetable}
\begin{deluxetable*}{lcccrr@{\,\(\pm\)\,}lccr@{\,\(\pm\)\,}lr@{\,\(\pm\)\,}lc}
\centering
%\singlespace
\tablewidth{0pt}
%\tabletypesize{\scriptsize}
\tablecaption{Survey Targets\label{tab:targets}}
\tablehead{
\colhead{Field} & \colhead{Project} & \colhead{R.A.}       & \colhead{Decl.}      & \colhead{\(R_{\rm IR}\)}   & \multicolumn{2}{c}{\(S_{9\ghz,\,L}\)}   & \colhead{Telescope\tablenotemark{a}} & \colhead{RRL}                     & \multicolumn{2}{c}{\(S_L/S_C\)} & \multicolumn{2}{c}{\(T_e\)} & \colhead{\(T_e\)} \\
\colhead{}      & \colhead{}        & \colhead{J2000}      & \colhead{J2000}      & \colhead{(arcsec)}         & \multicolumn{2}{c}{(mJy}                & \colhead{}                           & \colhead{Author\tablenotemark{b}} & \multicolumn{2}{c}{}            & \multicolumn{2}{c}{(K)}     & \colhead{Author\tablenotemark{c}} \\
\colhead{}      & \colhead{}        & \colhead{(hh:mm:ss)} & \colhead{(dd:mm:ss)} & \colhead{}                 & \multicolumn{2}{c}{beam\(^{-1}\))}      & \colhead{}                           & \colhead{}                        & \multicolumn{2}{c}{}            & \multicolumn{2}{c}{}        & \colhead{}
}
\startdata
G005.883$-$0.399 & 15B-178 & 18:00:31.5 & $-24$:04:18.9 & $22.35$ & $844.15$ & $5.77$ & 140 Foot & Q06 & \multicolumn{2}{c}{\nodata} & \multicolumn{2}{c}{\nodata} & \nodata \\
G009.598+0.199 & 15B-178 & 18:06:11.1 & $-20$:32:36.5 & $34.09$ & $226.92$ & $20.00$ & 140 Foot & L89 & \multicolumn{2}{c}{\nodata} & \multicolumn{2}{c}{\nodata} & \nodata \\
G010.596$-$0.381 & 15B-178 & 18:10:24.6 & $-19$:57:08.4 & $60.00$ & $586.69$ & $4.23$ & 140 Foot & Q06 & $0.0686$ & $0.0006$ & $9810$ & $90$ & Q06b;B15 \\
G012.804$-$0.207 & 15B-178 & 18:14:15.0 & $-17$:55:56.4 & $21.15$ & $3034.62$ & $23.85$ & 140 Foot & Q06 & $0.0808$ & $0.0007$ & $7620$ & $100$ & Q06b;B15 \\
G013.880+0.285 & 15B-178 & 18:14:35.7 & $-16$:45:09.7 & $144.31$ & $587.42$ & $5.00$ & 140 Foot & Q06 & $0.1210$ & $0.0012$ & $6960$ & $80$ & Q06b;B15 \\
G015.212+0.167 & 15B-178 & 18:17:40.0 & $-15$:38:13.8 & $176.72$ & $10.95$ & $0.11$ & GBT & A15b & \multicolumn{2}{c}{\nodata} & \multicolumn{2}{c}{\nodata} & \nodata \\
G017.336$-$0.146 & 15B-178 & 18:22:57.2 & $-13$:54:41.0 & $102.77$ & $6.70$ & $0.18$ & GBT & A11 & \multicolumn{2}{c}{\nodata} & \multicolumn{2}{c}{\nodata} & \nodata \\
G017.928$-$0.677 & 15B-178 & 18:26:01.7 & $-13$:38:14.6 & $164.84$ & $13.00$ & $0.30$ & GBT & A11 & \multicolumn{2}{c}{\nodata} & \multicolumn{2}{c}{\nodata} & \nodata \\
\enddata
\tablenotetext{a}{Original RRL detection telescope}
\tablenotetext{b}{Original RRL detection reference}
\tablenotetext{c}{RRL-to-continuum flux ratio measurement and electron temperature derivation reference}
\tablerefs{(L89) \citet{lockman1989}; (L96) \citet{lockman1996}; (Q06a) \citet{quireza2006a}; (Q06b) \citet{quireza2006b}; (A11) \citet{anderson2011}; (B11) \citet{balser2011}; (B12) \citet{bania2012}; (A15a) \citet{anderson2015a}; (A15b) \citet{anderson2015b}; (B15) \citet{balser2015}}
\end{deluxetable*}
\end{longrotatetable}

  \clearpage
}

Finally, we select our observing targets from this sample of 461
nebulae to maximize our coverage of the Galactic disk. We divide the
Galaxy into 120 bins of size \(12^\circ\) in Galactocentric azimuth,
over the azimuth range \(-30^\circ\) to \(150^\circ\), and \(2\kpc\)
in Galactocentric radius, up to \(18\kpc\). Using the maser parallax
distance, when available, or the \textit{WISE} Catalog kinematic
distance to compute the Galactocentric radii and azimuths of the
nebulae, we identify the two brightest and most compact \hii\ regions
in each bin. Some bins only have one (or zero) nebulae that meet our
distance accuracy and predicted RRL flux density requirements.
Figure~\ref{fig:candidates} shows the Galactocentric positions
of the 128 \hii\ regions we select using these criteria as well as the
20 nebulae observed in the pilot survey. One \hii\ region,
G032.272\(-\)0.226, is observed in both the pilot survey and main
survey. Of the 120 position bins, 78 (65\%) contain at least one
\hii\ region that meets our selection criteria.

Our final \hii\ region target catalog contains 147 unique
nebulae. Table~\ref{tab:targets} lists information about these
\hii\ regions: the \textit{WISE} Catalog name; the VLA project in
which it was observed (13A\(-\)030 is the pilot survey and 15B\(-\)178
is the main survey); the \textit{WISE} infrared position; the
\textit{WISE} infrared radius, \(R_{\rm IR}\); the estimated
\(9\,\ghz\) RRL flux density, \(S_{9\ghz,\,L}\); the telescope and
reference for the previous RRL detection; the previously measured
RRL-to-continuum brightness ratio, \(S_L/S_C\), and derived electron
temperature, \(T_e\); and the reference for the RRL-to-continuum
brightness ratio and electron temperature.

\begin{deluxetable*}{lcc}
  \centering
  \tablewidth{0pt}
%\tabletypesize{\normalsize}
\tablecaption{Observation Summary\label{tab:observations}}
\tablehead{
   & \colhead{13A-030} & \colhead{15B-178}
} 
\startdata
Dates & 2013 Feb and Apr & 2015 Oct and Nov \\
Observing Time (hr)  & 5  & 30 \\
\hii\ Region Targets & 20 & 128 \\
Primary Calibrators & 3C286 & 3C286, 3C48 \\
Secondary Calibrators  & J1733\(-\)1304, J1822\(-\)0938 & J0019+7327, J0102+5824 \\
& J1824+1044, J1922+1530 & J0244+6228, J0349+4609 \\
& & J0358+5606, J0625+1440 \\
& & J0653\(-\)0625, J0735\(-\)1735  \\
& & J0804\(-\)2749, J1604\(-\)4441 \\
& & J1744\(-\)3116, J1820\(-\)2528 \\
& & J1822\(-\)0938, J1824+1044 \\
& & J1922+1530, J1924+3329 \\
& & J1925+2106, J2007+4029 \\
& & J2025+3343, J2137+5101 \\
& & J2137+5101 \\
& & J2148+6107 \\
\enddata
\end{deluxetable*}

\begin{deluxetable*}{ccccccc}
%\tabletypesize{\normalsize}
\tablecaption{Correlator Setup\label{tab:correlator}}
\tablehead{
  \colhead{Window} & \colhead{\(\nu_{\rm center}\)} & \colhead{Bandwidth} & \colhead{Channels} & \colhead{\(\Delta\nu\)} & \colhead{RRL} & \colhead{\(\nu_{\rm RRL}\)} \\
  & \colhead{(\mhz)}            & \colhead{(\mhz)}    &          & \colhead{(\khz)}        &     & \colhead{(\mhz)}
} 
\startdata
0 & 7949.3 & 128 & 128 & 1000 & \nodata & \nodata \\
1 & 8049.1 & 128 & 128 & 1000 & \nodata & \nodata \\
2 & 8049.1 & 16  & 512 & 31.25 & H93\(\alpha\) & 8045.605 \\
3 & 8205.3 & 128 & 128 & 1000 & \nodata & \nodata \\
4 & 8333.3 & 128 & 128 & 1000 & \nodata & \nodata \\
5 & 8313.0 & 16  & 512 & 31.25 & H92\(\alpha\) & 8309.385 \\
6 & 8461.3 & 128 & 128 & 1000 & \nodata & \nodata \\
7 & 8589.3 & 128 & 128 & 1000 & \nodata & \nodata \\
8 & 8588.5 & 16  & 512 & 31.25 & H91\(\alpha\) & 8584.823 \\
9 & 8717.3 & 128 & 128 & 1000 & \nodata & \nodata \\
10 & 8845.3 & 128 & 128 & 1000 & \nodata & \nodata \\
11 & 8876.4 & 16  & 512 & 31.25 & H90\(\alpha\) & 8872.571 \\
12 & 9082.3 & 128 & 128 & 1000 & \nodata & \nodata \\
13 & 9210.3 & 128 & 128 & 1000 & \nodata & \nodata \\
14 & 9177.3 & 16  & 512 & 1000 & H89\(\alpha\) & 9173.323 \\
15 & 9338.3 & 128 & 128 & 1000 & \nodata & \nodata \\
16 & 9466.3 & 128 & 128 & 1000 & \nodata & \nodata \\
17 & 9491.9 & 16  & 512 & 31.25 & H88\(\alpha\) & 9487.824 \\
18 & 9594.3 & 128 & 128 & 1000 & \nodata & \nodata \\
19 & 9722.3 & 128 & 128 & 1000 & \nodata & \nodata \\
20 & 9850.3 & 128 & 128 & 1000 & \nodata & \nodata \\
21\tablenotemark{a} & 9821.1 & 16 & 512 & 31.25 & H87\(\alpha\) & 9816.867 \\
22 & 9887.3 & 16  & 512 & 31.25 & H109\(\beta\) & 9883.083 \\
23 & 9978.3 & 128 & 128 & 1000 & \nodata & \nodata \\
\enddata
\tablenotetext{a}{Spectral window 21 was mistuned for one observing session in 13A-030.}
\end{deluxetable*}

\section{Observations and Data Reduction}

We used the VLA to simultaneously observe radio continuum and RRL
emission toward our sample of 147 Galactic \hii\ regions. The data
were acquired in the most compact (D) antenna configuration as part of
two projects: the pilot survey (13A-030; 5 hours) in Feb and Apr 2013,
and the main survey (15B-178; 30 hours) in Oct and Nov 2015. A summary
of the observations is in Table~\ref{tab:observations}.

The VLA X-band receiver covers the frequency range \({\sim}\)8--12
\ghz. We used the Wideband Interferometric Digital ARchitecture
(WIDAR) correlator in the 8-bit sampler mode to simultaneously measure
\({\sim}\)8--10 \ghz\ radio continuum emission and 8 hydrogen RRL
transitions in both linear polarizations. The continuum data were
measured by 16 low spectral resolution spectral windows (hereafter,
continuum windows) covering 7.8--8.9 \ghz\ and 9--10
\ghz\ continuously. The RRL spectra were measured by 8 high spectral
resolution (31.25\,\khz) spectral windows (hereafter, spectral line
windows), each with 16\,\mhz\ of frequency coverage. There are only 7
H\(\alpha\) RRL transitions in this frequency range (H87\(\alpha\) to
H93\(\alpha\)), so we tuned one of the spectral line windows to
H109\(\beta\). The native velocity resolution ranges from
0.9\,\kms\ at H87\(\alpha\) to 1.2\,\kms\ at H93\(\alpha\), with a
velocity coverage ranging from 488\,\kms\ to 600\,\kms\ for these
transitions, respectively.  In one observing session of the pilot
survey, the spectral line window for H88\(\alpha\) was mistuned, so we
exclude that spectral window from these
analyses. Table~\ref{tab:correlator} lists the following properties
for each spectral window: the center frequency, \(\nu_{\rm center}\);
the bandwidth; the number of channels; the channel width,
\(\Delta\nu\); the targeted RRL transition; and the RRL rest
frequency, \(\nu_{\rm RRL}\).

Our targets are clustered into 12 observing sessions based on
position, with \({\sim}10\) \hii\ regions per group. Every observing
session begins with a \({\sim}15\) minute integration on a primary
calibrator, which is used for the absolute flux, delay, and bandpass
calibration, followed by a \({\sim}10\) minute integration on a
secondary calibrator located near the \hii\ region science targets,
which is used for the complex gain calibration. These calibrators are
listed in Table~\ref{tab:observations}. We observe each science target
for 10--15 minutes to reach the necessary spectral sensitivity, then
we return to the secondary calibrator for \({\sim}5\) more
minutes. During each observing session, we repeat this process for
each science target.

We use the Wenger Interferometry Software Package (WISP) to calibrate,
reduce, and analyze these data \citep{wisp2018}. WISP is a
\textit{Python} wrapper for the Common Astronomy Software Applications
package \citep[CASA;][]{casa2007}. Although WISP was developed to
reduce Australia Telescope Compact Array data for the Southern
\hii\ Region Discovery Survey \citep{wenger2019}, its modular
framework can be applied to any radio interferometric dataset. We
follow the \citet{wenger2019} data reduction process, which we briefly
describe here.

\subsection{Calibration}

The WISP calibration pipeline derives calibration solution tables
using the calibrator source data, flags radio frequency interference
(RFI) and other bad data, and applies the calibration solutions to the
science target data. We inspect both the calibration solutions and
calibrated data to assess the quality of the calibration solutions and
to manually flag bad data that was missed by the WISP automatic
flagging routines. The most common issues we flag are (1) antennas
with poor calibration solutions, (2) broad frequency RFI that
contaminates an entire spectral window, and (3) shadowed antennas. In
rare cases, RFI can compromise nearly half of all of our spectral
windows.

\subsection{Imaging}

We use the WISP imaging pipeline to automatically generate and clean
images from the calibrated visibility data. We begin by regridding all
of the data to a common kinematic local standard of rest (LSR)
velocity frame with a channel width \(\Delta v_{\rm LSR} =
1.2\kms\). Using the \textit{TCLEAN} task in \textit{CASA}, we
generate several images and data cubes: (1) a multiscale,
multi-frequency synthesis (MS-MFS) continuum image of the combined
continuum spectral windows, (2) an MS-MFS image of each continuum and
spectral line window, and (3) a multiscale data cube of each spectral
line window. Following the strategy of \citet{wenger2019}, we use
\textit{CLEAN} masks from each spectral line window MS-MFS image to
\textit{CLEAN} the data cube for that spectral window.

Many of our observed \hii\ regions are spatially resolved. We increase
our surface brightness sensitivity to resolved emission by
\textit{uv}-tapering our visibilities when generating images. This
process, however, reduces our point source sensitivity and worsens our
angular resolution. Therefore, we generate both non-tapered and
\textit{uv}-tapered images/data cubes for each field. The latter are
tapered to a synthesized half-power beam width (HPBW) of 15 arcsec,
which is about twice the native VLA resolution at X-band.

\section{Data Analysis} \label{sec:analysis}

The data analysis process for this survey closely follows the
\citet{wenger2019} strategy. Because multiple nebulae may be observed
in a single VLA pointing, we first identify unique \textit{WISE}
Catalog sources in each 8--10 \ghz MS-MFS continuum image. Emission is
associated with the \textit{WISE} Catalog nebulae as long as the peak
continuum brightness pixel is within a circle centered on the
\textit{WISE} Catalog position with a radius equal to the
\textit{WISE} Catalog infrared radius. We manually locate these peak
continuum brightness pixels for each nebula with detected radio
continuum emission.

Unlike \citet{wenger2019}, we wish to derive the total fluxes of
extended sources in addition to their peak fluxes. We use a watershed
segmentation algorithm to identify the pixels associated with the
manually identified continuum peaks in our images and data cubes. This
algorithm considers an image as a three dimensional topological
surface, where the image brightness corresponds to the ``depth'' of
the surface. The algorithm identifies the basins that would be filled
by flooding the surface from a given starting point
\citep[see][]{bertrand2005}. In cases where multiple starting points
will flood the same basin (i.e., in confused fields), the algorithm
divides the basin into separate regions for each flooding
source. Hereafter, we will use ``watershed region'' to describe the
regions identified by the watershed segmentation algorithm.

\begin{figure}
  \centering
  \includegraphics[width=\linewidth]{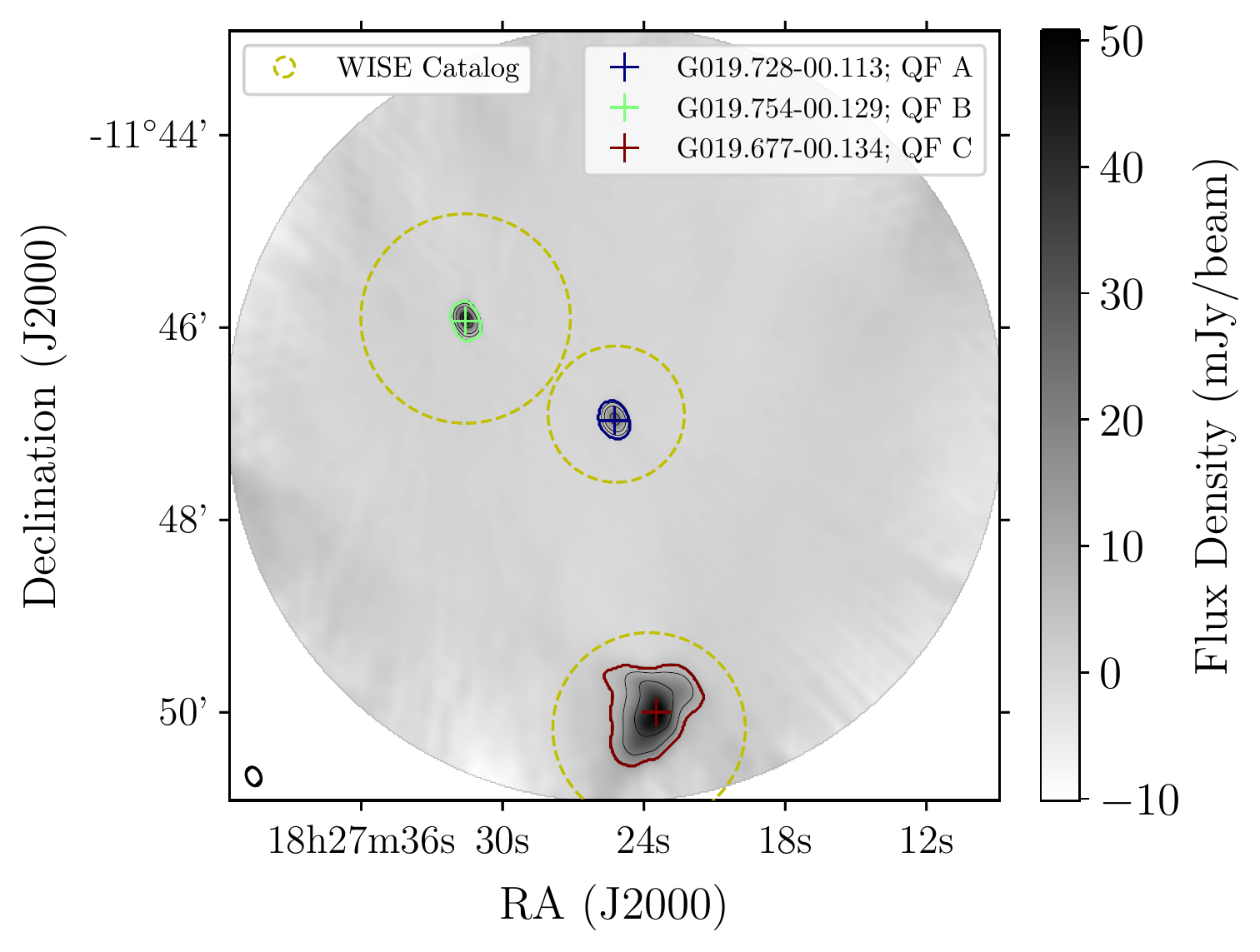}
  \caption{Watershed regions in a \({\sim}2\ghz\) combined MS-MFS
    continuum image. This field is centered on G019.728\(-\)0.113 and
    contains three \textit{WISE} Catalog \hii\ regions. The black
    contours are at 5, 10, 20, and 50 times the spatial rms noise
    (\({\sim}0.6\,\text{mJy beam}^{-1}\) at the field center), and the
    yellow dashed circles represent the position and infrared radii of
    the \textit{WISE} Catalog nebulae. The manually identified peak
    continuum brightness pixels are indicated by the colored plus
    symbols, and the watershed regions by the colored contours. These
    regions were created using the MS-MFS image clipped at 5 times the
    spatial rms noise to avoid including noise spikes in the watershed
    regions. These nebulae are examples of continuum quality factors
    (QF) A, B, and C, as indicated in the legend (see
    Section~\ref{sec:products}).}
  \label{fig:watershed}
\end{figure}

We set the manually identified continuum brightness peak locations as
the flooding sources for the watershed segmentation algorithm. Using
the MS-MFS images clipped at 5 times the spatial rms noise, we run the
algorithm to identify the watershed regions associated with each
continuum source. Figure~\ref{fig:watershed} shows an example region
identified by this algorithm. We use the clipped continuum images to
avoid low-brightness noise spikes in the watershed regions, but, as a
result, we also miss faint emission associated with the
nebulae. Therefore, our total continuum fluxes are systematically
under-estimated, especially for faint sources.

For each continuum source we measure the brightness and total flux at
the location of the peak brightness and within the watershed region,
respectively. The uncertainty of the peak continuum brightness is
derived as the spatial rms of the \textit{CLEAN} residual image
divided by the VLA primary beam response at the peak continuum
brightness position. To compute the uncertainty on the total continuum
flux, we must consider that the spatial noise in an interferometric
image is correlated on the scale of the synthesized beam. The variance
in the sum of the brightnesses of \(N\) pixels within a region is
\begin{equation}
  \sigma_T^2 = \sum_i^N\sum_j^N\rho_{ij}\sigma_i\sigma_j\,, \label{eq:noise}
\end{equation}
where \(\sigma_i\) is the spatial rms of the \textit{CLEAN} residual
image divided by the VLA primary beam response at the position of the
\(i\)th pixel, \(\rho_{ij}\) is the correlation coefficient between
the \(i\)th and \(j\)th pixels, and the sums are taken over all \(N\)
pixels within the region. We use the two-dimensional Gaussian
synthesized beam to define the correlation coefficient:
\begin{equation}
  \rho_{ij} = \exp\left[-A\Delta x^2 - 2B\Delta x\Delta y - C\Delta y^2\right]\,,
\end{equation}
where
\begin{align}
  A & = \frac{\cos^2\phi}{2\theta_{\rm maj}^2} + \frac{\sin^2\phi}{2\theta_{\rm min}^2}\,, \nonumber \\
  B & = -\frac{\sin(2\phi)}{4\theta_{\rm maj}^2} + \frac{\sin(2\phi)}{4\theta_{\rm min}^2}\,, \nonumber \\
  C & = \frac{\sin^2\phi}{2\theta_{\rm maj}^2} + \frac{\cos^2\phi}{2\theta_{\rm min}^2} \,, \nonumber
\end{align}
\(\Delta x\) and \(\Delta y\) are the angular separations between the
\(i\)th and \(j\)th pixels in the east-west and north-south
directions, respectively, and \(\theta_{\rm maj}\), \(\theta_{\rm
  min}\), and \(\phi\) are the synthesized beam major axis, minor
axis, and north-through-east position angle, respectively. In the
simple case where \(\sigma_i \simeq \sigma_j \simeq \sigma\) (i.e.,
the noise is constant across the source), equation~\ref{eq:noise}
reduces to
\begin{equation}
  \sigma_T^2 \simeq \sigma^2\sum_i^N\sum_j^N\rho_{ij} \simeq \sigma^2N_{\rm beam}\,, \label{eq:noise_approx}
\end{equation}
where \(N_{\rm beam}\) is the number of synthesized beams contained
within the region. Many of our sources are extended or located near
the edge of the primary beam, such that the primary beam response and
noise varies across the source. Therefore, we use
equation~\ref{eq:noise} to derive the total continuum flux
uncertainties.

We maximize our sensitivity to the faint RRL emission by averaging
each observed Hn\(\alpha\) RRL transition and both polarizations.
This average spectrum is denoted by \(<\)Hn\(\alpha\)\(>\). For
non-tapered images, we extract spectra from each line spectral window
data cube at the location of the peak continuum brightness. The
\(<\)Hn\(\alpha\)\(>\) spectrum is computed as the weighted average of the
individual RRL transitions. The weights are given by \(w_i =
S_{C,i}/{\rm rms}_i^2\) where \(S_{C,i}\) is the continuum brightness
and rms\(_i\) is the spectral rms noise of the \(i\)th spectral
window, both measured in the line-free region of the spectrum. For
\textit{uv}-tapered images, we spatially smooth the data cubes to a
common beam size, then extract the spectra and compute the
\(<\)Hn\(\alpha\)\(>\) spectrum in the same fashion.

The total RRL emission within the watershed regions is extracted from
the data cubes differently than for the peak position. For each pixel
in the region, we measure the median continuum brightness in the
line-free region of the spectrum, \(S_{C,i}\). Then, we sum each
pixel's spectrum, \(S_{L,i}\), weighted by the median continuum
brightness in that pixel. The final extracted spectrum for this
spectral window is normalized by the ratio of the median non-weighted
sum and median weighted sum:
\begin{equation}
  \begin{split}
    S_{L}(\nu) = \left(\sum_i S_{L,i}(\nu)S_{C,i}\right)\times \\
    \frac{{\rm median}\left(\sum_i S_{L,i}\right)}{{\rm median}\left(\sum_i S_{L,i}(\nu)S_{C,i}\right)}.
  \end{split}
\end{equation}
This complicated procedure correctly weights the final spectrum by the
continuum level in each pixels' spectra, thereby maximizing the
signal-to-noise ratio of the RRL and ensuring that the final spectrum
has the correct flux density. The watershed region
\(<\)Hn\(\alpha\)\(>\) spectrum is then computed using the same
weighted average of the individual RRL transitions as for the peak
positions.

Finally, we measure the \(<\)Hn\(\alpha\)\(>\) RRL properties. We first
identify the line-free regions of the spectrum to estimate the
spectral rms noise and to fit and remove a third-order polynomial
baseline. Then, we fit a Gaussian to the baseline-subtracted spectrum
and measure the RRL brightness, the FWHM line width, and the LSR
velocity.

\section{Results}

\subsection{VLA Data Products} \label{sec:products}

Our goal is to derive an accurate nebular electron temperature for as
many of the observed Galactic \hii\ regions as possible. Given that
some of these nebulae will be extremely faint, spatially resolved,
and/or in confusing fields, no single data analysis method will work
for each nebula. For each source, we therefore employ a suite of
different analysis methods and then we pick the combination of
non-tapered or \textit{uv}-tapered images and peak position
\(<\)Hn\(\alpha\)\(>\) or watershed region \(<\)Hn\(\alpha\)\(>\)
spectra that maximizes our RRL sensitivity and minimizes our electron
temperature uncertainty.

We detect radio continuum emission in 88 (59\%) of the 148 observed
fields. This low detection rate is a result of the relatively poor
surface brightness sensitivity of the VLA. Many of the fields,
however, contain multiple \textit{WISE} Catalog \hii\ regions and/or
\hii\ region candidates. We detect radio continuum emission toward 114
known or candidate \hii\ regions.  Table~\ref{tab:cont_products} lists
the measured radio continuum properties of these nebulae: the
\textit{WISE} Catalog source name; the MS-MFS synthesized frequency of
the combined continuum spectral windows, \(\nu_C\); the peak continuum
flux density, \(S_C^P\); a quality factor for the peak flux density,
QF\(_C^P\); a column indicating whether the peak flux density was
measured using the non-tapered (N) or \textit{uv}-tapered (Y) image;
the total flux density within the watershed region, \(S_C^T\); a
quality factor for the total flux density, QF\(_C^T\); and a column
indicating whether the peak flux density was measured using the
non-tapered or \textit{uv}-tapered image. The MS-MFS synthesized
frequency varies slightly for each field due to differences in data
flagging. We select either non-tapered or \textit{uv}-tapered based on
which gives the smallest fractional uncertainty in the final electron
temperature derivation (if the source also has a RRL detection), or
which has the smallest fractional uncertainty in the continuum flux
density. For resolved nebulae, the \textit{uv}-tapered images
typically have a smaller fractional electron temperature or continuum
flux density uncertainty.

\startlongtable
\begin{deluxetable*}{lcr@{\,\(\pm\)\,}lccr@{\,\(\pm\)\,}lcc}
%\centering
\singlespace
%\tabletypesize{\scriptsize}
\tablewidth{0pt}
\tablecaption{Continuum Data Products\label{tab:cont_products}}
\tablehead{
\colhead{Name} & \colhead{\(\nu_C\)} & \multicolumn{2}{c}{\(S_C^{\rm P}\)} & \colhead{QF\(_C^{\rm P}\)} & \colhead{Taper\(^{\rm P}\)\tablenotemark{a}} & \multicolumn{2}{c}{\(S_C^{\rm T}\)} & \colhead{QF\(_C^{\rm T}\)} & \colhead{Taper\(^{\rm P}\)\tablenotemark{a}} \\
\colhead{}     & \colhead{(MHz)}     & \multicolumn{2}{c}{(mJy beam\(^{-1}\))} & \colhead{}                 & \colhead{}                  & \multicolumn{2}{c}{(mJy)}           & \colhead{}                 & \colhead{}                  
}
\startdata
G005.885$-$00.393 & 8962.2 & 4516.01 & 13.31 & A & N & 5254.49 & 35.45 & A & N \\ 
G010.596$-$00.381 & 8962.2 & 395.02 & 6.66 & A & Y & 907.08 & 18.66 & A & Y \\ 
G013.880+00.285 & 8962.2 & 1696.64 & 3.26 & A & Y & 3368.06 & 10.64 & B & N \\ 
G017.336$-$00.146 & 8962.1 & 10.91 & 0.29 & B & Y & 51.38 & 0.84 & B & N \\ 
G017.928$-$00.677 & 8962.1 & 14.48 & 0.37 & B & Y & 57.76 & 1.07 & B & N \\ 
G018.584+00.344 & 8962.1 & 22.53 & 0.82 & A & Y & 46.79 & 1.20 & A & N \\ 
G018.630+00.309 & 8962.1 & 13.01 & 4.37 & C & Y & 0.04 & 0.18 & C & N \\ 
G019.677$-$00.134 & 8962.1 & 163.19 & 3.36 & C & Y & 469.63 & 7.10 & C & N \\ 
G019.728$-$00.113 & 8962.1 & 24.23 & 0.40 & A & N & 27.24 & 0.68 & A & N \\ 
G019.754$-$00.129 & 8962.1 & 46.45 & 0.59 & B & N & 45.73 & 1.01 & B & N \\ 
G020.227+00.110 & 8962.1 & 8.61 & 0.13 & B & Y & 41.72 & 0.36 & B & N \\ 
G020.363$-$00.014 & 8962.1 & 50.30 & 0.09 & A & N & 58.28 & 0.23 & A & N \\ 
G020.387$-$00.018 & 8962.1 & 8.58 & 0.16 & B & Y & 26.85 & 0.46 & B & Y \\ 
G021.386$-$00.255 & 8962.1 & 122.94 & 0.12 & A & N & 136.88 & 0.45 & A & N \\ 
G021.596$-$00.161 & 8962.2 & 5.70 & 0.16 & A & N & 6.77 & 0.26 & A & N \\ 
G021.603$-$00.169 & 8962.2 & 19.32 & 0.15 & A & N & 27.62 & 0.34 & A & N \\ 
G023.661$-$00.252 & 8962.2 & 30.11 & 0.46 & B & Y & 152.51 & 1.16 & B & N \\ 
G024.153+00.163 & 8962.2 & 10.94 & 1.62 & C & N & 4.60 & 1.14 & C & N \\ 
G024.166+00.250 & 8962.2 & 16.56 & 0.79 & B & N & 17.44 & 1.11 & B & N \\ 
G024.195+00.242 & 8962.2 & 9.77 & 0.57 & B & N & 47.78 & 1.69 & B & N \\ 
G024.713$-$00.125 & 8962.2 & 32.51 & 2.11 & C & N & 138.58 & 5.73 & C & N \\ 
G025.397+00.033 & 8962.2 & 229.49 & 0.56 & B & N & 494.08 & 2.57 & B & N \\ 
G025.398+00.562 & 8962.1 & 203.74 & 0.36 & A & Y & 221.10 & 1.21 & A & N \\ 
G025.401+00.021 & 8962.2 & 54.54 & 0.60 & B & N & 150.97 & 1.87 & B & N \\ 
G027.562+00.084 & 8898.2 & 47.71 & 0.23 & A & N & 111.74 & 0.71 & A & N \\ 
G028.320+01.243 & 8962.1 & 21.17 & 0.04 & A & N & 30.22 & 0.11 & A & N \\ 
G028.438+00.014 & 8962.2 & 4.01 & 0.35 & A & N & 11.73 & 0.74 & A & N \\ 
G028.451+00.001 & 8962.2 & 36.09 & 0.30 & A & N & 84.81 & 1.30 & A & N \\ 
G028.581+00.145 & 8962.2 & 25.87 & 0.18 & A & N & 39.68 & 0.41 & A & N \\ 
G029.770+00.219 & 8962.2 & 35.40 & 0.16 & A & N & 72.53 & 0.45 & A & N \\ 
G029.956$-$00.020 & 8962.2 & 1770.38 & 4.48 & A & N & 4299.65 & 22.54 & A & N \\ 
G030.211+00.428 & 8962.2 & 15.61 & 0.04 & A & N & 25.81 & 0.12 & A & N \\ 
G031.269+00.064 & 8962.2 & 2.70 & 0.34 & A & N & 1.00 & 0.22 & A & N \\ 
G031.279+00.061 & 8962.2 & 125.29 & 0.35 & A & N & 306.72 & 1.15 & A & N \\ 
G031.580+00.074 & 8962.2 & 13.41 & 0.21 & B & N & 15.17 & 0.38 & B & N \\ 
G032.030+00.048 & 8962.2 & 17.10 & 0.19 & A & N & 25.83 & 0.40 & A & N \\ 
G032.057+00.077 & 8962.2 & 13.36 & 1.03 & C & Y & 94.17 & 2.13 & C & N \\ 
G032.272$-$00.226 & 8962.2 & 147.87 & 0.18 & A & N & 330.84 & 0.75 & A & N \\ 
G032.928+00.606 & 8898.2 & 173.86 & 0.27 & A & N & 336.64 & 1.38 & A & N \\ 
G033.643$-$00.229 & 8962.2 & 6.37 & 0.09 & A & Y & 10.85 & 0.15 & A & N \\ 
G034.041+00.052 & 8962.2 & 25.97 & 0.48 & A & Y & 83.27 & 1.08 & A & N \\ 
G034.089+00.438 & 8962.2 & 34.42 & 2.87 & C & N & 83.68 & 7.38 & C & Y \\ 
G034.133+00.471 & 8962.2 & 378.58 & 1.10 & A & Y & 517.00 & 2.35 & A & N \\ 
G034.686+00.068 & 8962.2 & 55.42 & 0.60 & A & Y & 107.24 & 0.95 & A & N \\ 
G035.126$-$00.755 & 8962.2 & 123.85 & 0.43 & A & Y & 241.91 & 0.71 & A & N \\ 
G035.948$-$00.149 & 8962.2 & 12.05 & 0.03 & A & N & 26.66 & 0.08 & A & N \\ 
G036.870+00.462 & 8962.2 & 3.03 & 0.31 & C & N & 9.77 & 0.67 & C & N \\ 
G036.877+00.498 & 8962.2 & 1.22 & 0.17 & C & N & 1.90 & 0.22 & C & N \\ 
G036.918+00.482 & 8962.2 & 6.21 & 0.08 & A & N & 7.56 & 0.15 & A & N \\ 
G038.550+00.163 & 8962.2 & 54.11 & 0.25 & A & N & 122.86 & 0.77 & A & N \\ 
G038.643$-$00.227 & 8962.3 & 18.70 & 0.05 & A & N & 24.79 & 0.18 & A & N \\ 
G038.652+00.087 & 8962.2 & 19.77 & 0.24 & A & N & 49.91 & 0.96 & A & N \\ 
G038.840+00.495 & 8962.2 & 4.49 & 0.09 & B & N & 84.27 & 0.66 & B & N \\ 
G038.875+00.308 & 8962.2 & 279.04 & 0.43 & A & N & 320.84 & 1.04 & A & N \\ 
G039.183$-$01.422 & 8962.3 & 20.75 & 0.15 & A & Y & 57.82 & 0.30 & A & N \\ 
G039.196+00.224 & 8962.3 & 62.54 & 0.06 & A & N & 67.11 & 0.19 & A & N \\ 
G039.213+00.202 & 8962.3 & 5.13 & 0.09 & B & N & 5.85 & 0.16 & B & N \\ 
G039.864+00.645 & 8962.3 & 67.52 & 0.51 & A & Y & 103.48 & 0.79 & A & N \\ 
G043.146+00.013 & 8962.3 & 1434.45 & 126.49 & B & Y & 1427.47 & 158.00 & B & Y \\ 
G043.165$-$00.031 & 8962.3 & 2330.17 & 78.58 & C & N & 3341.55 & 144.37 & C & N \\ 
G043.168+00.019 & 8962.3 & 332.86 & 17.06 & B & N & 600.24 & 31.78 & B & N \\ 
G043.170$-$00.004 & 8962.3 & 4331.44 & 149.88 & B & Y & 11158.53 & 398.42 & B & Y \\ 
G043.432+00.516 & 8962.3 & 11.08 & 0.35 & B & Y & 82.31 & 0.83 & B & N \\ 
G043.523$-$00.648 & 8962.3 & 5.84 & 0.04 & A & Y & 13.22 & 0.09 & A & N \\ 
G043.818+00.395 & 8962.3 & 21.54 & 0.97 & B & Y & 94.69 & 1.89 & B & N \\ 
G043.968+00.993 & 8962.2 & 47.26 & 0.07 & A & N & 49.81 & 0.20 & A & N \\ 
G043.999+00.978 & 8962.2 & 22.23 & 0.22 & C & N & 25.05 & 0.42 & C & N \\ 
G044.501+00.332 & 8962.3 & 21.92 & 1.24 & B & Y & 135.68 & 2.23 & B & N \\ 
G044.503+00.349 & 8962.3 & 7.16 & 0.36 & A & N & 8.58 & 0.54 & A & N \\ 
G045.197+00.740 & 8962.3 & 7.76 & 0.18 & B & N & 140.36 & 1.42 & B & N \\ 
G048.719+01.147 & 8962.4 & 37.12 & 0.10 & A & Y & 69.80 & 0.29 & A & N \\ 
G049.399$-$00.490 & 8962.4 & 166.98 & 7.12 & A & Y & 232.47 & 8.75 & A & N \\ 
G052.098+01.042 & 8962.3 & 287.77 & 0.50 & A & Y & 432.07 & 0.89 & A & N \\ 
G052.232+00.735 & 8962.4 & 68.65 & 4.32 & C & Y & 162.56 & 5.42 & C & N \\ 
G054.093+01.748 & 8962.3 & 18.84 & 0.03 & A & Y & 34.60 & 0.08 & A & Y \\ 
G054.490+01.579 & 8962.3 & 24.30 & 0.06 & A & Y & 44.24 & 0.13 & A & N \\ 
G054.543+01.560 & 8962.3 & 3.73 & 0.26 & C & Y & 3.52 & 0.21 & C & N \\ 
G055.114+02.422 & 8962.3 & 138.55 & 1.25 & A & Y & 618.76 & 2.79 & B & N \\ 
G060.592+01.572 & 8962.3 & 55.66 & 0.22 & A & Y & 166.35 & 0.51 & A & N \\ 
G061.720+00.863 & 8962.7 & 90.28 & 0.16 & A & N & 97.51 & 0.38 & A & N \\ 
G062.577+02.389 & 8962.7 & 51.31 & 0.28 & B & N & 359.10 & 1.71 & B & N \\ 
G068.144+00.915 & 8962.7 & 42.25 & 1.61 & B & Y & 302.02 & 4.24 & B & N \\ 
G070.280+01.583 & 8962.6 & 542.61 & 15.70 & A & Y & 1930.78 & 37.24 & A & N \\ 
G070.293+01.599 & 8962.6 & 3550.27 & 9.03 & A & N & 5690.67 & 39.20 & A & N \\ 
G070.304+01.595 & 8962.6 & 245.05 & 9.37 & A & N & 1829.40 & 41.28 & A & N \\ 
G070.329+01.589 & 8962.6 & 1067.39 & 23.78 & B & N & 2670.68 & 65.51 & B & N \\ 
G070.673+01.190 & 8962.6 & 260.40 & 0.72 & A & Y & 407.26 & 1.43 & A & N \\ 
G070.765+01.820 & 8962.6 & 28.79 & 0.52 & A & Y & 173.85 & 1.36 & B & N \\ 
G071.150+00.397 & 8962.7 & 208.43 & 0.25 & A & N & 392.62 & 0.92 & A & N \\ 
G073.878+01.023 & 8962.6 & 75.77 & 0.09 & A & N & 120.61 & 0.26 & A & N \\ 
G074.155+01.646 & 8962.6 & 10.25 & 0.04 & A & N & 37.97 & 0.21 & A & N \\ 
G074.753+00.912 & 8962.6 & 55.70 & 0.07 & A & N & 74.69 & 0.20 & A & N \\ 
G075.768+00.344 & 8962.6 & 1059.58 & 10.20 & A & Y & 4104.53 & 20.67 & B & N \\ 
G078.114$-$00.550 & 8962.6 & 14.17 & 3.03 & C & Y & 0.04 & 0.10 & C & N \\ 
G078.174$-$00.550 & 8962.6 & 4.21 & 0.19 & B & N & 23.01 & 0.72 & B & N \\ 
G078.886+00.709 & 8962.6 & 83.02 & 0.08 & A & N & 110.14 & 0.25 & A & N \\ 
G080.191+00.534 & 8962.6 & 5.14 & 0.09 & A & N & 40.72 & 0.45 & A & N \\ 
G094.263$-$00.414 & 8963.1 & 4.40 & 0.04 & B & Y & 18.73 & 0.14 & B & N \\ 
G096.289+02.593 & 8963.1 & 27.93 & 0.32 & B & N & 442.24 & 2.68 & B & N \\ 
G096.434+01.324 & 8963.1 & 23.34 & 0.10 & A & N & 36.94 & 0.25 & A & N \\ 
G097.515+03.173 & 8963.1 & 131.05 & 0.65 & A & Y & 508.47 & 1.73 & B & N \\ 
G097.528+03.184 & 8963.1 & 41.23 & 0.29 & A & N & 49.32 & 0.55 & A & N \\ 
G101.016+02.590 & 8963.0 & 17.71 & 0.06 & A & Y & 21.24 & 0.14 & A & N \\ 
G104.700+02.784 & 8963.0 & 9.00 & 0.17 & A & Y & 39.99 & 0.36 & A & N \\ 
G109.104$-$00.347 & 8963.0 & 7.10 & 0.07 & A & N & 19.36 & 0.20 & A & N \\ 
G124.637+02.535 & 8963.4 & 252.56 & 0.20 & A & N & 293.16 & 0.62 & A & N \\ 
G125.092+00.778 & 8963.5 & 6.70 & 0.02 & A & Y & 20.65 & 0.07 & B & N \\ 
G135.188+02.701 & 8963.4 & 19.82 & 0.09 & A & Y & 65.47 & 0.18 & B & N \\ 
G141.084$-$01.063 & 8963.8 & 12.17 & 0.19 & A & Y & 62.35 & 0.37 & B & N \\ 
G150.859$-$01.115 & 8963.8 & 11.73 & 0.10 & A & Y & 18.02 & 0.15 & A & N \\ 
G196.448$-$01.673 & 8964.0 & 10.93 & 0.47 & B & N & 350.13 & 4.13 & B & N \\ 
G218.737+01.850 & 8964.1 & 202.41 & 0.69 & A & Y & 554.43 & 1.73 & A & N \\ 
G351.246+00.673 & 8962.2 & 7191.06 & 24.43 & A & Y & 11722.85 & 66.92 & A & N \\ 
G351.311+00.663 & 8962.2 & 2809.89 & 29.27 & A & Y & 5561.39 & 64.30 & A & N \\ 
\enddata
\tablenotetext{a}{"N" if non-tapered image measurement; "Y" if \textit{uv}-tapered image measurement}
\end{deluxetable*}

The quality factor (QF) is a qualitative assessment of the accuracy of
the continuum flux measurement. QF A detections are isolated,
unresolved, and near the center of the primary beam, QF B detections
are slightly resolved, in crowded fields, and/or are located
off-center from the primary beam, QF C detections are well-resolved,
in very crowded fields, and/or are located near the edge of the
primary beam.  Any continuum sources that are confused/blended are
assigned QF D. These nebulae are excluded from the tables and all
subsequent analysis since we are unable to measure their continuum
fluxes accurately. The three nebulae in Figure~\ref{fig:watershed} are
examples of each continuum QF: G019.728\(-\)00.113 is a QF A
detection, G019.754\(-\)00.129 is a QF B detection because it is
off-center, and G019.677\(-\)00.134 is a QF C detection because it is
resolved and near the edge of the primary beam.

\begin{figure*}
  \centering
  \includegraphics[width=0.49\linewidth]{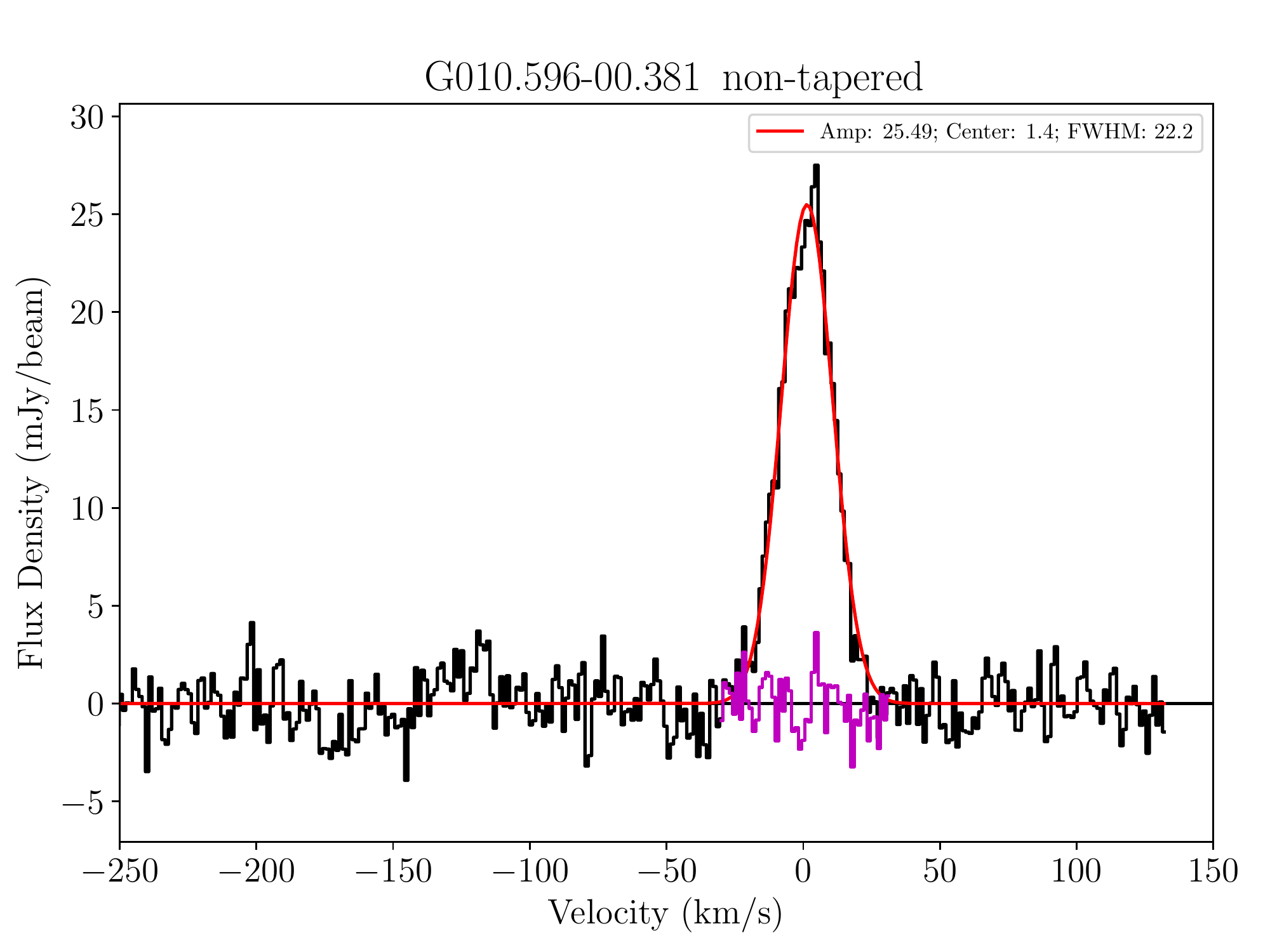}
  \includegraphics[width=0.49\linewidth]{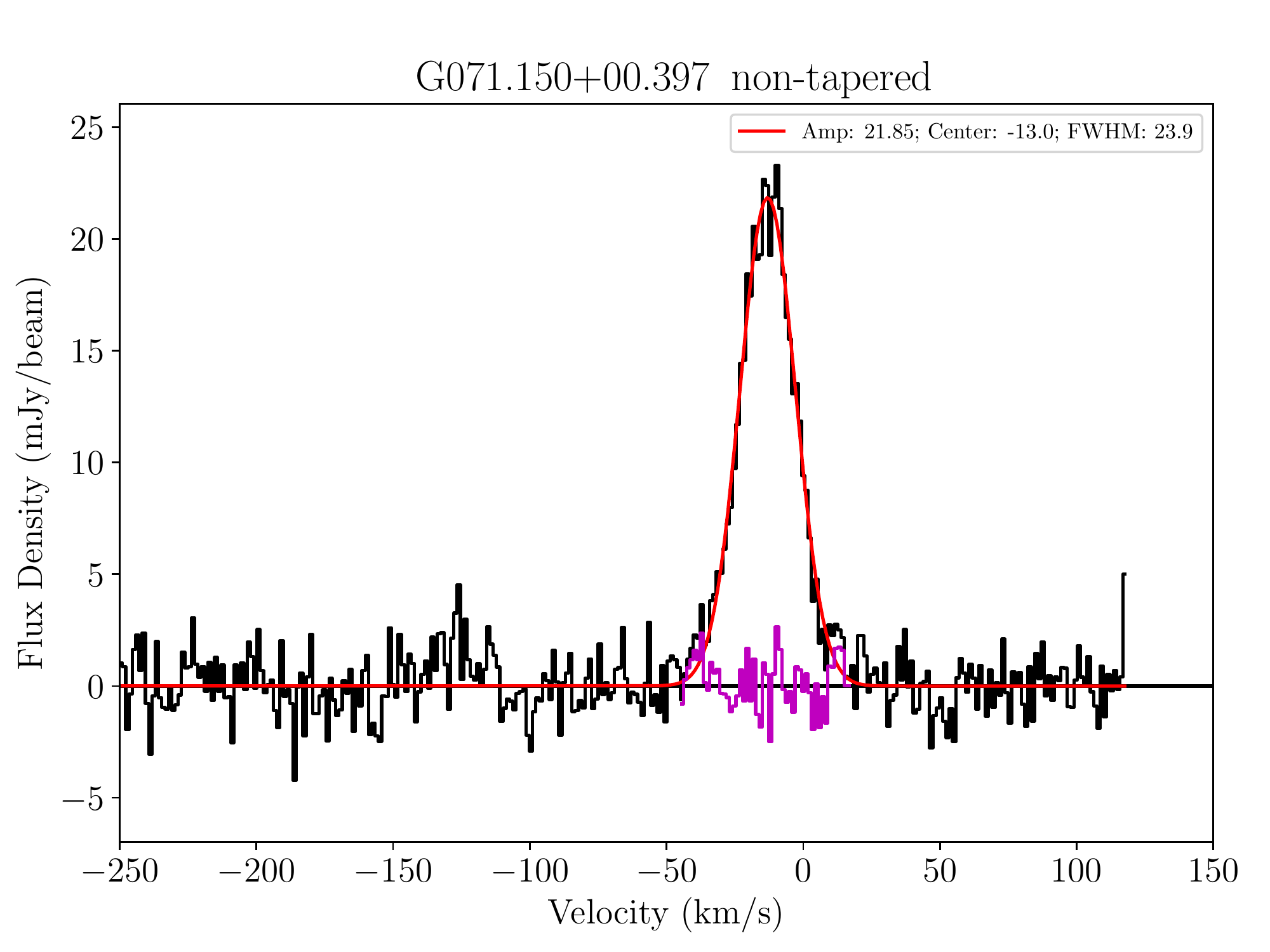} \\
  \includegraphics[width=0.49\linewidth]{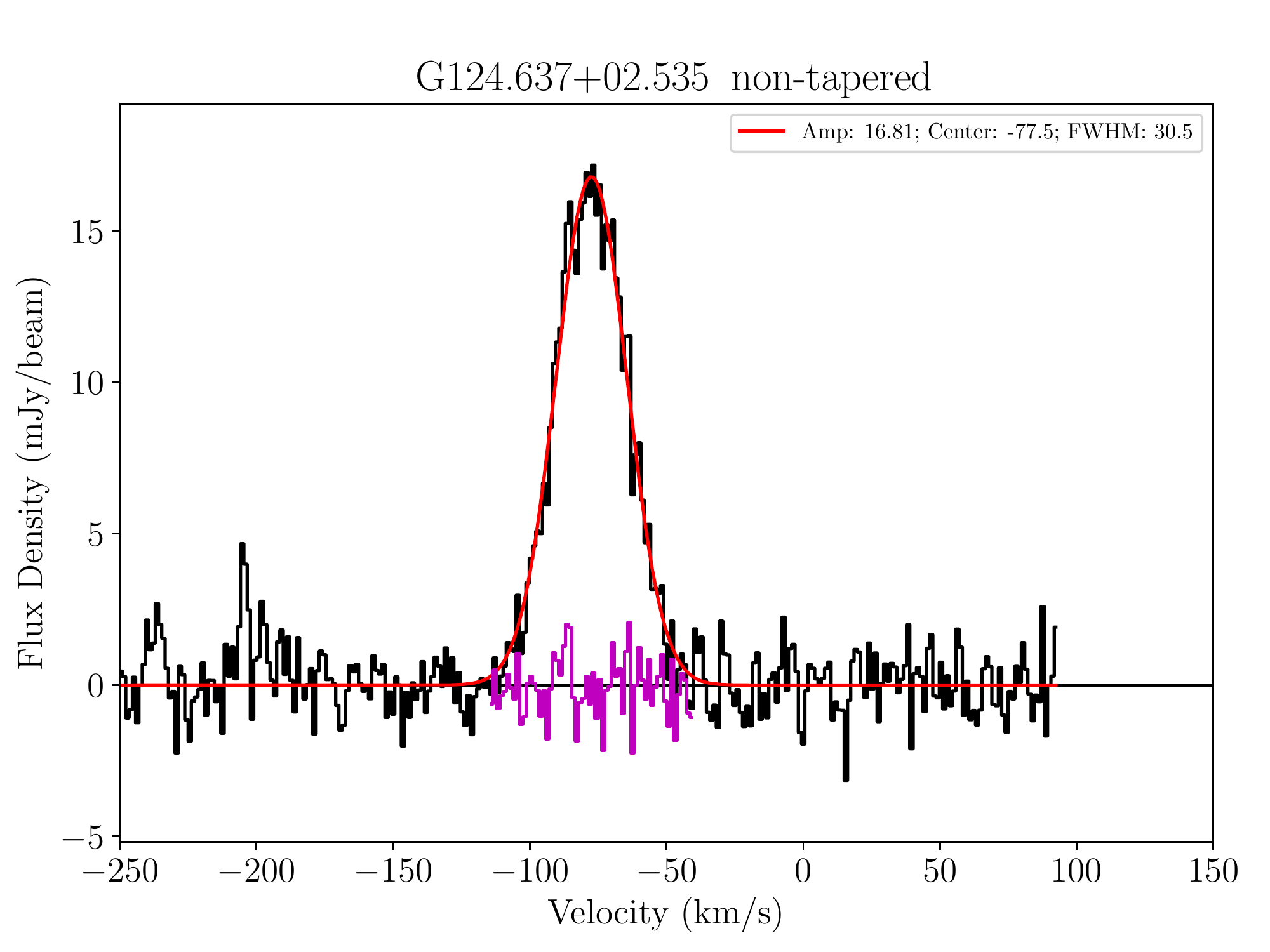}
  \includegraphics[width=0.49\linewidth]{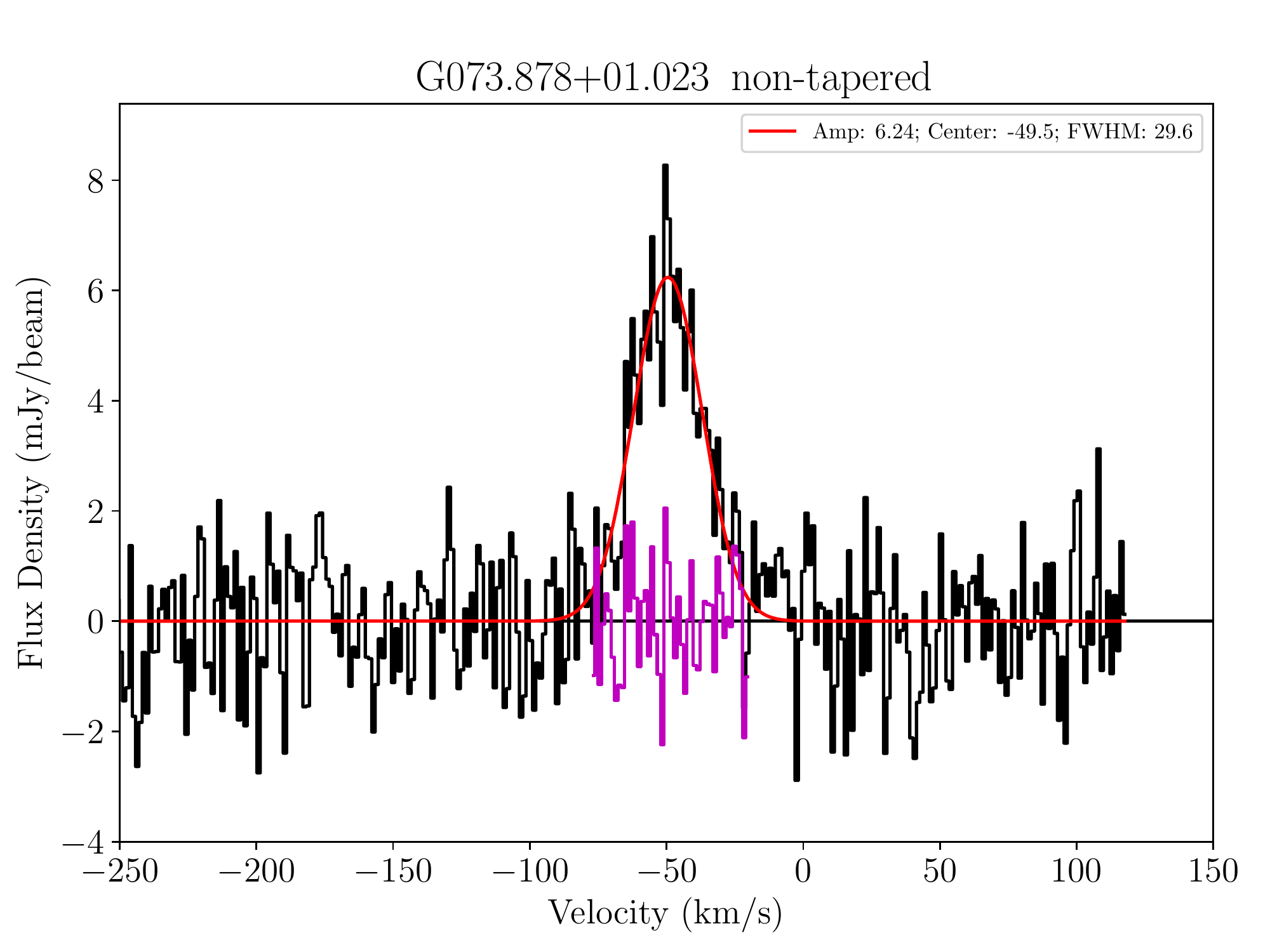} \\
  \caption{Representative \(<\)Hn\(\alpha\)\(>\) stacked spectra. The
    spectra for G010.596$-$00.381 (top-left), G071.150+00.397
    (top-right), G124.637+02.535 (bottom-left), and G073.878+01.023
    (bottom-right) span the range of typical RRL detection
    signal-to-noise ratios. The black histogram is the data, the red
    curve is the Gaussian fit with parameters listed in the legend,
    and the magenta curve is the fit residuals. These spectra were
    extracted from the non-tapered data cubes at the location of the
    peak continuum brightness.}
  \label{fig:rrls}
\end{figure*}

We detect \(<\)Hn\(\alpha\)\(>\) RRL emission toward 82 (72\%) of our
114 continuum sources. All RRL detections are toward previously-known
\hii\ regions.  Figure~\ref{fig:rrls} shows representative
\(<\)Hn\(\alpha\)\(>\) RRL detections with different signal-to-noise
ratios. Our typical spectral rms noise is \({\sim}1\,\text{mJy
  beam\(^{-1}\)}\), about three times greater than what we estimated
using the VLA sensitivity calculator. This decease in sensitivity is
likely due to RFI that compromised entire spectral line spectral
windows. We may be able to further increase our spectral line
sensitivity by self-calibration.

Table~\ref{tab:line_products} lists the measured
\(<\)Hn\(\alpha\)\(>\) RRL properties of our detections: the
\textit{WISE} Catalog source name; the weighted average frequency of
the \(<\)Hn\(\alpha\)\(>\) spectrum, \(\nu_L\), where the weights are
the same as those used to average the individual RRL transitions (see
Section~\ref{sec:analysis}); the amplitude of the Gaussian fit to the
spectrum extracted from the location of peak continuum brightness,
\(S_L^P\); the spectral rms at this position, rms\(^P\); the center
LSR velocity of the fitted Gaussian, \(v_{\rm LSR}^P\); the FWHM line
width of the fitted Gaussian, \(\Delta V^P\); a column indicating
whether the spectrum was extracted from the non-tapered (N) or
\textit{uv}-tapered (Y) image; the amplitude of the Gaussian fit to
the spectrum summed within the watershed region, \(S_L^T\); the
spectral rms in this region, rms\(^T\); the center LSR velocity of the
fitted Gaussian, \(v_{\rm LSR}^T\); the FWHM line width of the fitted
Gaussian, \(\Delta V^T\); and a column indicating whether the spectrum
was extracted from the non-tapered or \textit{uv}-tapered image. As
before, we use either the non-tapered or \textit{uv}-tapered image
depending on which gives the smallest fractional uncertainty in the
derived electron temperature. Unlike B15, we do not assign quality
factors to our RRL detections. Our spectral baselines are always flat
and well-modeled by a third-order polynomial, therefore no qualitative
assessment is necessary. Two nebulae, G005.885$-$00.393 and
G070.293+01.599, are excluded from Table~\ref{tab:line_products}
because they have blended, non-Gaussian line profiles.

\subsection{Electron Temperatures}

Thermal bremsstrahlung (free-free) emission is the primary source of
\hii\ region radio continuum emission. Its intensity depends on the
plasma electron temperature, the plasma electron density, and the
stellar ionizing photon rate. The free-free opacity of an \hii\ region
in LTE is well-approximated by
\begin{equation}
  \begin{split}
    \tau_C \simeq 3.28\times10^{-7}\left(\frac{T_e}{10^4\,\text{K}}\right)^{-1.35}\left(\frac{\nu}{\text{GHz}}\right)^{-2.1}\times \\
    \left(\frac{\text{EM}}{\text{pc cm\(^{-6}\)}}\right)
  \end{split}
\end{equation}
where \(T_e\) is the plasma electron temperature, EM is the emission
measure, and \(\nu\) is the frequency \citep{mezger1967a}. The
emission measure is the integral of the squared electron number
density, \(n_e^2\), along the line of sight path through the nebula:
\(\text{EM} = \int n_e^2\,dl\).  An optically thin \hii\ region has a
continuum brightness temperature \(T_C \simeq \tau_CT_e\).  Without an
independent determination of the emission measure, we are unable to
use the continuum emission alone to derive the nebular electron
temperature.

The RRL intensity and line width reveal the physical characteristics
of an \hii\ region. The line center opacity of an \hii\ region in LTE
is approximated by
\begin{equation}
  \tau_L \simeq 1.92\times10^3\left(\frac{T_e}{\text{K}}\right)^{-2.5}\left(\frac{\text{EM}}{\text{pc cm\(^{-6}\)}}\right)\left(\frac{\Delta \nu}{\text{kHz}}\right)^{-1}
\end{equation}
where \(\Delta \nu\) is the full-width half-maximum (FWHM) line width
in frequency units \citep{kardashev1959,mezger1967b}. Similar to the
continuum, we need an independent measurement of the emission measure
in order to use the RRL properties to derive the electron
temperature.

The typical hydrogen RRL line width for Galactic \hii\ regions is
\({\sim}25\kms\) \citep{wenger2019}. There are four physical effects
that contribute to the RRL FWHM line width: (1) intrinsic broadening,
due to the uncertainty principle; (2) collisional broadening, due to
the collisions of the emitting atoms; (3) thermal Doppler broadening,
due to the Maxwellian velocity distribution of emitting atoms in the
plasma; and (4) non-thermal Doppler broadening.  Of these, thermal and
non-thermal Doppler broadening are the most significant contributors
to the width of RRLs.  The non-thermal (i.e., turbulent) components
can only be constrained with additional information. RRL line width
measurements for nebular plasma atoms other than hydrogen are needed,
since atoms with different masses have different Maxwellian velocity
distributions. Alternatively, the thermal contribution to the RRL line
width can be determined by deriving the plasma temperature.

\begin{figure}
  \centering
  \includegraphics[width=\linewidth]{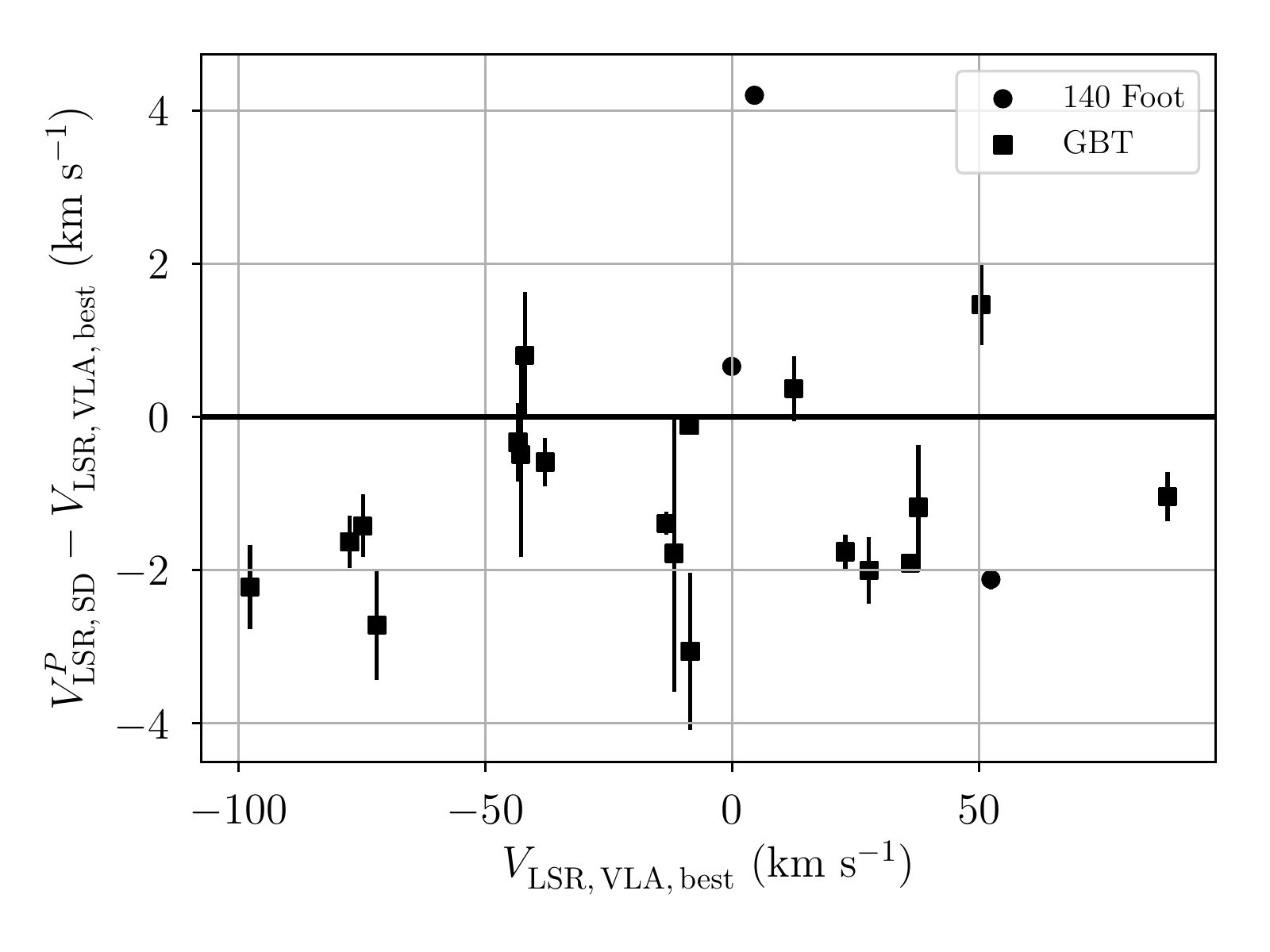}
  \includegraphics[width=\linewidth]{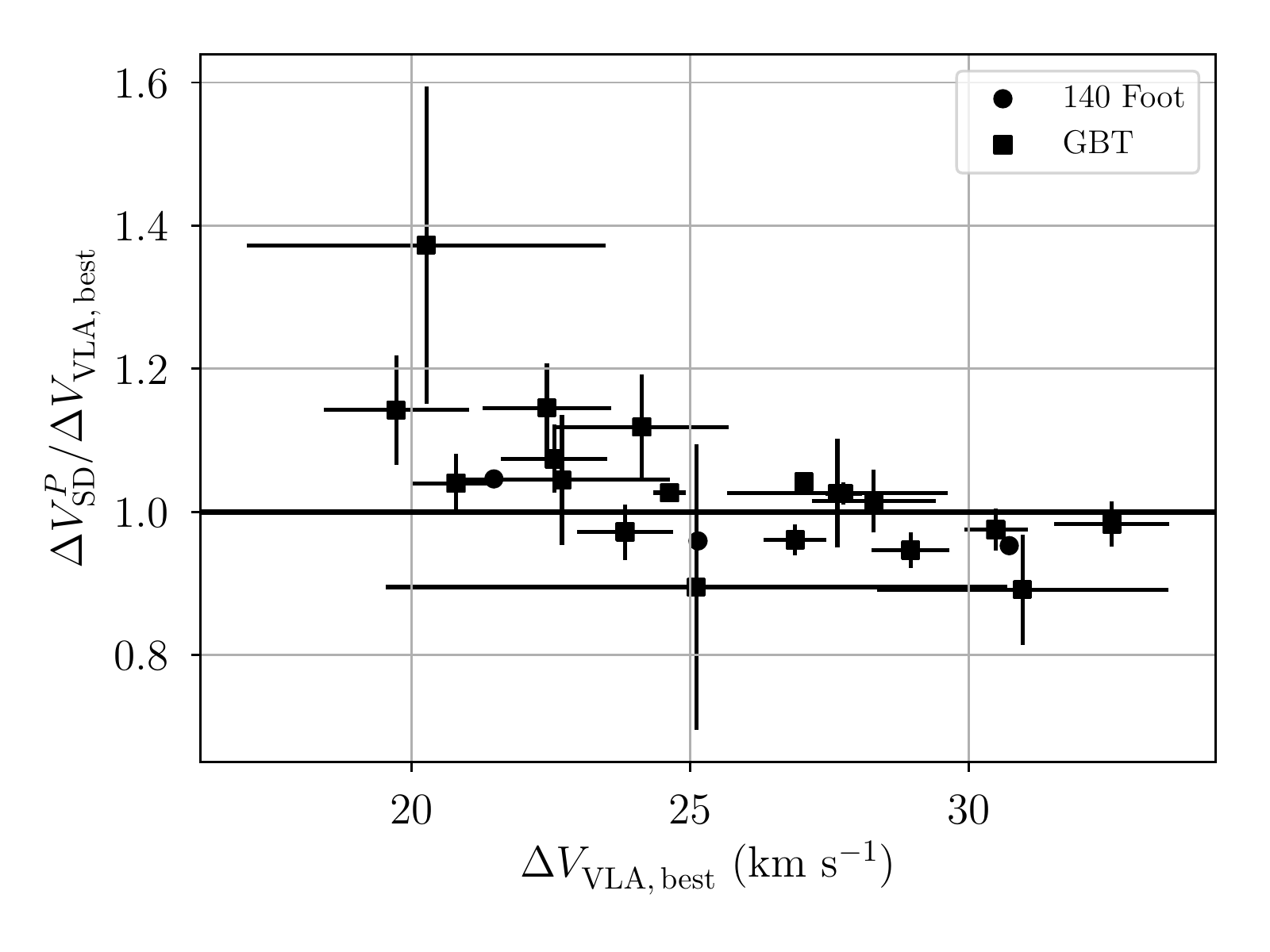}
  \caption{Difference between single dish and VLA RRL LSR velocities
    (top) and ratio of single dish to VLA RRL FWHM line widths
    (bottom) as a function of the VLA values for 22 nebulae also
    observed by the GBT (squares) or 140 Foot (circles). We use the
    ``best'' VLA images and spectral extraction technique, which
    minimizes the fractional uncertainty of the derived electron
    temperature. The weighted mean LSR velocity difference is
    \(-0.09\pm0.34\kms\) and the weighted mean FWHM line width ratio
    is \(0.99\pm0.02\), where the weights are the reciprocal variances
    in the differences or ratios derived from the fitted Gaussian
    parameter uncertainties.}
  \label{fig:rrl_compare}
\end{figure}

\begin{figure}
  \centering
  \includegraphics[width=\linewidth]{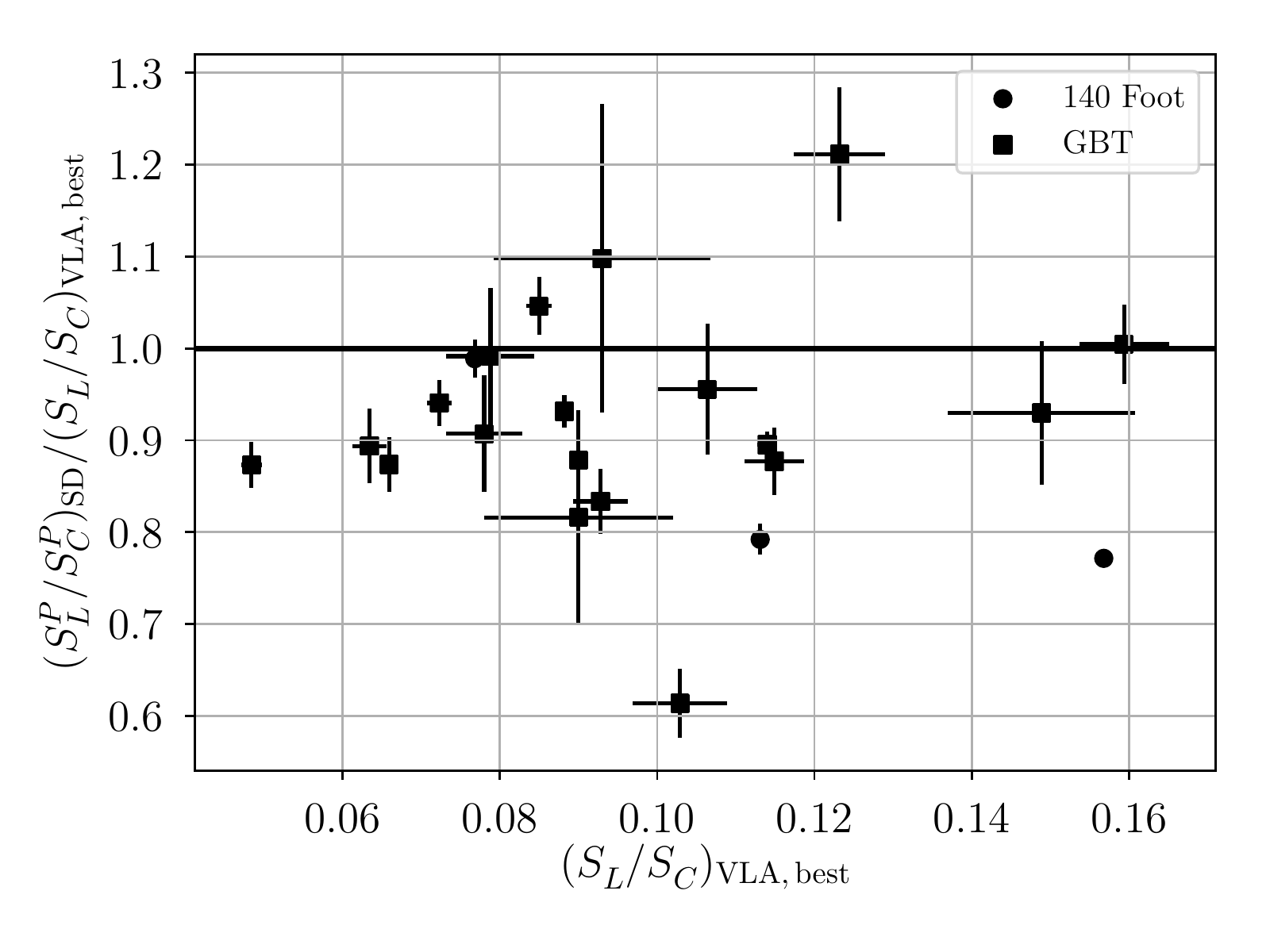}
  \includegraphics[width=\linewidth]{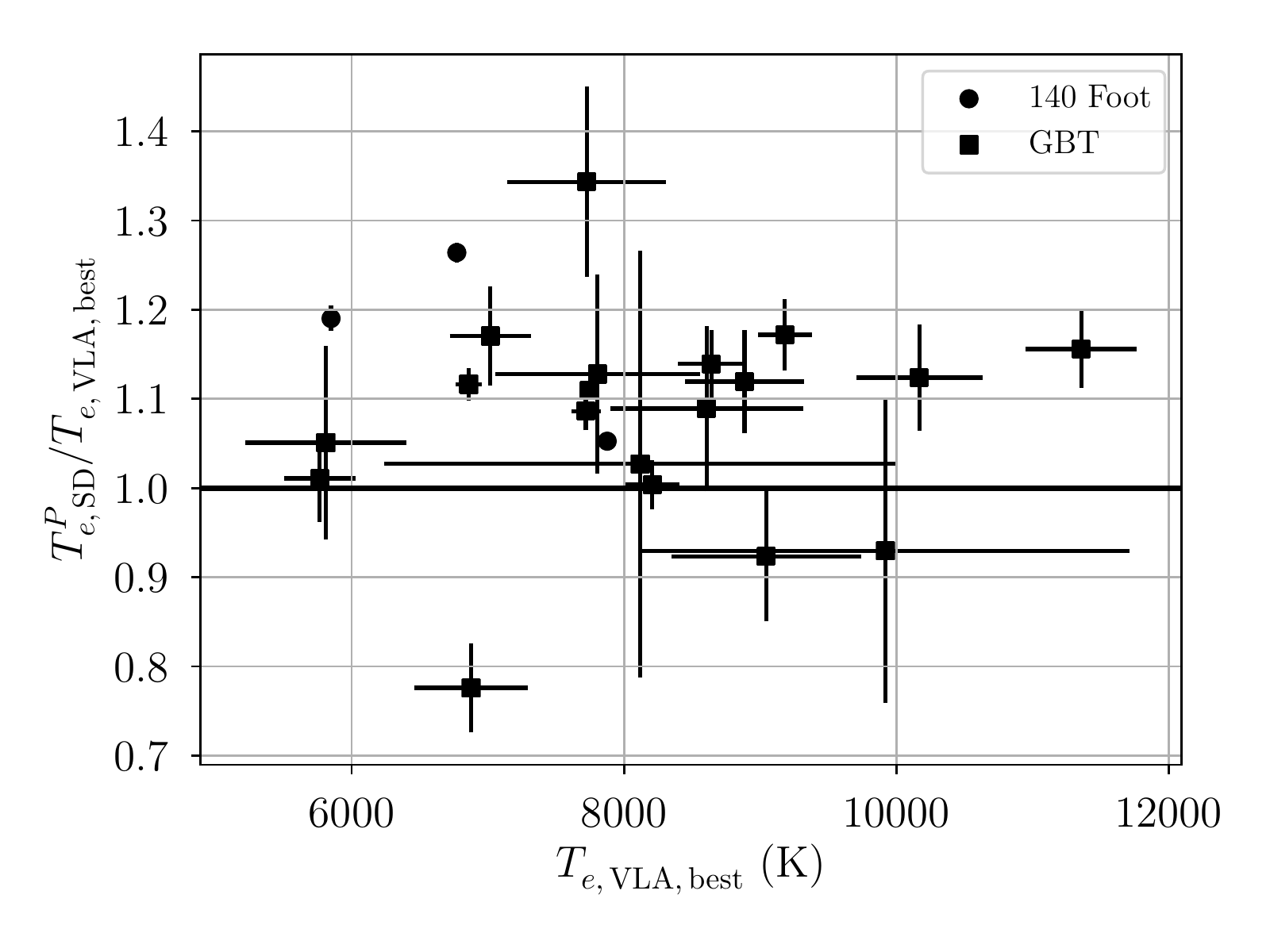}
  \caption{Ratio of single dish to VLA RRL-to-continuum brightness
    ratios (top) and electron temperatures (bottom) as a function of
    the VLA values for the same nebulae as in
    Figure~\ref{fig:rrl_compare}. The weighted mean ratio of the
    single dish and VLA RRL-to-continuum brightness ratios is
    \(0.86\pm0.03\) and the weighted mean electron temperature ratio
    is \(1.12\pm0.03\), where the weights are the reciprocal variances
    in the ratios derived from the fitted Gaussian parameter
    uncertainties.}
  \label{fig:te_compare}
\end{figure}

We derive the nebular electron temperature from the RRL-to-continuum
brightness ratio. For an \hii\ region in LTE that is optically thin to
both continuum and RRL emission, the ratio of the radio continuum
brightness temperature to the RRL peak brightness temperature is equal
to the ratio of the continuum opacity to the line center opacity. This
ratio is independent of the emission measure. A complete derivation of
the electron temperature equation is in
Appendix~\ref{sec:app_electron_temperature}. For RRLs near
H90\(\alpha\), assuming the continuum and RRL emission originate in
the same volume of gas, we find
\begin{equation}
  \begin{split}
    \frac{T_e}{\rm K} \simeq \left[7.100\times10^3\left(\frac{\nu_L}{\rm GHz}\right)^{1.1} \left(\frac{S_C}{S_L}\right)\times \right. \\
      \left. \left(\frac{\Delta V}{\text{km s\(^{-1}\)}}\right)^{-1}\left(1 + y^+\right)^{-1}\right]^{0.87} \label{eq:te}
  \end{split}
\end{equation}
where \(\nu_L\) is the RRL frequency, \(S_C\) is the continuum flux
density, \(S_L\) is the RRL center flux density, \(\Delta V\) is the
RRL FWHM line width in velocity units, and \(y^+\) is the ratio of the
number density of singly ionized helium to hydrogen.

We use Equation~\ref{eq:te} to derive the electron temperatures of the
72 nebulae in our sample with a VLA \(<\)Hn\(\alpha\)\(>\) RRL
detection and a continuum quality factor A, B, or C. We only detect
helium RRLs in a few, bright sources, so we assume \(y^+ = 0.08\) for
all VLA detections, following \citet{balser2011} and B15.
Equation~\ref{eq:te} is only weakly dependent on \(y^+\). A 10\%
increase from \(y^+ = 0.08\) results in a mere \(0.6\%\) increase in
\(T_e\). We do not consider uncertainties in \(y^+\) in the subsequent
analyses because the electron temperature uncertainties ares typically
much greater than \(0.6\%\).  Furthermore, we assume non-LTE effects
and collisional broadening are negligible at these frequencies
\citep[see][]{balser1999}. The RRL flux density, RRL FWHM line width,
and continuum flux density are measured in the \(<\)Hn\(\alpha\)\(>\)
stacked spectrum, and the RRL frequency is the weighted average
frequency of the individual RRL transitions.  Again, the frequency
weights are the same as those used to average the individual RRL
transitions (see Section~\ref{sec:analysis}). In
Appendix~\ref{sec:app_electron_temperature}, we show that this
strategy can produce accurate electron temperatures.

Table~\ref{tab:catalog} lists the \textit{WISE} Catalog source name,
the telescope used for the observation, the measured RRL-to-continuum
flux ratios, the RRL FWHM line widths, and the derived electron
temperatures for the B15 single dish and our VLA \hii\ region
samples. This table only lists the highest quality electron
temperature derivations; we remove all QF D sources from the B15 and
VLA samples. The electron temperature uncertainties are computed by
propagating the RRL-to-continuum flux ratio and FWHM line width
uncertainties through Equation~\ref{eq:te}. For VLA sources, the
``Type'' column indicates whether the position of peak continuum
brightness (P) or watershed region (T) is used to measure the
RRL-to-continuum flux ratio. The ``Taper'' column identifies which
data cube is used (N for non-tapered and Y for
\textit{uv}-tapered). We select the combination of ``Type'' and
``Taper'' that minimizes the fractional uncertainty in the derived
electron temperature.  In cases where the same source is detected in
multiple surveys, we only list the VLA values, if available. If the
source is not observed or detected in the VLA survey, we list the GBT
values. If the source is not in the VLA survey nor the GBT survey, we
list the 140 Foot values. Table~\ref{tab:catalog} also includes
information about the \hii\ region distances, which is discussed in
Section~\ref{sec:distances}.

In total, there are now 189 Galactic \hii\ regions with accurate
electron temperature determinations. This is an increase of 99 nebulae
(110\%) over the B15 sample. A fraction of these nebulae have
inaccurate distances, however, and can not be used to investigate
Galactic metallicity structure.

\subsection{Comparison with Single Dish}

Our sample combines measurements from three telescopes: the 140 Foot,
the GBT, and the VLA. Each of these telescopes may be affected by
systematics that lead to discrepancies between the derived electron
temperatures because each is sampling a different volume of gas within
and surrounding the \hii\ regions. For example, diffuse foreground and
background emission may affect the single dish observations, but such
extended emission is filtered out by the VLA. In principle, there may
be differences between the different single dish measurements as
well. \citet{balser2011} find no significant difference between the
derived electron temperatures for 16 nebulae observed by both the 140
Foot and the GBT. Here we compare the single dish and VLA observations
of 22 nebulae in common between the B15 single dish catalog and our
VLA catalog.

We first compare the fitted LSR velocity of these nebulae. The top
panel of Figure~\ref{fig:rrl_compare} shows the difference between the
single dish RRL LSR velocity and that measured by the VLA for the 22
nebulae observed by the VLA and either the GBT or the 140 Foot. Here
and in all subsequent analyses, we use the ``best'' combination of
non-tapered or \textit{uv}-tapered data cubes and continuum peak
brightness location or watershed region for spectral
extraction. ``Best'' means the combination of tapering and spectral
extraction technique that minimizes the fractional uncertainty in the
derived electron temperature. The single dish and VLA LSR velocities
are in good agreement, with a weighted mean difference of
\(-0.09\pm0.34\kms\) (the error here is the uncertainty of the mean),
a median difference of \(-1.28\kms\), and a standard deviation of
1.59\kms. Throughout these analyses, we use the reciprocal variances
of the fitted Gaussian parameters as the weights in the averages.

Next we compare the single dish and VLA RRL FWHMs. The bottom panel of
Figure~\ref{fig:rrl_compare} shows the ratio of the single dish RRL
line width to that measured by the VLA for the overlapping
nebulae. The weighted mean of the line width ratios is
\(0.99\pm0.02\), the median ratio is \(1.03\), and the standard
deviation is \(0.10\). For the narrowest RRLs, the VLA line widths are
\({\sim}\)5-\(10\%\) smaller than those measured by the single dish
telescopes. This trend is likely due to the fact that the VLA is
probing a denser and less turbulent component of the nebulae.

\afterpage{
  \begin{longrotatetable}
\begin{deluxetable*}{lcr@{\,\(\pm\)\,}lcr@{\,\(\pm\)\,}lr@{\,\(\pm\)\,}lcr@{\,\(\pm\)\,}lcr@{\,\(\pm\)\,}lr@{\,\(\pm\)\,}lc}
\centering
%\singlespace
\tablewidth{0pt}
%\tabletypesize{\scriptsize}
\tablecaption{RRL Data Products\label{tab:line_products}}
\tablehead{
\colhead{Name} & \colhead{\(\nu_L\)} & \multicolumn{2}{c}{\(S_L^{\rm P}\)} & \colhead{rms\(^{\rm P}\)} & \multicolumn{2}{c}{\(v_{\rm LSR}^{\rm P}\)} & \multicolumn{2}{c}{\(\Delta V^{\rm P}\)} & \colhead{Taper\(^{\rm P}\)\tablenotemark{a}} & \multicolumn{2}{c}{\(S_L^{\rm T}\)} & \colhead{rms\(^{\rm T}\)} & \multicolumn{2}{c}{\(v_{\rm LSR}^{\rm T}\)} & \multicolumn{2}{c}{\(\Delta V^{\rm T}\)} & \colhead{Taper\(^{\rm T}\)\tablenotemark{a}} \\ 
\colhead{}     & \colhead{(MHz)}     & \multicolumn{2}{c}{(mJy }           & \colhead{(mJy}            & \multicolumn{2}{c}{(km s\(^{-1}\))}         & \multicolumn{2}{c}{(km s\(^{-1}\))}      & \colhead{}                                   & \multicolumn{2}{c}{(mJy)}           & \colhead{(mJy)}           & \multicolumn{2}{c}{(km s\(^{-1}\))}         & \multicolumn{2}{c}{(km s\(^{-1}\))}      & \colhead{}                                   \\ 
\colhead{}     & \colhead{}          & \multicolumn{2}{c}{beam\(^{-1}\))}  & \colhead{beam\(^{-1}\))}  & \multicolumn{2}{c}{}                        & \multicolumn{2}{c}{}                     & \colhead{}                                   & \multicolumn{2}{c}{}                & \colhead{}                & \multicolumn{2}{c}{}                        & \multicolumn{2}{c}{}                     & \colhead{}                                   
}
\startdata
G009.612+00.205 & 8862.2 & 5.53 & 0.39 & 1.19 & 2.5 & 0.8 & 22.2 & 1.9 & N & 2.71 & 0.08 & 0.24 & 3.2 & 0.3 & 22.1 & 0.8 & N \\ 
G009.613+00.200 & 8786.3 & 81.07 & 0.67 & 1.97 & 4.0 & 0.1 & 20.4 & 0.2 & Y & 133.99 & 1.12 & 3.29 & 3.8 & 0.1 & 20.5 & 0.2 & N \\ 
G010.596$-$00.381 & 8816.4 & 59.07 & 0.60 & 1.87 & 1.1 & 0.1 & 23.3 & 0.3 & Y & 107.40 & 0.98 & 3.06 & 1.1 & 0.1 & 23.1 & 0.2 & Y \\ 
G010.621$-$00.380 & 8789.6 & 80.22 & 0.52 & 1.67 & $-$0.5 & 0.1 & 24.8 & 0.2 & N & 3.61 & 0.03 & 0.10 & $-$0.7 & 0.1 & 24.9 & 0.2 & N \\ 
G010.623$-$00.385 & 8737.2 & 175.69 & 1.08 & 4.12 & 1.1 & 0.1 & 35.0 & 0.2 & Y & 182.71 & 0.99 & 4.03 & 0.9 & 0.1 & 39.9 & 0.2 & N \\ 
G012.805$-$00.196 & 8779.1 & 1097.15 & 3.02 & 11.78 & 36.3 & 0.0 & 36.4 & 0.1 & Y & 2079.31 & 5.80 & 22.02 & 36.7 & 0.0 & 34.5 & 0.1 & Y \\ 
G012.813$-$00.200 & 8767.2 & 199.79 & 1.44 & 4.85 & 30.2 & 0.1 & 27.2 & 0.2 & N & 74.74 & 0.65 & 2.20 & 30.4 & 0.1 & 27.7 & 0.3 & N \\ 
G013.880+00.285 & 8806.2 & 267.42 & 0.64 & 1.90 & 52.4 & 0.0 & 21.5 & 0.1 & Y & 530.66 & 1.35 & 4.06 & 52.0 & 0.0 & 21.6 & 0.1 & Y \\
\enddata
\tablenotetext{a}{"N" if non-tapered image measurement; "Y" if \textit{uv}-tapered image measurement}
\end{deluxetable*}
\end{longrotatetable}

  \clearpage
  \begin{longrotatetable}
\begin{deluxetable*}{lcr@{\,\(\pm\)\,}lr@{\,\(\pm\)\,}lr@{\,\(\pm\)\,}lccr@{\,}lr@{\,}lcc}
\centering
%\singlespace
\tablewidth{0pt}
%\tabletypesize{\scriptsize}
\tablecaption{\hii\ Region Distances and Properties\label{tab:catalog}}
\tablehead{
\colhead{Name} & \colhead{Telescope} & \multicolumn{2}{c}{\(S_L/S_C\)} & \multicolumn{2}{c}{\(\Delta V\)}    & \multicolumn{2}{c}{\(T_e\)} & \colhead{Type\tablenotemark{a}} & \colhead{Taper\tablenotemark{b}} & \multicolumn{2}{c}{\(d\)} & \multicolumn{2}{c}{\(R\)} & \colhead{Distance\tablenotemark{c}} & \colhead{Distance} \\ 
\colhead{}     & \colhead{}          & \multicolumn{2}{c}{}            & \multicolumn{2}{c}{(km s\(^{-1}\))} & \multicolumn{2}{c}{(K)}     & \colhead{}                      & \colhead{}                       & \multicolumn{2}{c}{(kpc)} & \multicolumn{2}{c}{(kpc)} & \colhead{Method}                    & \colhead{Reference}   
}
\startdata
G000.666$-$00.036 & 140 Foot & 0.0569 & 0.0033 & 40.5 & 0.4 & 8170 & 180 & \nodata & \nodata & 7.59 & $^{+0.84}_{-0.65}$ & 0.21 & $^{+0.91}_{-0.11}$ & P & R09c \\ 
G001.125$-$00.106 & 140 Foot & 0.1070 & 0.0018 & 24.5 & 0.2 & 7130 & 70 & \nodata & \nodata & \multicolumn{2}{c}{\nodata\tablenotemark{d}} & \multicolumn{2}{c}{\nodata\tablenotemark{d}} & K & \nodata \\
G003.266$-$00.061 & 140 Foot & 0.0978 & 0.0100 & 25.3 & 0.4 & 7440 & 280 & \nodata & \nodata & \multicolumn{2}{c}{\nodata\tablenotemark{d}} & \multicolumn{2}{c}{\nodata\tablenotemark{d}} & K & \nodata \\
G005.900$-$00.431 & 140 Foot & 0.0691 & 0.0006 & 22.5 & 0.2 & 11130 & 170 & \nodata & \nodata & 2.99 & $^{+0.17}_{-0.20}$ & 5.38 & $^{+0.19}_{-0.16}$ & P & S14 \\ 
G005.987$-$01.191 & 140 Foot & 0.0840 & 0.0008 & 26.5 & 0.2 & 8180 & 70 & \nodata & \nodata & \multicolumn{2}{c}{\nodata\tablenotemark{d}} & \multicolumn{2}{c}{\nodata\tablenotemark{d}} & K & \nodata \\
G008.137+00.232 & 140 Foot & 0.1019 & 0.0007 & 25.4 & 0.1 & 7090 & 60 & \nodata & \nodata & \multicolumn{2}{c}{\nodata\tablenotemark{d}} & \multicolumn{2}{c}{\nodata\tablenotemark{d}} & K & \nodata \\
G010.160$-$00.350 & 140 Foot & 0.0911 & 0.0005 & 31.2 & 0.2 & 6830 & 30 & \nodata & \nodata & \multicolumn{2}{c}{\nodata\tablenotemark{d}} & \multicolumn{2}{c}{\nodata\tablenotemark{d}} & K & \nodata \\
G010.308$-$00.150 & 140 Foot & 0.0874 & 0.0005 & 31.6 & 0.2 & 6800 & 40 & \nodata & \nodata & \multicolumn{2}{c}{\nodata\tablenotemark{d}} & \multicolumn{2}{c}{\nodata\tablenotemark{d}} & K & \nodata \\
\enddata
\tablenotetext{a}{"P" if measured at the location of peak continuum brightness; "T" if measured within the watershed segmentation region}
\tablenotetext{b}{"N" if non-tapered image measurement; "Y" if \textit{uv}-tapered image measurement}
\tablenotetext{c}{"K" for Monte Carlo kinematic distance; "P" for parallax distance}
\tablenotetext{d}{Kinematic distances are unreliable in the direction of the Galactic center and anti-center}
\tablerefs{(A11) \citet{ando2011}; (B09) \citet{brunthaler2009}; (C14) \citet{choi2014}; (H06) \citet{hachisuka2006}; (H07) \citet{honma2007}; (H15) \citet{hachisuka2015};(I13) \citet{immer2013}; (K08) \citet{kim2008}; (M07) \citet{menten2007}; (M09) \citet{moscadelli2009}; (O10) \citet{oh2010}; (RD09) \citet{roman-duval2009}; (R09a) \citet{reid2009a}; (R09c) \citet{reid2009c}; (R10) \citet{rygl2010}; (R12) \citet{rygl2012}; (S07) \citet{sandstrom2007}; (S10) \citet{sato2010}; (S14) \citet{sato2014}; (Sa14) \citet{sanna2014}; (U12) \citet{urquhart2012}; (W12) \citet{wu2012}; (W14) \citet{wu2014}; (X06) \citet{xu2006}; (X11) \citet{xu2011}; (X13) \citet{xu2013}; (Z09) \citet{zhang2009}; (Z13) \citet{zhang2013}; (Z14) \citet{zhang2014}}
\end{deluxetable*}
\end{longrotatetable}

  \clearpage
}

Finally we compare the measured RRL-to-continuum brightness ratios and
derived electron temperatures between the single dish and VLA
surveys. Figure~\ref{fig:te_compare} shows the ratio of the single
dish and VLA measured RRL-to-continuum flux ratios (top) and electron
temperatures (bottom). The single dish RRL-to-continuum brightness
ratios are systematically \({\sim}10\%\) less than the VLA brightness
ratios. The weighted mean of these ratios is \(0.86\pm0.03\) with a
median of \(0.90\) and a standard deviation of \(0.12\). Consequently,
the single dish electron temperatures are \({\sim}10\%\) greater than
the VLA electron temperatures. The weighted mean of the electron
temperature ratios is \(1.12\pm0.03\) with a median of \(1.10\) and a
standard deviation of {\(0.12\).

The cause of the systematic difference between the single dish and VLA
RRL-to-continuum brightness ratios and electron temperatures is
unclear. The difference may be due to a problem with the derivation of
the RRL-to-continuum brightness ratio or perhaps due to a fundamental
difference in the RRL and/or continuum emission measured by the
different telescopes. We know that there are a few issues with how the
single dish RRL-to-continuum ratios are derived. B15 measured the
continuum flux densities of their nebulae at \(\nu_C = 8556\,\mhz\),
whereas the average frequency of their observed RRL transitions is
\(\langle\nu_L\rangle = 8902\,\mhz\). In
Appendix~\ref{sec:app_electron_temperature}, we show that the B15
strategy overestimates the true electron temperature by \({\sim}6\%\).
Furthermore, we do not scale the single dish and VLA RRL-to-continuum
brightness ratios to a common frequency because each survey observed
similar RRL transitions. The typical VLA \(<\)Hn\(\alpha\)\(>\)
weighted frequency is within \(2\%\) of the B15 average RRL
frequency. Neither of these two effects can fully explain the observed
\(10\%\) difference between the single dish and VLA RRL-to-continuum
brightness ratios.

There are several factors that might affect the measured continuum
and/or RRL flux densities: the single dish continuum flux densities
are uncertain due to poor continuum background subtraction; the single
dish telescopes are not pointed at the center of the continuum source
during the RRL observation; the VLA is not sensitive to extended
emission associated with the \hii\ region; and/or the VLA is seeing
more optically thick gas. (1) The continuum flux densities are the
largest source of uncertainty in the single dish electron temperature
derivation (see B15). If the continuum background level is poorly
constrained, then the single dish continuum flux densities will be
inaccurate. We limit our analysis to high continuum QF single dish
nebulae, however, so these problems should be minimal. Furthermore,
random errors in the single dish continuum background levels would not
cause the observed systematic difference in single dish
vs. interferometric electron temperatures. (2) The single dish RRL
spectra must be measured at the location of the peak continuum
brightness. If the telescope is not pointed properly, then the RRL
flux densities will be underestimated. This is also not a likely
explanation for the discrepancy, because B15 peaked on source for
their RRL observations. (3) The VLA is not sensitive to diffuse
emission. If the source of such emission has a different density
and/or temperature, the VLA electron temperatures will differ from the
single dish values. (4) Finally, the nebulae may be optically thick,
and/or the compact emission visible to the VLA is more optically thick
than the diffuse emission missed by the VLA.  Optical depth effects
such as these would lead to an underestimation of the VLA continuum
flux densities and electron temperatures. Some or all of these issues
may be contributing to the remaining \(4\%\) discrepancy between the
single dish and VLA RRL-to-continuum brightness ratios.

We wish to use as much data as possible to constrain the metallicity
structure of the Galactic disk. Therefore, in subsequent analyses that
combine the single dish and VLA electron temperatures, we multiply the
single dish electron temperatures by \(0.9\) to accommodate the
systematic offset between the VLA and single dish data.

\subsection{Distances} \label{sec:distances}

Distances to Galactic \hii\ regions are derived in three main ways:
(1) spectrophotometrically, (2) geometrically, and (3) kinematically.
Spectrophotometric distances are only available for optically
unobscured nebulae. Since most of the nebulae in our sample are very
distant with lines of sight passing through the Galactic plane, we do
not consider spectrophotometric distances in this analysis. The
extremely fine angular resolution provided by very long baseline
interferometry (VLBI) is used to measure the parallaxes and proper
motions of masers associated with high-mass star forming regions
\citep[e.g.,][]{reid2014rev}.  Several hundred maser parallax
measurements have been made as part of the Bar and Spiral Structure
Legacy (BeSSeL) Survey\footnote{http://bessel.vlbi-astrometry.org/},
the Japanese VLBI Exploration of Radio Astrometry
(VERA)\footnote{http://veraserver.mtk.nao.ac.jp/}, and various
European VLBI Network (EVN)\footnote{http://www.evlbi.org/}
projects. The vast majority of Galactic \hii\ regions, however, lack
parallax measurements. We must therefore rely on kinematic techniques
to derive the distances to nebulae without a geometric distance
determination.

Of the 189 Galactic \hii\ regions in our sample with accurate electron
temperature determinations, 46 (24\%) have a maser parallax
measurement. As in \citet{wenger2018b}, we derive the parallax
distance and distance uncertainties by Monte Carlo resampling the
measured parallax within its uncertainties. We generate 5000 samples
of the parallax distance, then we fit a kernel density estimator (KDE)
to the distance distribution to calculate a probability distribution
function (PDF). The peak of the PDF is the derived parallax distance,
and the width of the PDF characterizes the parallax distance
uncertainty \citep[see][]{wenger2018b}.

Kinematic distances are computed by measuring the line of sight
velocity of an object and assuming that object follows some Galactic
rotation model (GRM). We use the \citet{wenger2018b} Monte Carlo
kinematic distance method and the \citet{reid2014} GRM to derive the
kinematic distances to our sample of Galactic \hii\ regions. This
method computes the distances and distance uncertainties by resampling
the observed LSR velocities, the solar motion parameters, which define
the LSR, and the GRM parameters to determine the kinematic distance
PDFs. \citet{wenger2018b} find that the Monte Carlo kinematic
distances are reasonably accurate when compared to the parallax
distances for a sample of 75 Galactic high-mass star forming
regions. The median difference between the kinematic and parallax
distances for these nebulae is 17\% (0.42\,\text{kpc}).

Within the Solar circle, there exists a kinematic distance ambiguity
(KDA). An axisymmetric GRM yields the same LSR velocity at two
distances, and additional information must be used to break this
degeneracy. The \textit{WISE} Catalog lists the KDA resolutions (KDAR)
for a subset of the known Galactic \hii\ regions. As in
\citet{wenger2018b}, we use the \textit{WISE} Catalog KDARs for
nebulae with LSR velocities farther than 20\kms\ from the tangent
point velocity. All nebulae within 20\kms\ of the tangent point
velocity are assigned to the tangent point distance.

\begin{figure}
  \centering
  \includegraphics[width=\linewidth]{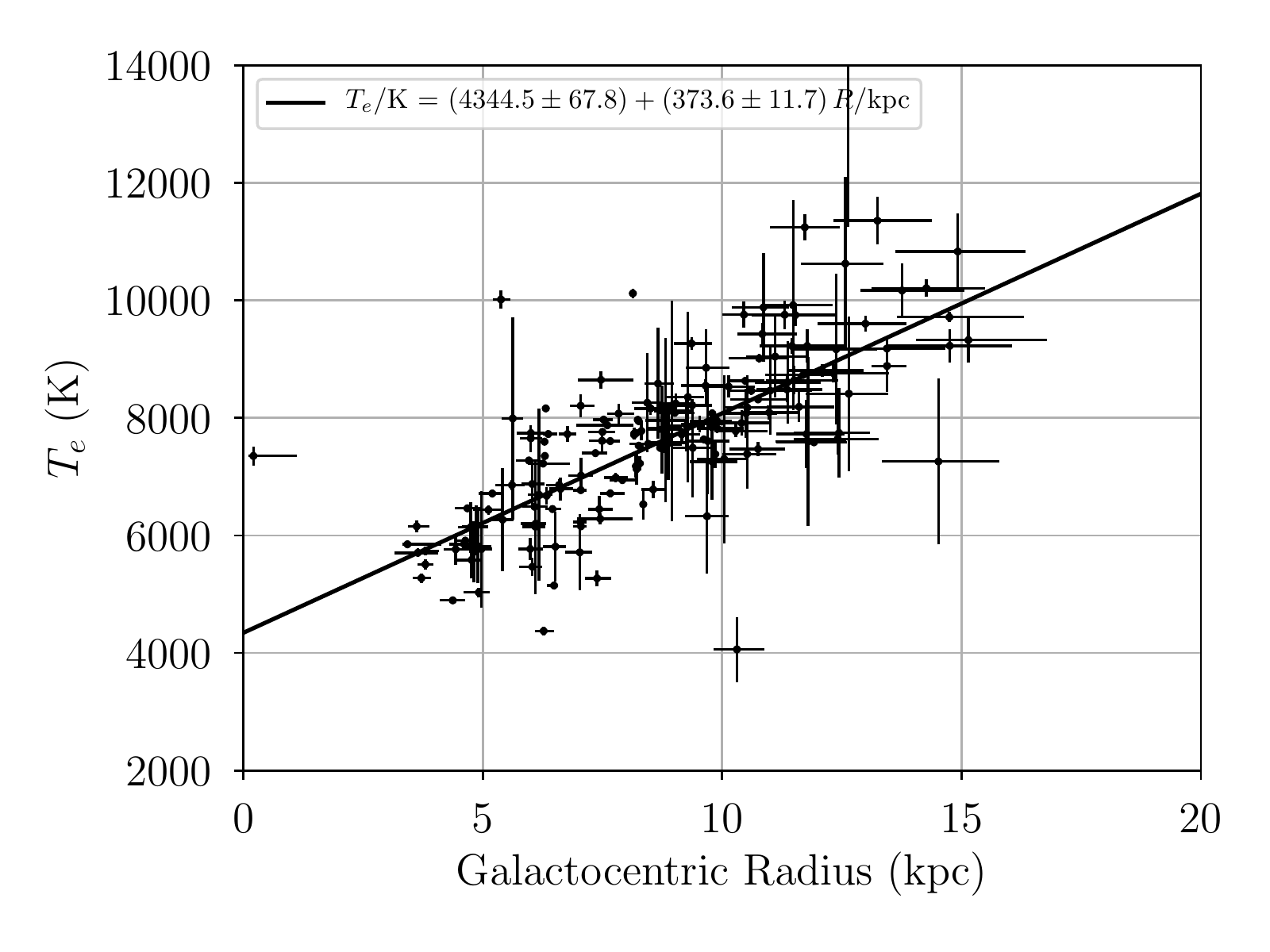}
  \includegraphics[width=\linewidth]{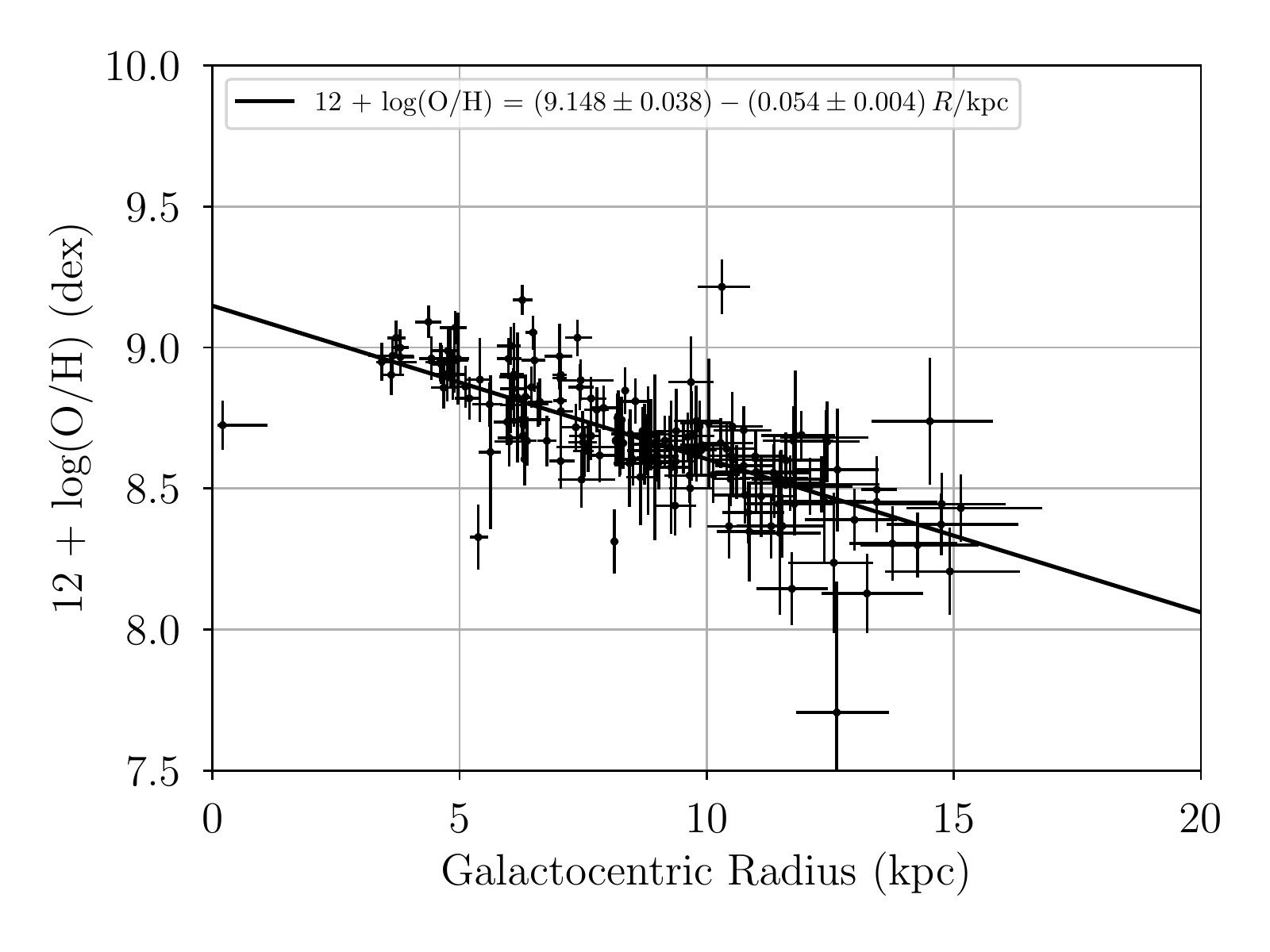}
  \caption{The nominal radial electron temperature (top) and
    metallicity (bottom) gradients. The abscissa error bars are the
    \(1\sigma\) uncertainties in the parallax or kinematic distances
    derived from our Monte Carlo distance analysis, and the ordinate
    error bars are the \(1\sigma\) uncertainties in the electron
    temperature or metallicity derived from the continuum and RRL
    uncertainties.  The lines are the robust least squares linear
    model fits to the data as defined in the legends.}
  \label{fig:gradient_nom}
\end{figure}

Due to line of sight velocity crowding, kinematic distances are
inaccurate in the direction of the Galactic center and anti-center.
Following \citet{wenger2018b}, we remove all kinematic distance
nebulae located within the zones \(-15^\circ < \ell < 15^\circ\) and
\(160^\circ < \ell < 200^\circ\). After removing these nebulae, we are
left with 121 Galactic \hii\ regions with kinematic distances. Our
final catalog contains 167 nebulae with accurate electron temperatures
and either a parallax (46) or kinematic (121)
distance. Table~\ref{tab:catalog} lists the relevant distance
parameters for each nebulae: the heliocentric distance \(d\); the
Galactocentric radius, \(R\); the distance method (``P'' for parallax
and ``K'' for kinematic); and the maser parallax observation
reference, if any. Nebulae with accurate electron temperatures,
without a parallax measurement, and in the direction of the Galactic
center/anti-center are also included in this table for
completeness. These nebulae are, however, excluded from all subsequent
analyses.

\subsection{Metallicity Structure}

\hii\ region electron temperatures are a proxy for their nebular
metallicities \citep[e.g.,][]{churchwell1975}. The \hii\ region
electron temperature structure across the Galactic disk thus reveals
structure in metallicity. \citet{shaver1983} derived an empirical
relationship between \hii\ region metallicities, determined using
optical collisionally excited lines to derive the oxygen and hydrogen
column densities, and electron temperatures, determined from RRLs:
\begin{equation}
  \begin{split}
    12+\text{log}_{10}(\text{O/H}) = \left(9.82\pm0.02\right) - \\
    \left(1.49\pm0.11\right)\frac{T_e}{10^4\,\text{K}} \label{eq:metallicity}
  \end{split}
\end{equation}
where \(T_e\) is the nebular electron temperature.

\begin{figure}
  \centering
  \includegraphics[width=\linewidth]{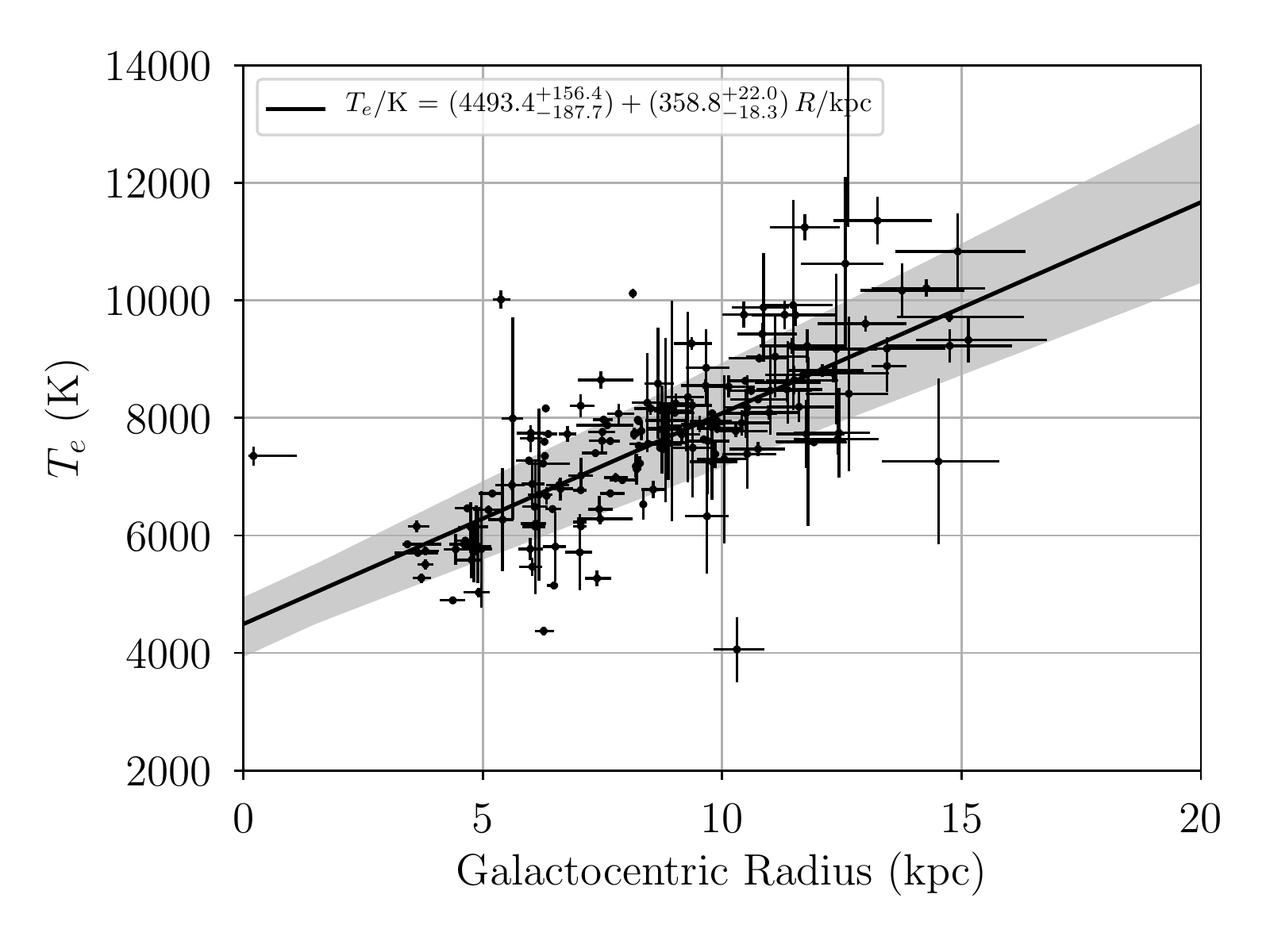}
  \includegraphics[width=0.75\linewidth]{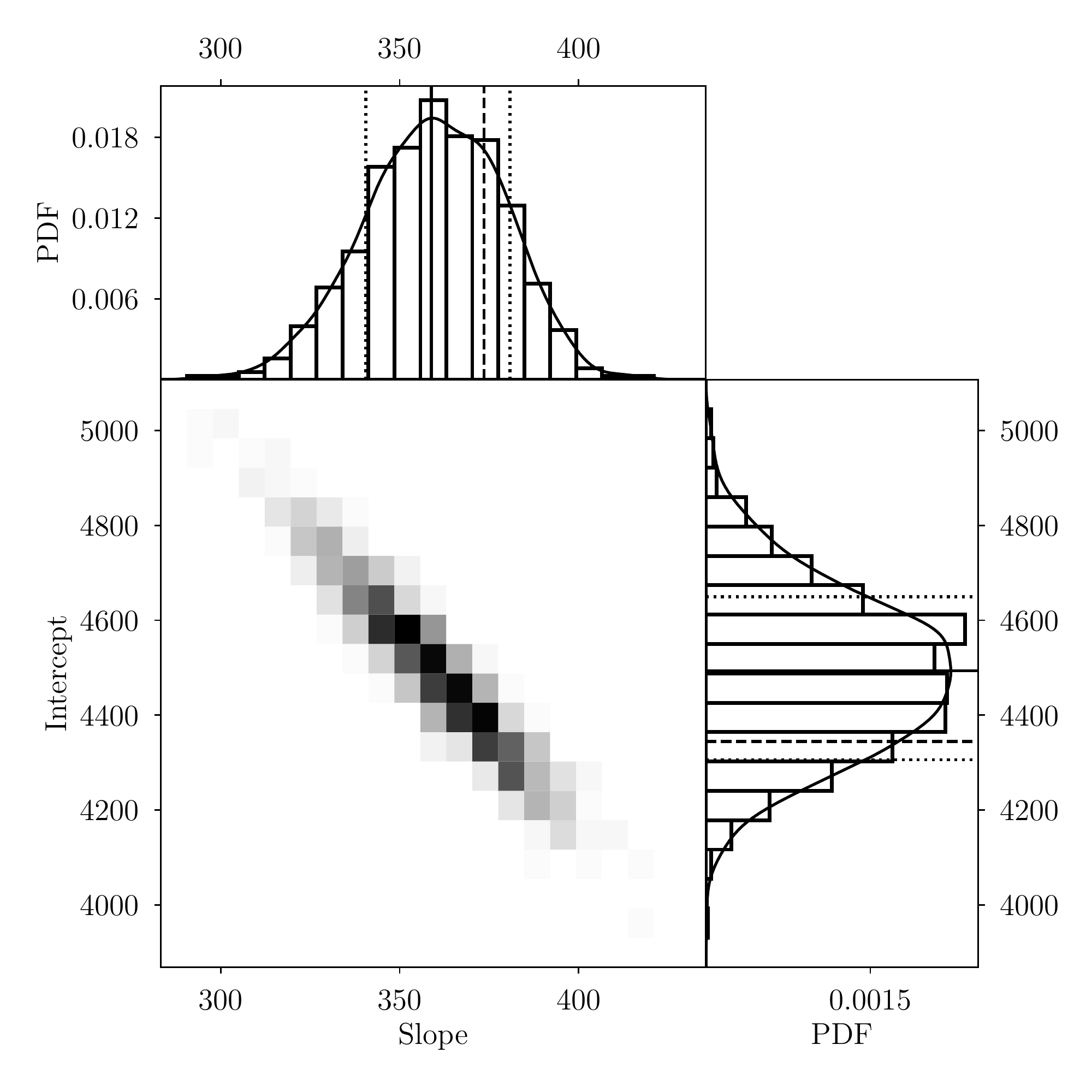}
  \caption{The most likely electron temperature gradient determined by
    Monte Carlo resampling the derived electron temperatures and
    Galactocentric radii. The top panel shows the data and the most
    likely linear model (black line) as defined in the legend. The
    error bars are the same as in Figure~\ref{fig:gradient_nom}. The
    shaded region represents the range of fits from 1000 Monte Carlo
    realizations of the data. The bottom panel shows the covariances
    between the linear model parameters (slope, with units of K
    kpc\(^{-1}\), and intercept, with units of K). The histograms are
    the PDFs of the Monte Carlo fit parameters, and the black curves
    are KDE fits to the PDFs. The solid lines are the peaks of the
    PDFs (the most likely fit parameters), and the dotted lines
    represent the \(1\sigma\) confidence intervals. The dashed lines
    are the nominal values of the fit parameters derived from the
    robust least squares fit to the data (i.e. without Monte Carlo
    resampling, as in Figure~\ref{fig:gradient_nom}).}
  \label{fig:te_gradient_mc}
\end{figure}

\begin{figure}
  \centering
  \includegraphics[width=\linewidth]{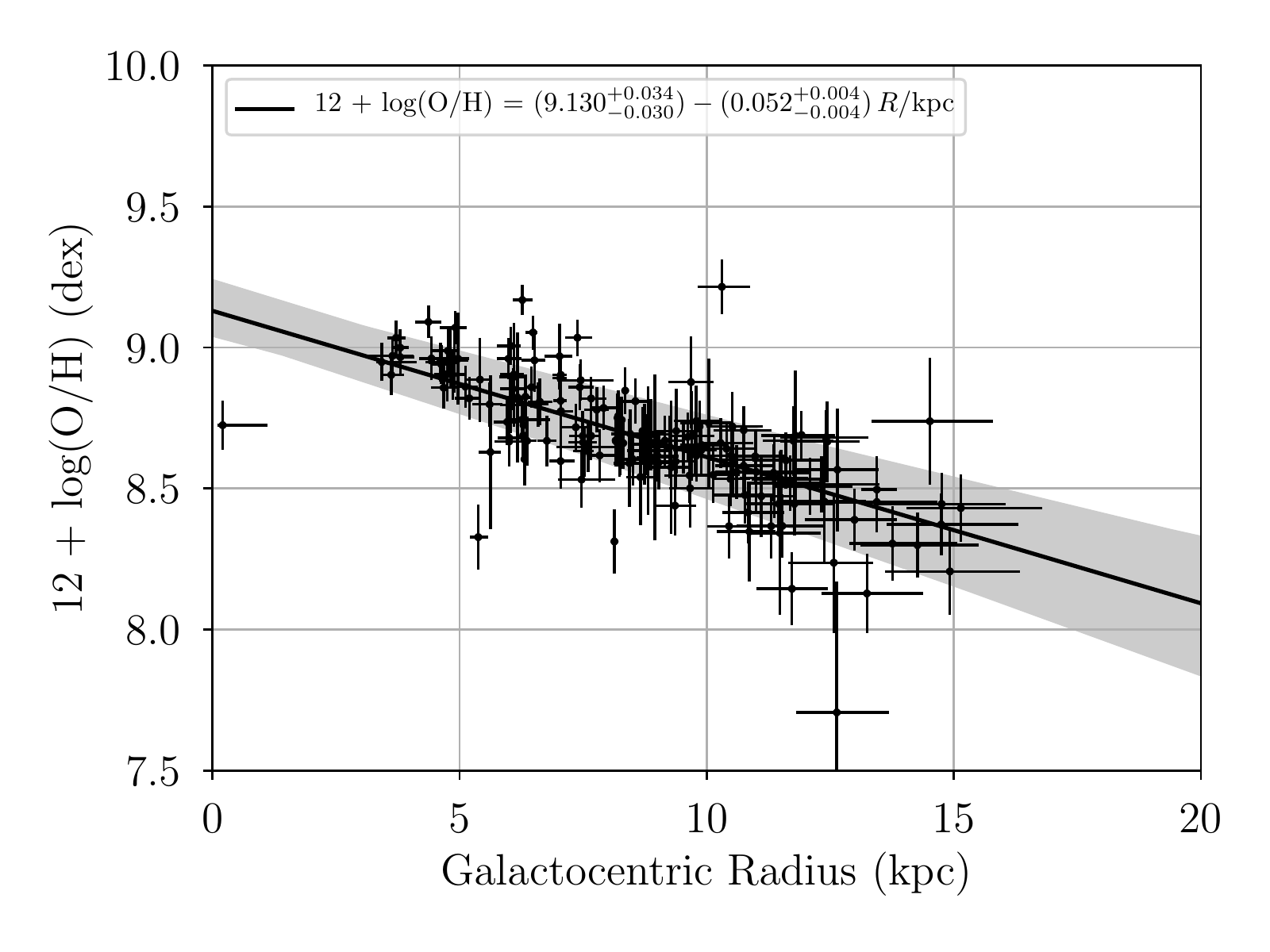}
  \includegraphics[width=0.75\linewidth]{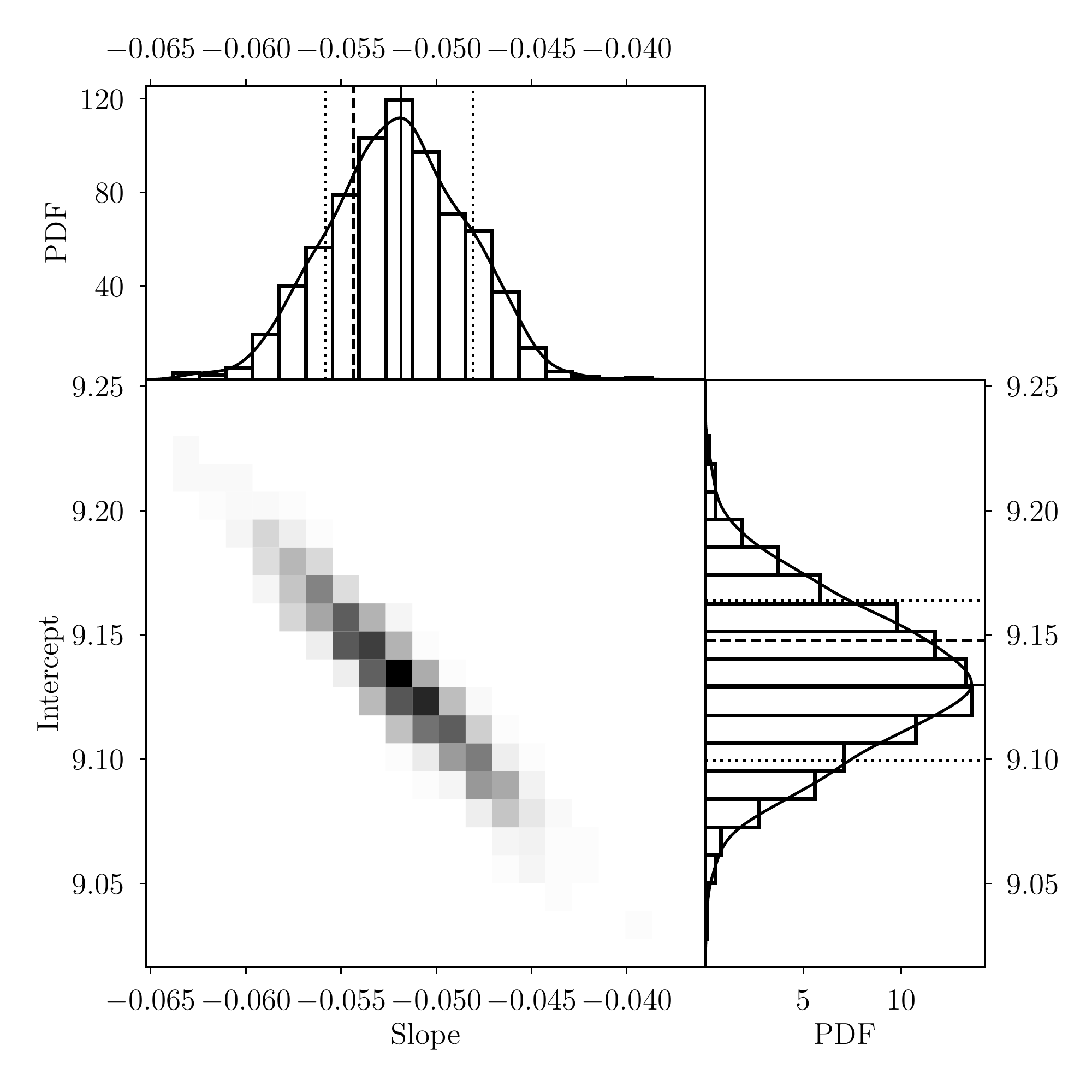}
  \caption{Same as Figure~\ref{fig:te_gradient_mc} for the radial
    metallicity gradient. The most likely linear model is defined in
    the legend. The covariance slope has units of dex kpc\(^{-1}\) and
    the intercept has units of dex.}
  \label{fig:o2h_gradient_mc}
\end{figure}

We begin our investigation of Galactic chemical structure by measuring
the radial electron temperature and metallicity
gradients. Figure~\ref{fig:gradient_nom} shows the nebular electron
temperature and metallicity gradients using the electron temperatures
and Galactocentric radii from Table~\ref{tab:catalog} and
metallicities derived using Equation~\ref{eq:metallicity}. The
metallicity uncertainties are determined by propagating the electron
temperature uncertainties through Equation~\ref{eq:metallicity}. We
use a robust least squares routine to fit a linear model to both
distributions. The least squares fit is robust because we dampen the
effect of outliers by minimizing a ``soft'' loss function, \(\rho(z) =
\sqrt{1 + z^2}-1\), where \(z\) is the squared residuals. This routine
does not consider the uncertainties of the data, because (1) there are
uncertainties in both the dependent and independent variables, and (2)
the Galactocentric radius uncertainties are asymmetric.  Nonetheless,
the best fit linear model to the nebular electron temperature
distribution is \(T_e/\text{K} = (4345\pm68) +
(374\pm12)\,R/\text{kpc}\), and the best fit for the nebular
metallicity distribution is \(12 + \text{log}_{10}(\text{O/H}) =
(9.148\pm0.038) - (0.054\pm0.004)\,R/\text{kpc}\). Within the errors,
these gradients are consistent with the gradients found by B15 using
their ``Best'' distances and Green Bank sample: \(T_e/\text{K} \propto
(402\pm33)\,R/\text{kpc}\) and \(12+\text{log}_{10}(\text{O/H})
\propto (-0.058\pm0.004)\,R/\text{kpc}\).

\begin{figure}
  \centering
  \includegraphics[width=\linewidth]{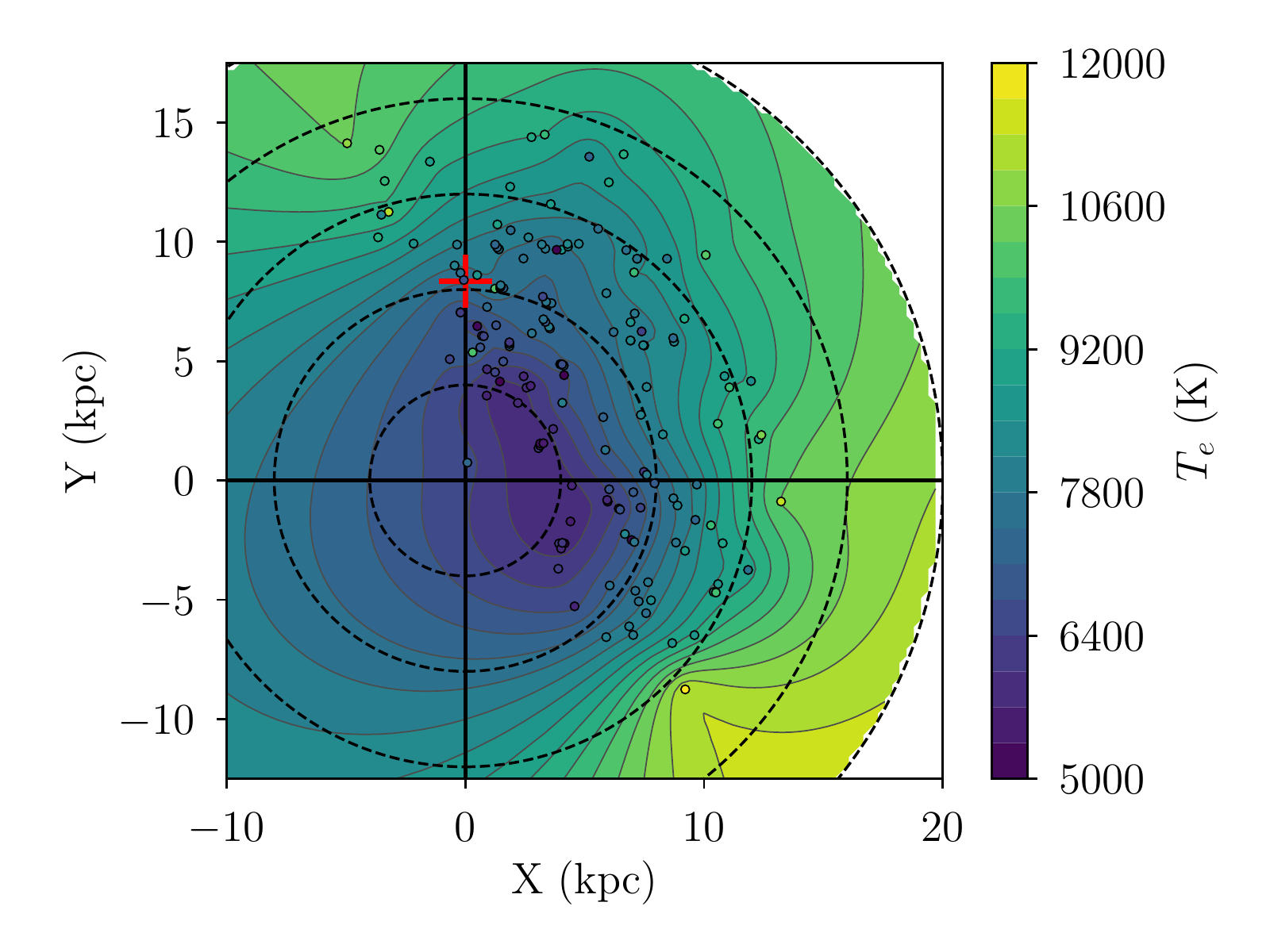}
  \includegraphics[width=\linewidth]{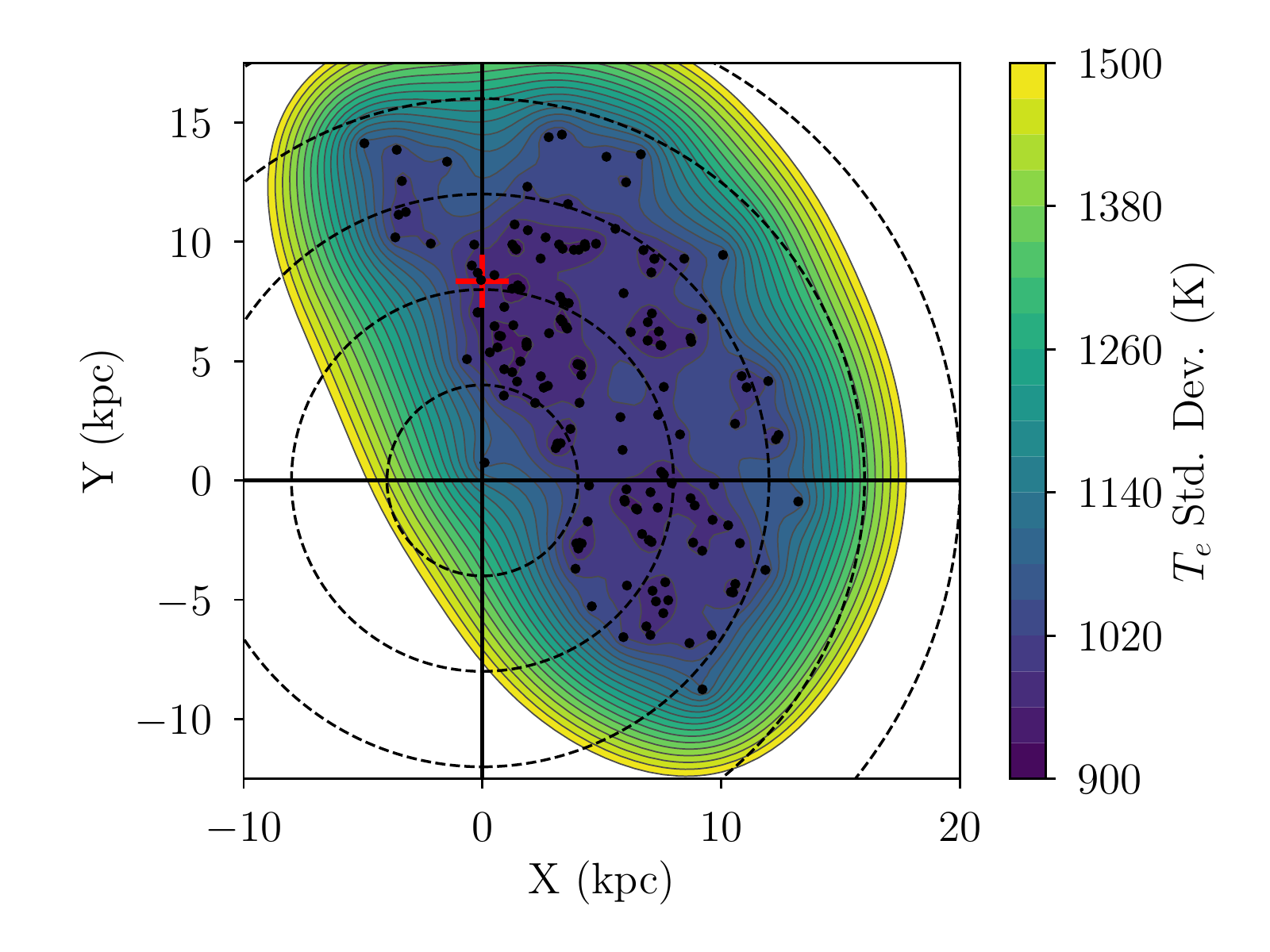}
  \caption{Kriging map of nebular electron temperatures. The top panel
    shows the Kriging interpolation in a face-on view of the Galactic
    disk. The points are the \hii\ regions in our sample, colored by
    their derived electron temperatures. The bottom panel shows the
    Kriging standard deviation. The Galactic Center is located at the
    origin and the Sun is located at the red cross. The dashed circles
    are 4, 8, 12, 16, and 20 kpc in radius. White areas are outside
    \(R=20\,\text{kpc}\) or have data values beyond the colorbar
    range.}
  \label{fig:te_nom}
\end{figure}

\begin{figure}
  \centering
  \includegraphics[width=\linewidth]{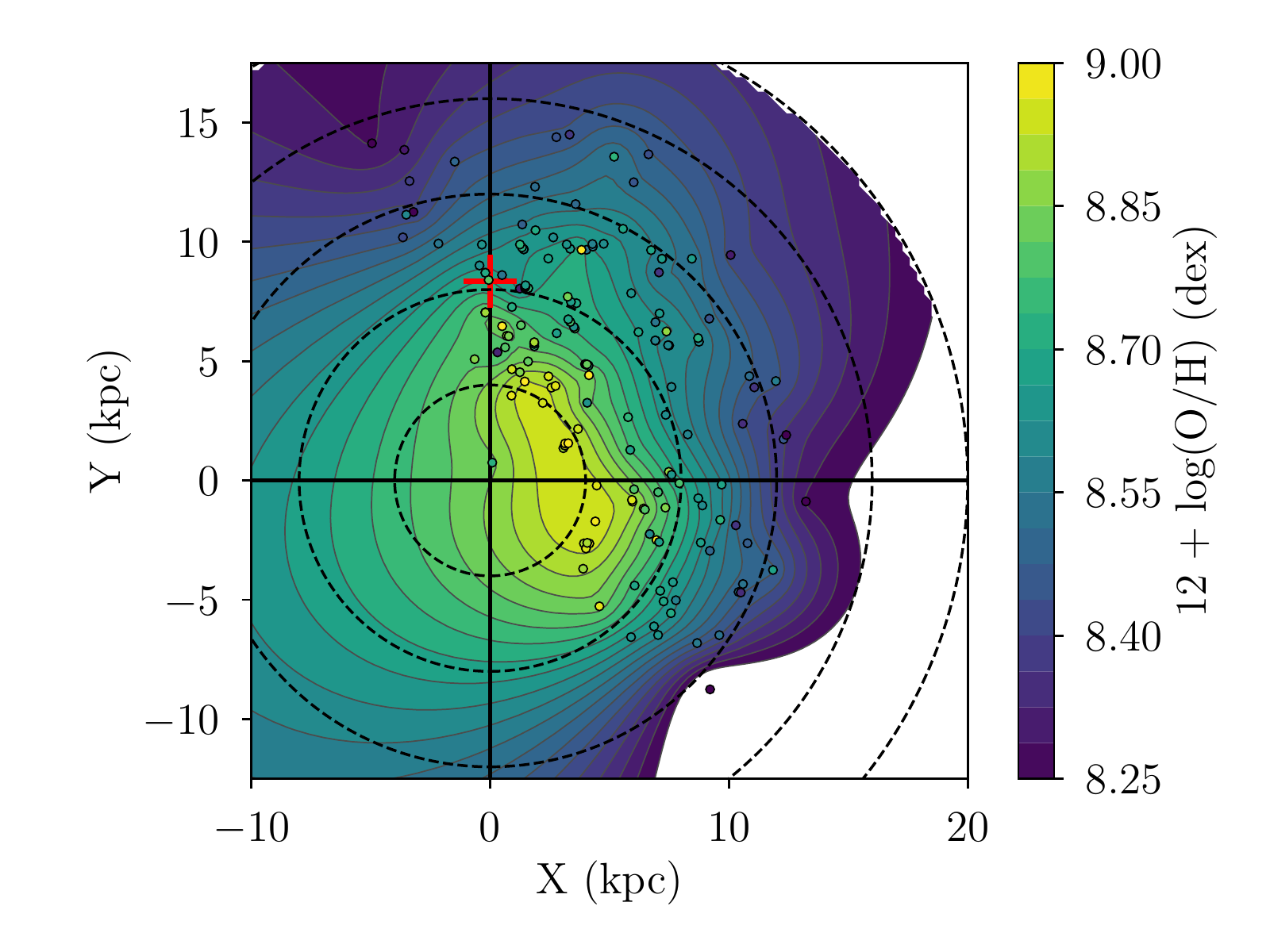}
  \includegraphics[width=\linewidth]{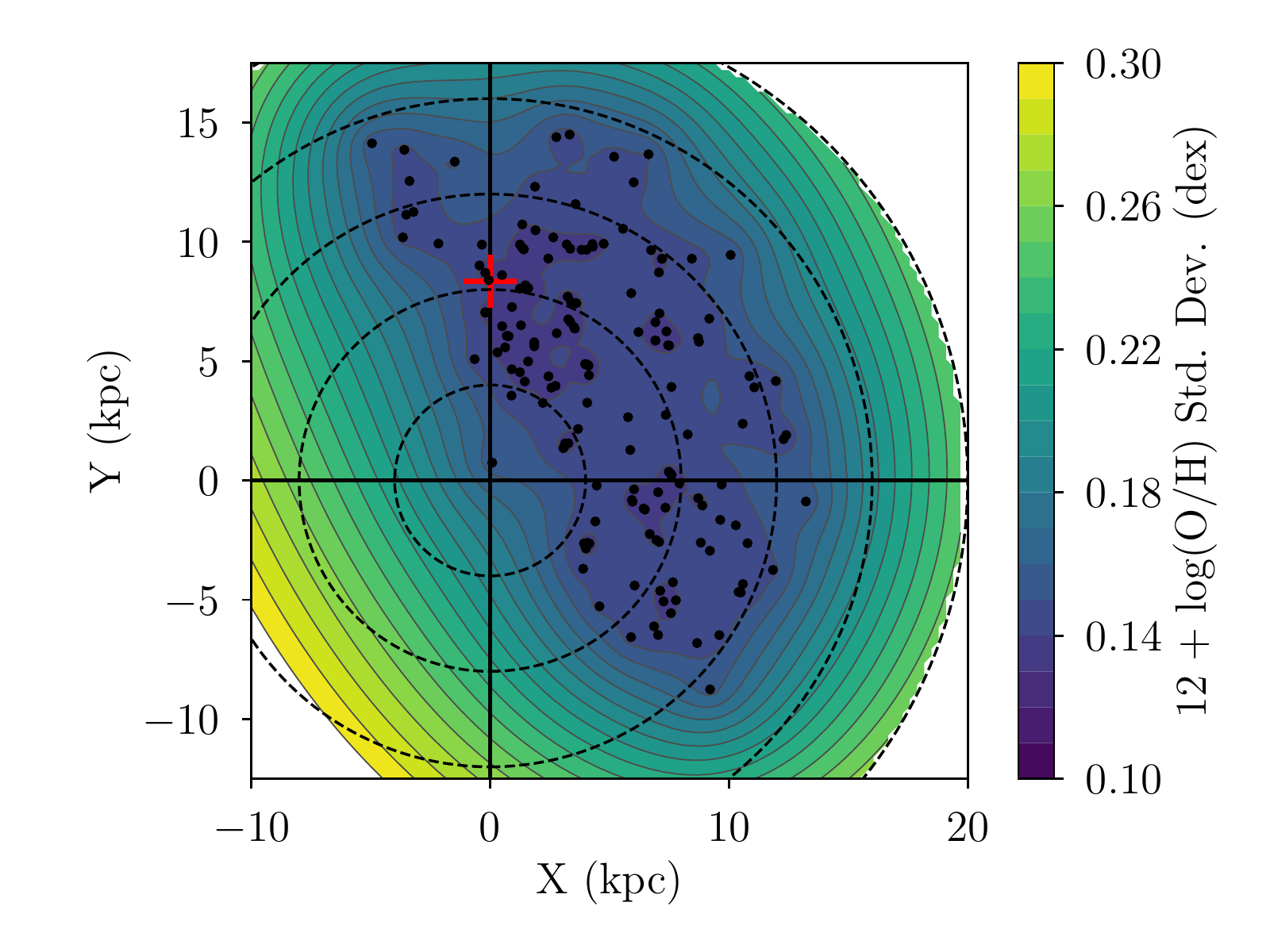}
  \caption{Same as Figure~\ref{fig:te_nom} for the nebular
    metallicities.}
  \label{fig:o2h_nom}
\end{figure}

A simple least squares fitting method cannot account for asymmetric
uncertainties in both the abscissas (i.e., Galactocentric radii) and
the ordinates (i.e., electron temperatures). Therefore, we estimate
the true variance of the linear model by Monte Carlo resampling the
data 1000 times. The electron temperatures are drawn from a Gaussian
distribution centered at the derived electron temperature and with a
width equal to the derived electron temperature uncertainty. The
Galactocentric radii are drawn from the parallax or kinematic distance
PDFs. For each realization of the data, we fit a robust least squares
linear model. Similar to the Monte Carlo kinematic distance method in
\citet{wenger2018b}, we estimate the most likely linear model
parameters by fitting a KDE to the PDFs of each model parameter. The
peak of this KDE is the most likely parameter, and the \(1\sigma\)
confidence interval is derived from the bounds of the PDF such that
(1) the PDF evaluated at the lower bound is equal to the PDF evaluated
at the upper bound and (2) the integral of the normalized PDF between
the bounds is \(68.3\%\). Figures~\ref{fig:te_gradient_mc} and
\ref{fig:o2h_gradient_mc} show, respectively, the most likely linear
model parameters derived from this Monte Carlo method and the
covariance between the model parameters for the electron temperature
and metallicity gradients. The most likely fits are \(T_e/\text{K} =
4493^{+156}_{-188} + 359^{+22}_{-18}\,R/\text{kpc}\) and \(12 +
\text{log}_{10}(\text{O/H}) = 9.130^{+0.034}_{-0.030} -
0.052^{+0.004}_{-0.004}\,R/\text{kpc}\). These gradients are within
\(1\sigma\) of the nominal least-squares values, and the asymmetric
uncertainties are more accurate given the uncertainties in the derived
electron temperatures and distances.

To visualize the variations in nebular electron temperature in the
Galactic disk, we use Kriging to spatially interpolate between
discrete nebulae (see also B15). The Kriging method computes the
average semivariance of the data as a function of the spatial
separation between the data points. The average semivariance is
measured in many separation bins, known as ``lags,'' and the
semivariogram (average semivariance as a function of lag) is fitted
with a model. The expected value of the data at any position is
derived from this semivariogram model \citep[see][]{feigelson2012}.

\begin{figure*}
  \centering
  \includegraphics[width=0.49\linewidth]{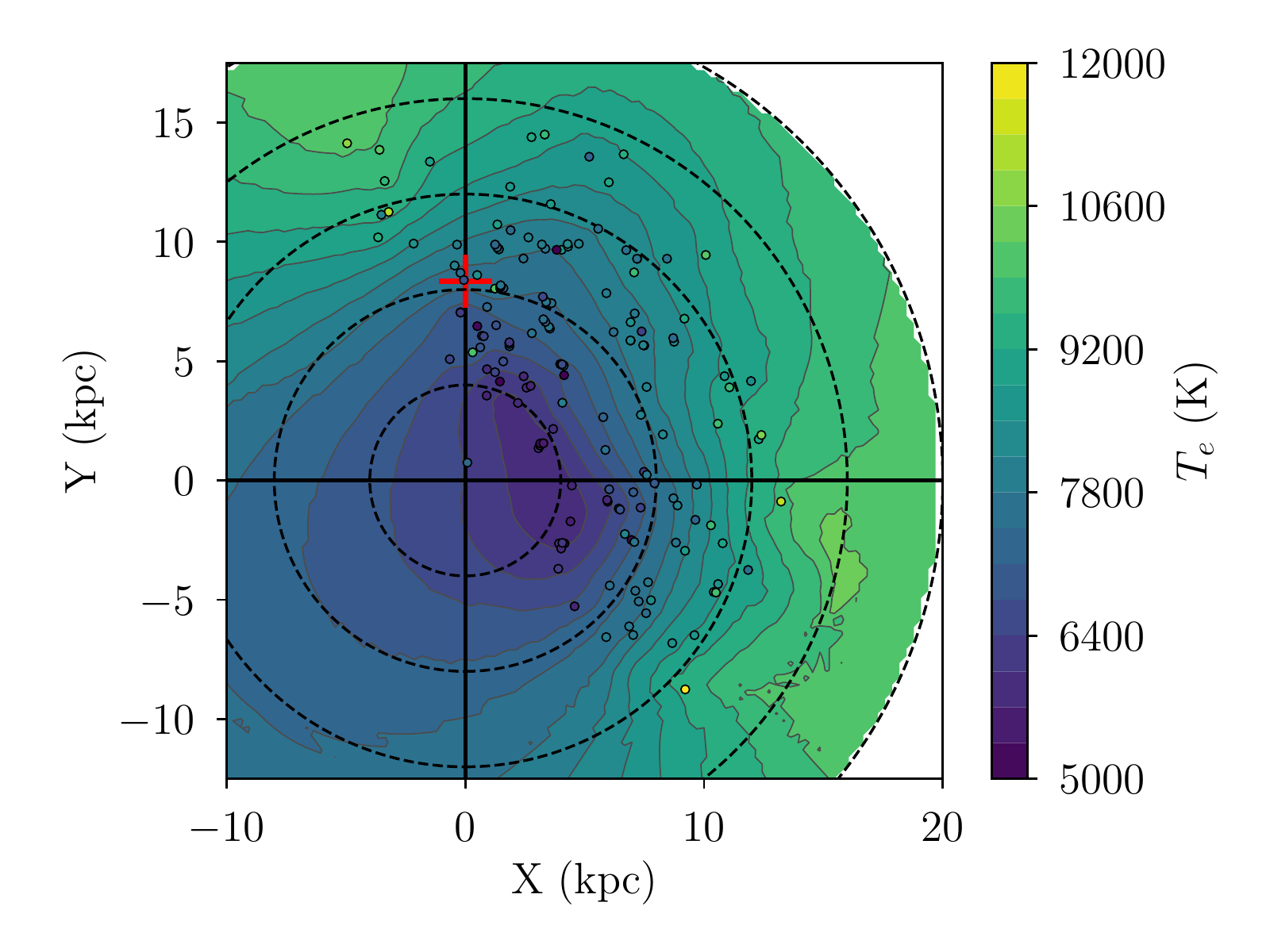}
  \includegraphics[width=0.49\linewidth]{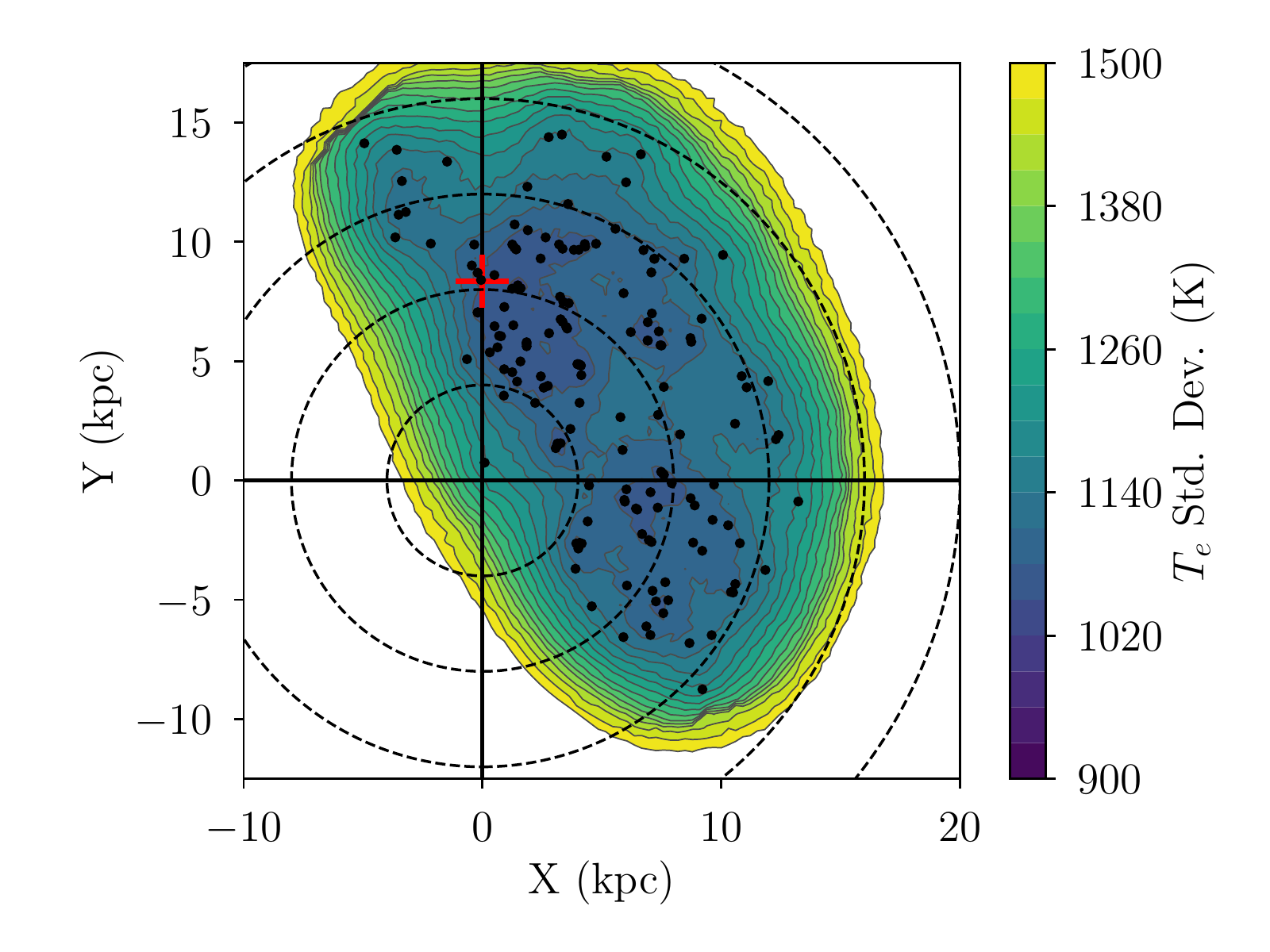} \\
  \includegraphics[width=0.49\linewidth]{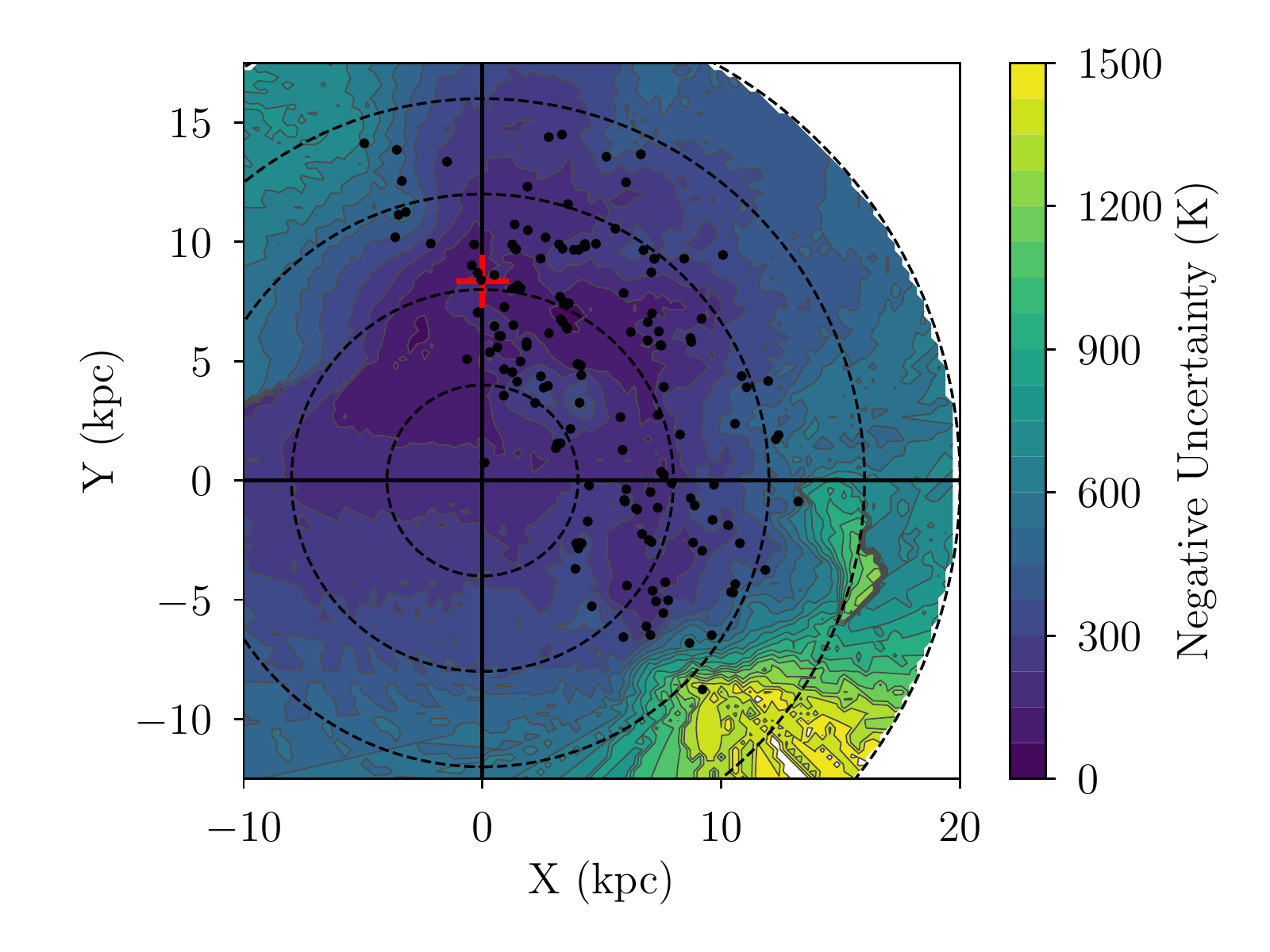}
  \includegraphics[width=0.49\linewidth]{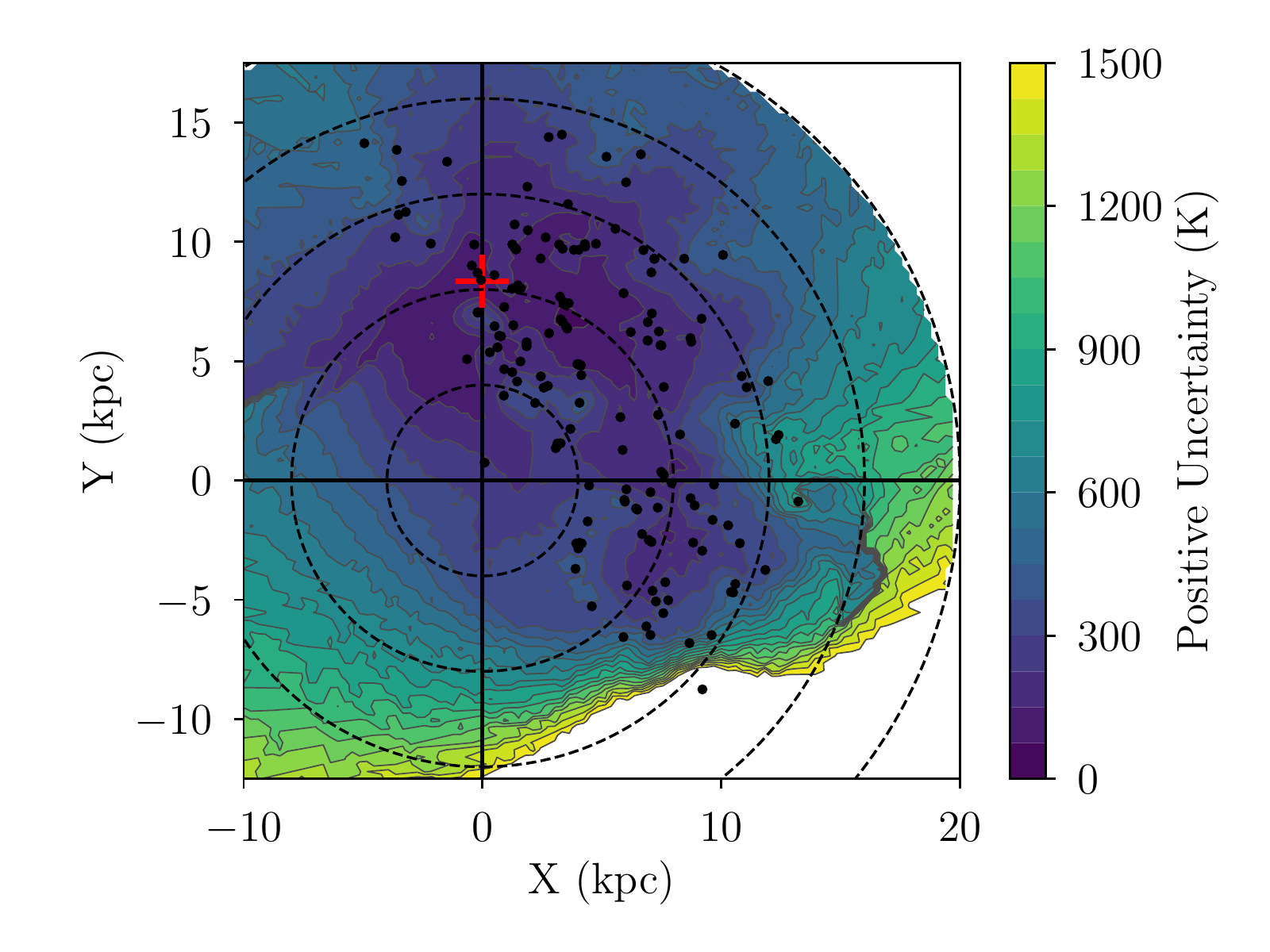}
  \caption{Most likely Kriging map of nebular electron temperatures
    determined by Monte Carlo resampling the derived electron
    temperatures and distances. Shown are the most likely Kriging
    interpolation values (top left), most likely Kriging standard
    deviation values (top right), lower \(1\sigma\) bounds (bottom
    left), and upper \(1\sigma\) bounds (bottom right) on the Kriging
    interpolation confidence intervals. The features in each plot are
    the same as in Figure~\ref{fig:te_nom}.}
  \label{fig:te_mc}
\end{figure*}

\begin{figure*}
  \centering
  \includegraphics[width=0.49\linewidth]{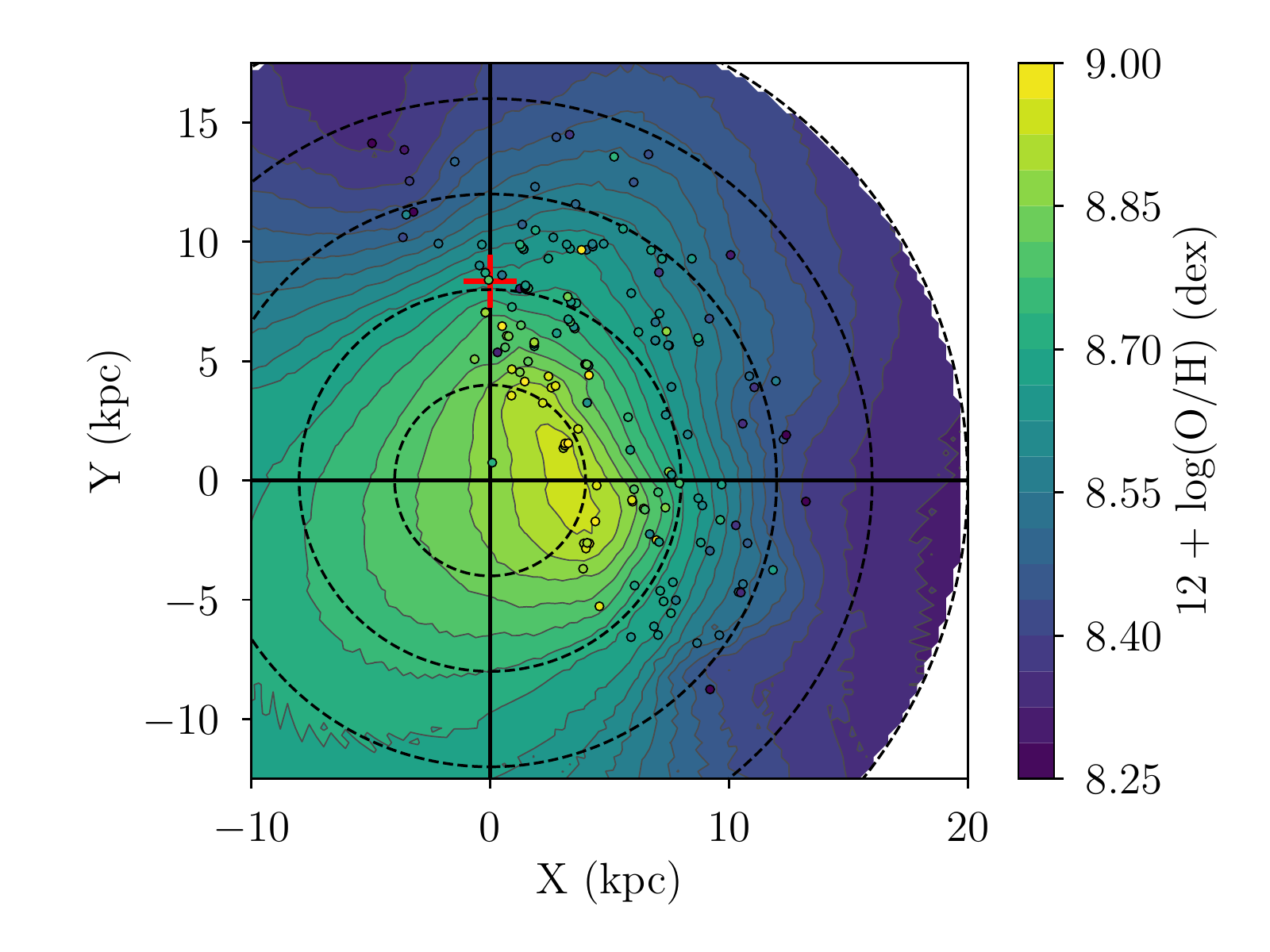}
  \includegraphics[width=0.49\linewidth]{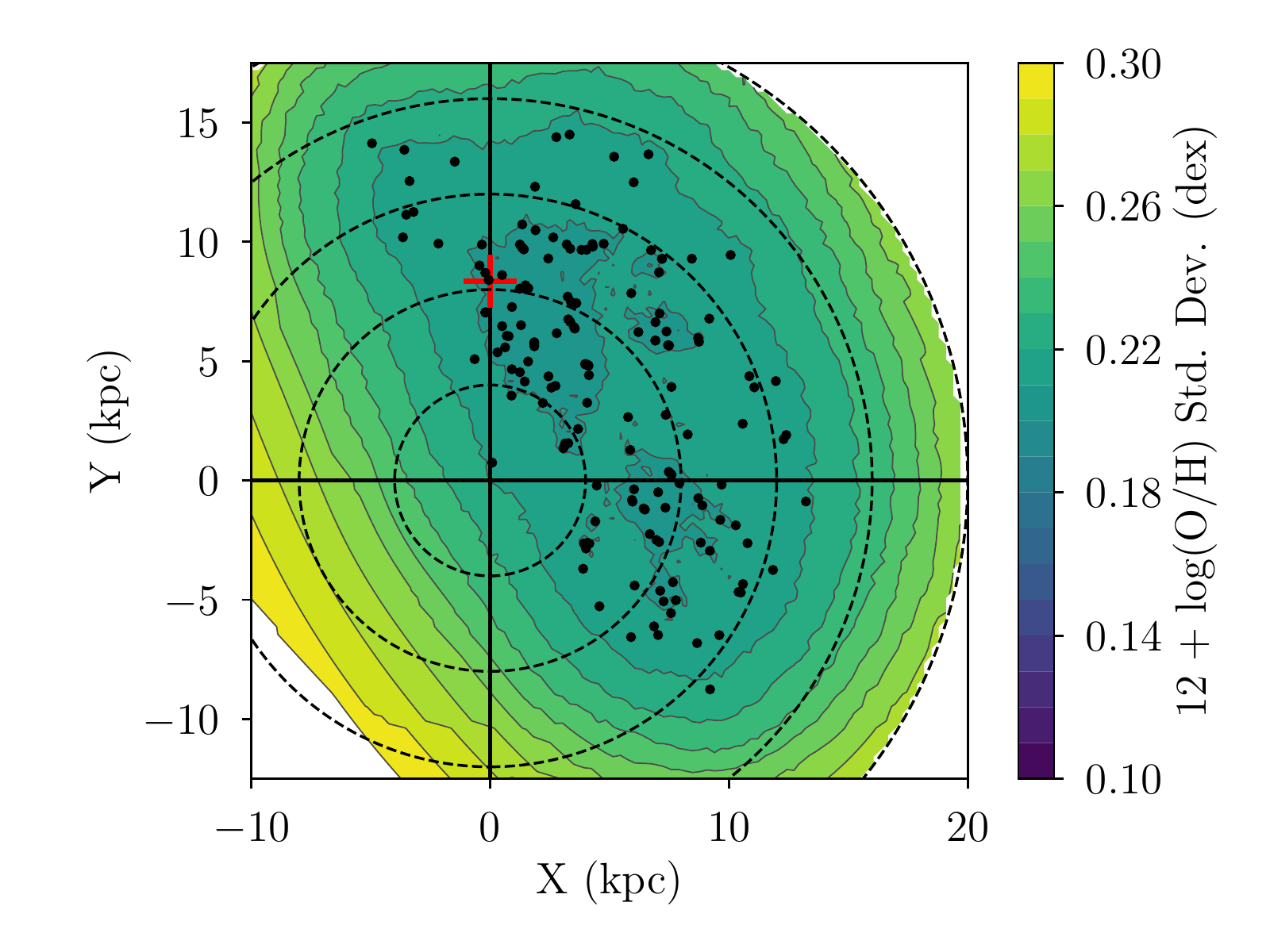} \\
  \includegraphics[width=0.49\linewidth]{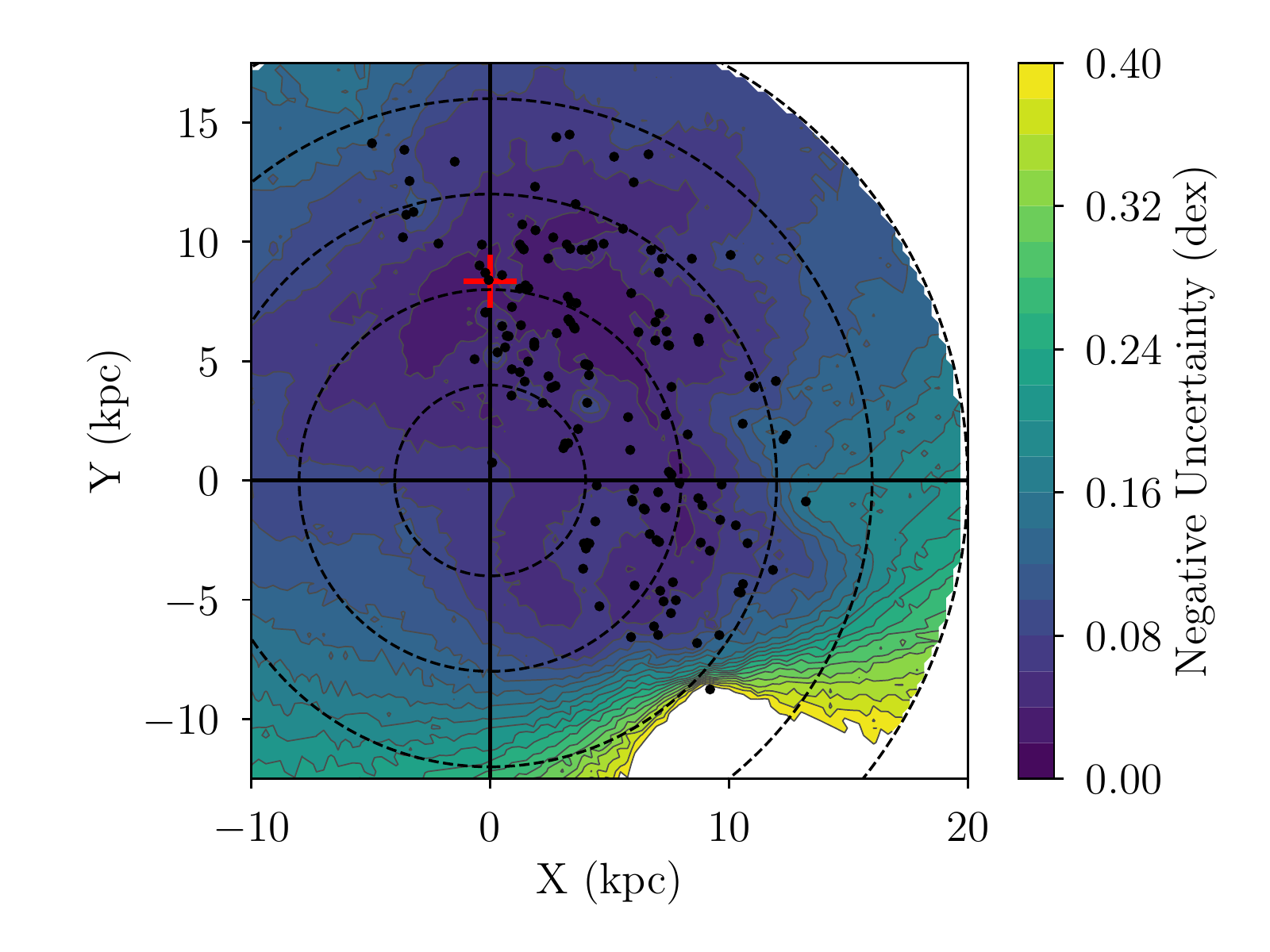}
  \includegraphics[width=0.49\linewidth]{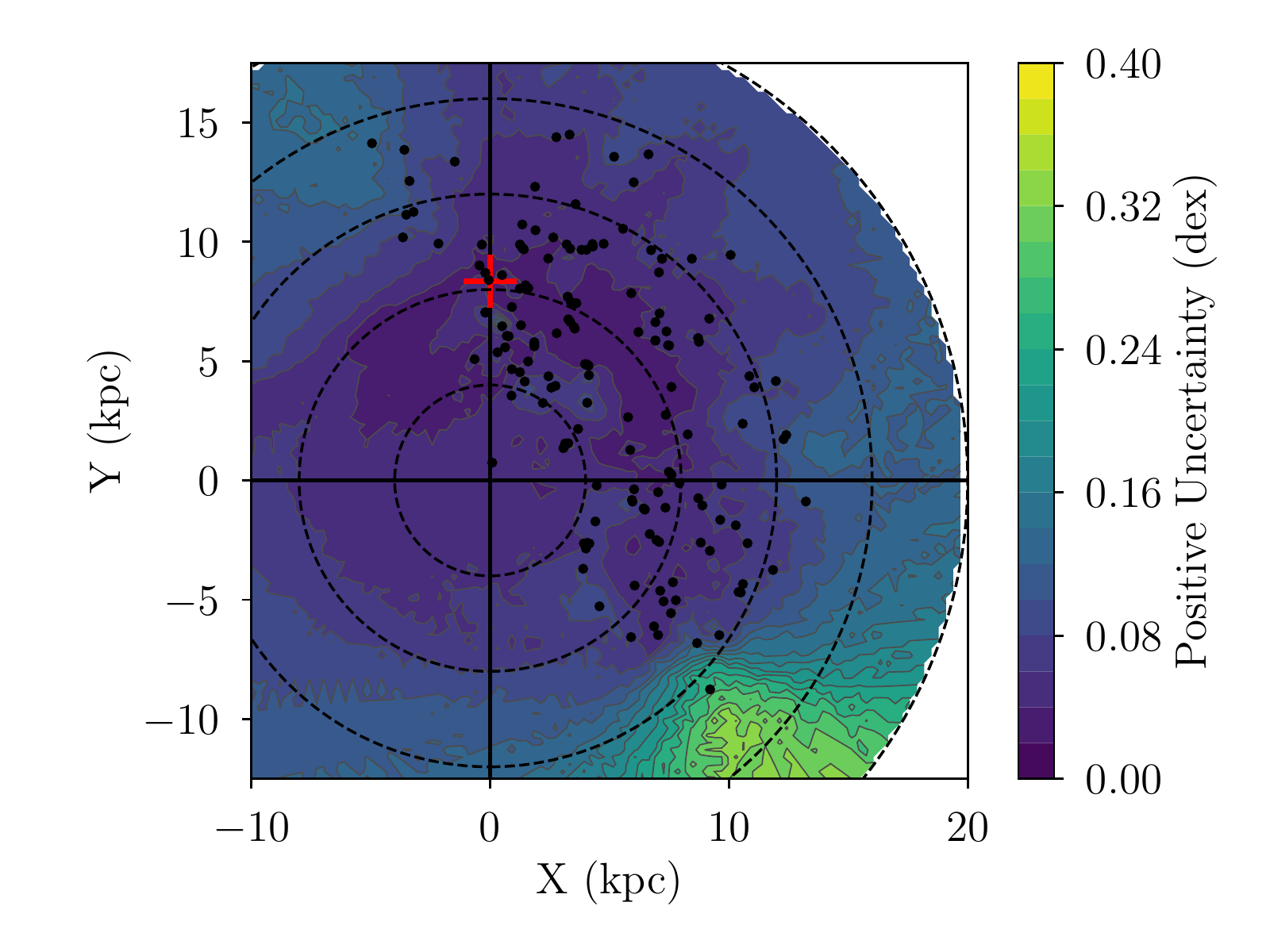}
  \caption{Same as Figure~\ref{fig:te_mc} for the nebular
    metallicities.}
  \label{fig:o2h_mc}
\end{figure*}

We compute the nominal Kriging map of nebular electron temperatures
using the Table~\ref{tab:catalog} electron temperatures and
distances. Figure~\ref{fig:te_nom} shows this electron temperature
map, where we use a linear semivariogram model to interpolate between
the discrete \hii\ region positions. The top panel is the Kriging
result and the bottom panel is the standard deviation of the Kriging
interpolation. This standard deviation map characterizes the intrinsic
scatter of the data across the Galactic disk. The \hii\ region points
are colored by their electron temperature to highlight the differences
between the actual nebular electron temperature and the interpolated
value at that position. Figure~\ref{fig:o2h_nom} shows the same
Kriging results with a linear semivariogram model for the \hii\ region
metallicities.  Qualitatively, these figures are similar to the
electron temperature and metallicity maps in B15.  It is clear from
these figures that the radial gradients have a strong dependence on
Galactocentric azimuth.

These Kriging results consider neither the uncertainties in the
nebular electron temperatures and metallicities nor the \hii\ region
distance uncertainties. We estimate the most likely Kriging map of
nebular electron temperatures and metallicities using a Monte Carlo
technique in the same way as we determined the most likely radial
gradients. We Monte Carlo resample the data within their uncertainties
1000 times, and, for each realization of the data, we generate a
Kriging map. At each pixel of the Kriging map, we construct a PDF of
the interpolation values, fit a KDE, and locate the peak and bounds of
the KDE. The peak is the most likely Kriging value at that position,
and the bounds represent the \(1\sigma\) confidence interval, as
before.

\begin{figure}
  \centering
  \includegraphics[width=\linewidth]{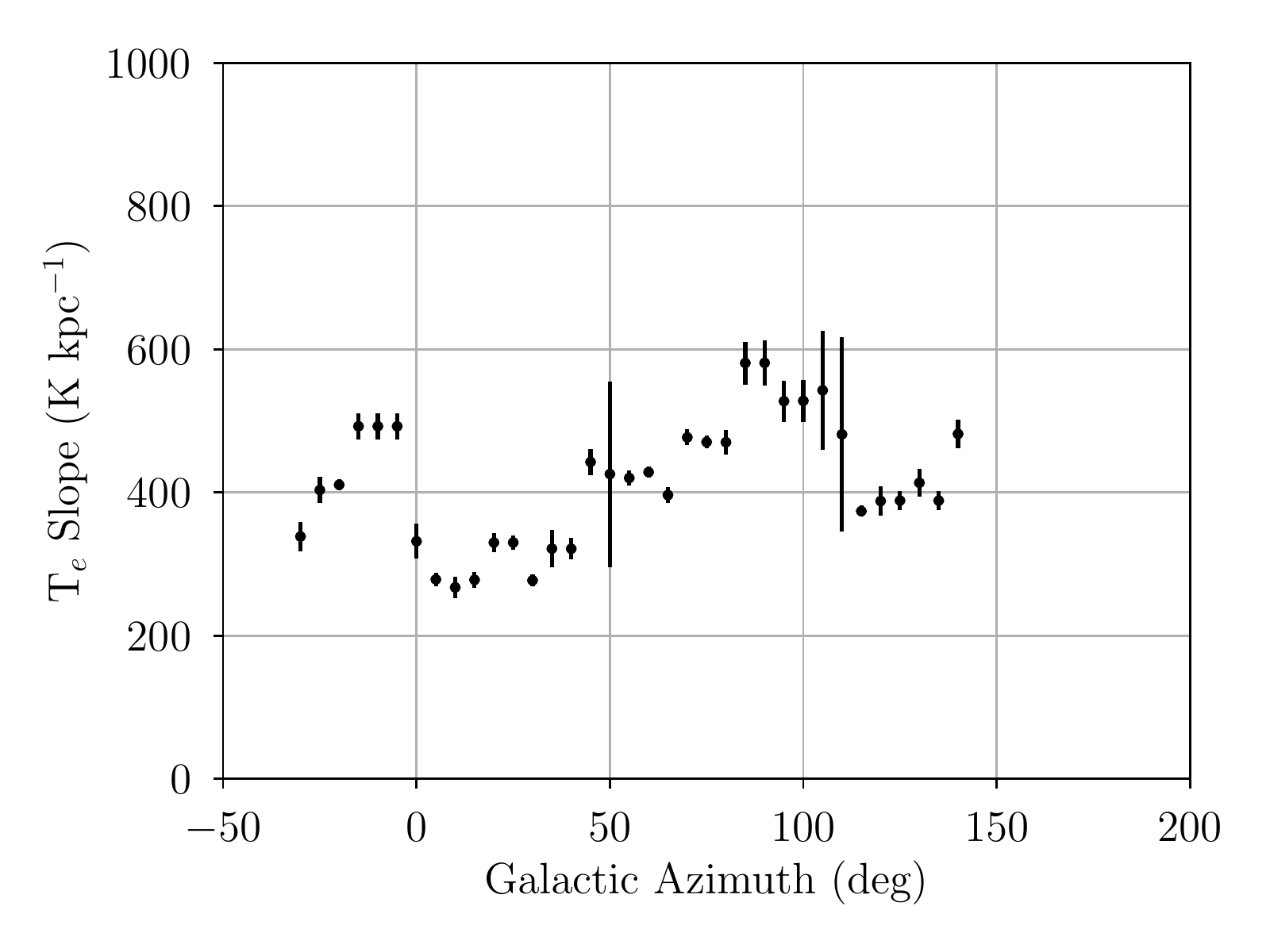}
  \includegraphics[width=\linewidth]{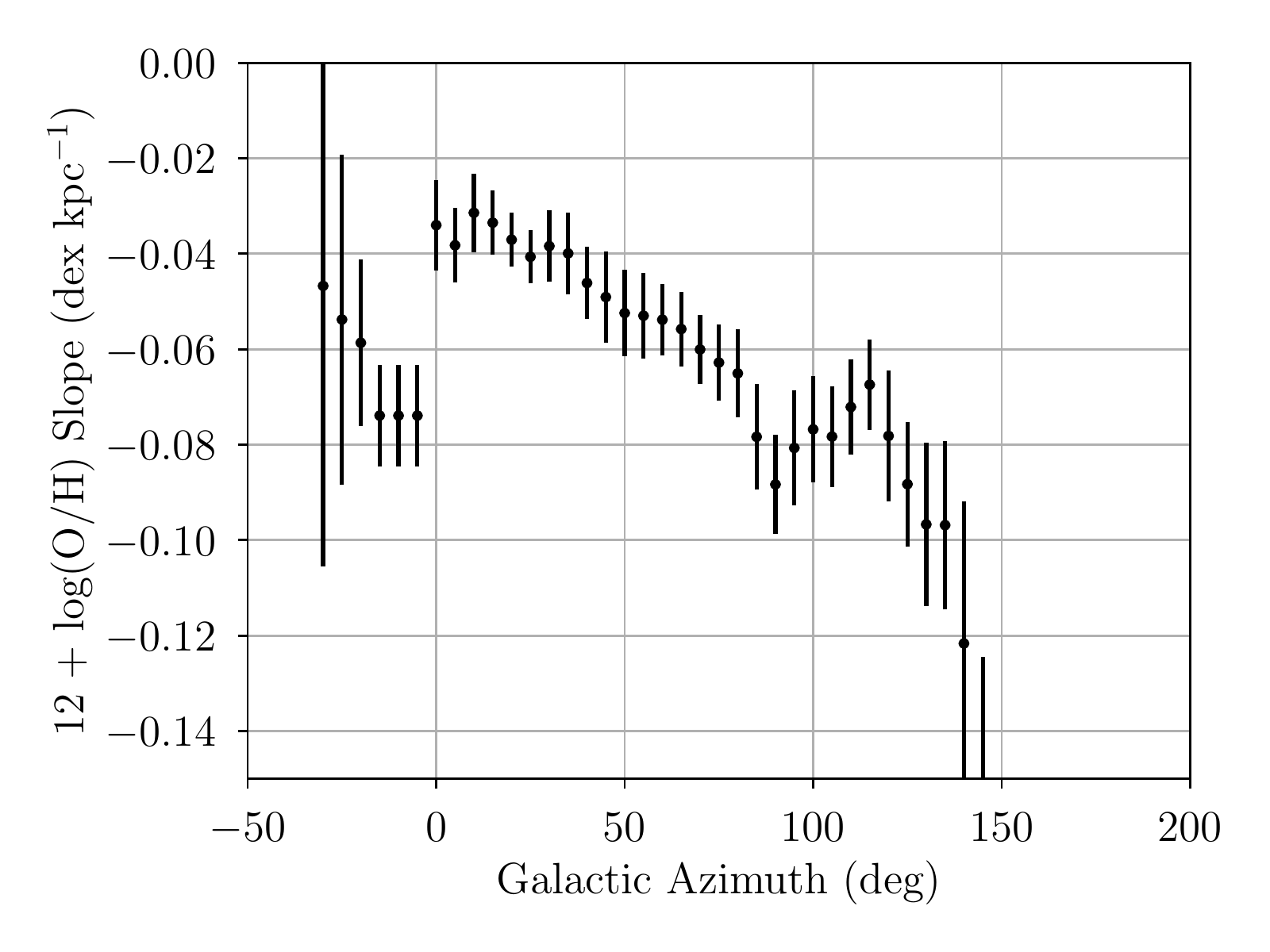} \\
  \caption{Nominal variations in the radial electron temperature (top)
    and metallcity (bottom) gradients as a function of Galactocentric
    azimuth. The Galaxy is divided into \(30^\circ\) bins spaced every
    \(5^\circ\) in Galactocentric azimuth. The points are the slopes
    of the robust least squares linear model fit to the data in each
    bin, and the error bars are the \(1\sigma\) uncertainties in the
    fitted slopes. Bins below \({\sim}0^\circ\) and above
    \({\sim}120^\circ\) are sparsely populated and their slopes are
    unreliable.}
  \label{fig:azimuthal_nom}
\end{figure}

\begin{figure}
  \centering
  \includegraphics[width=\linewidth]{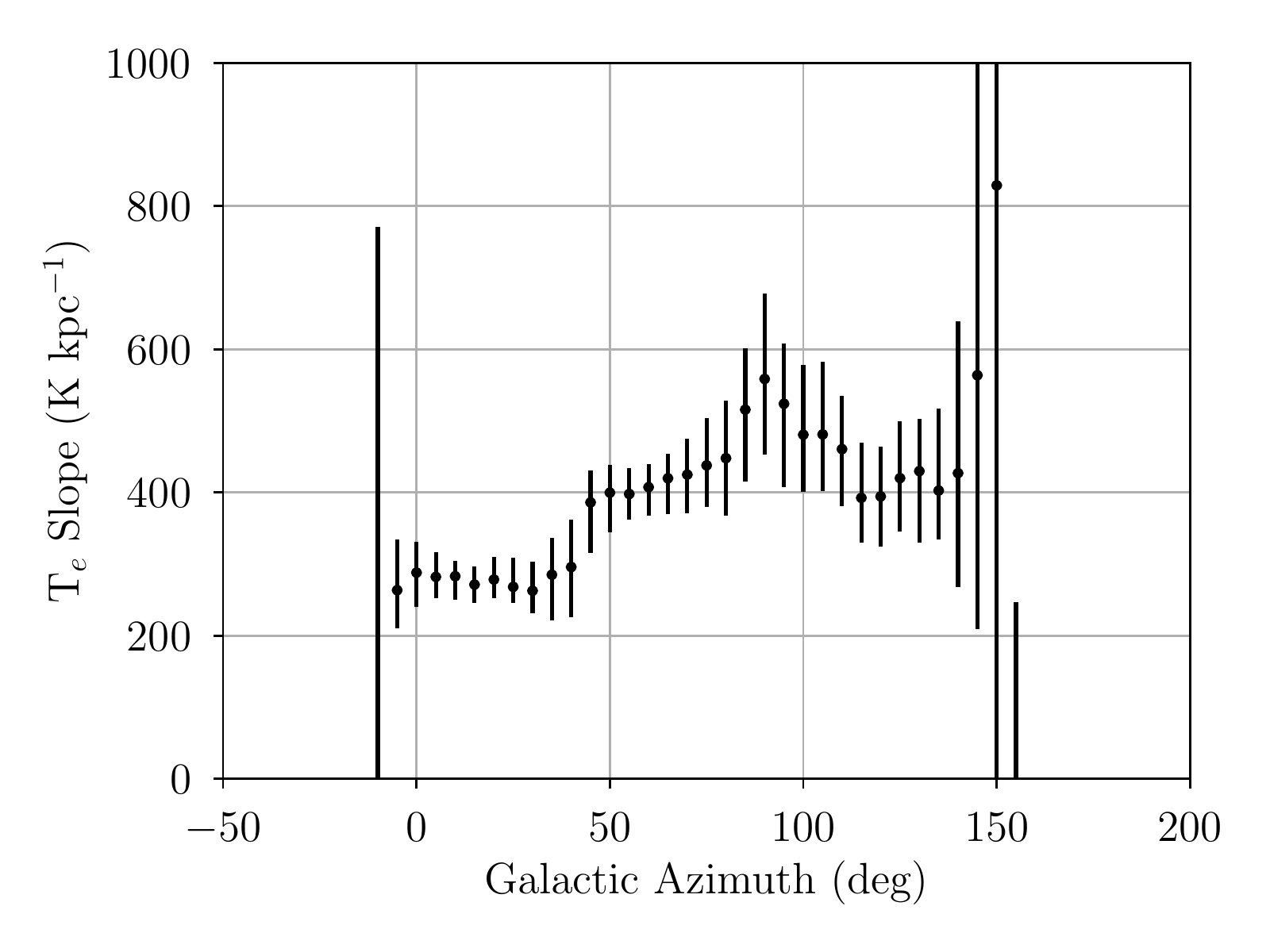}
  \includegraphics[width=\linewidth]{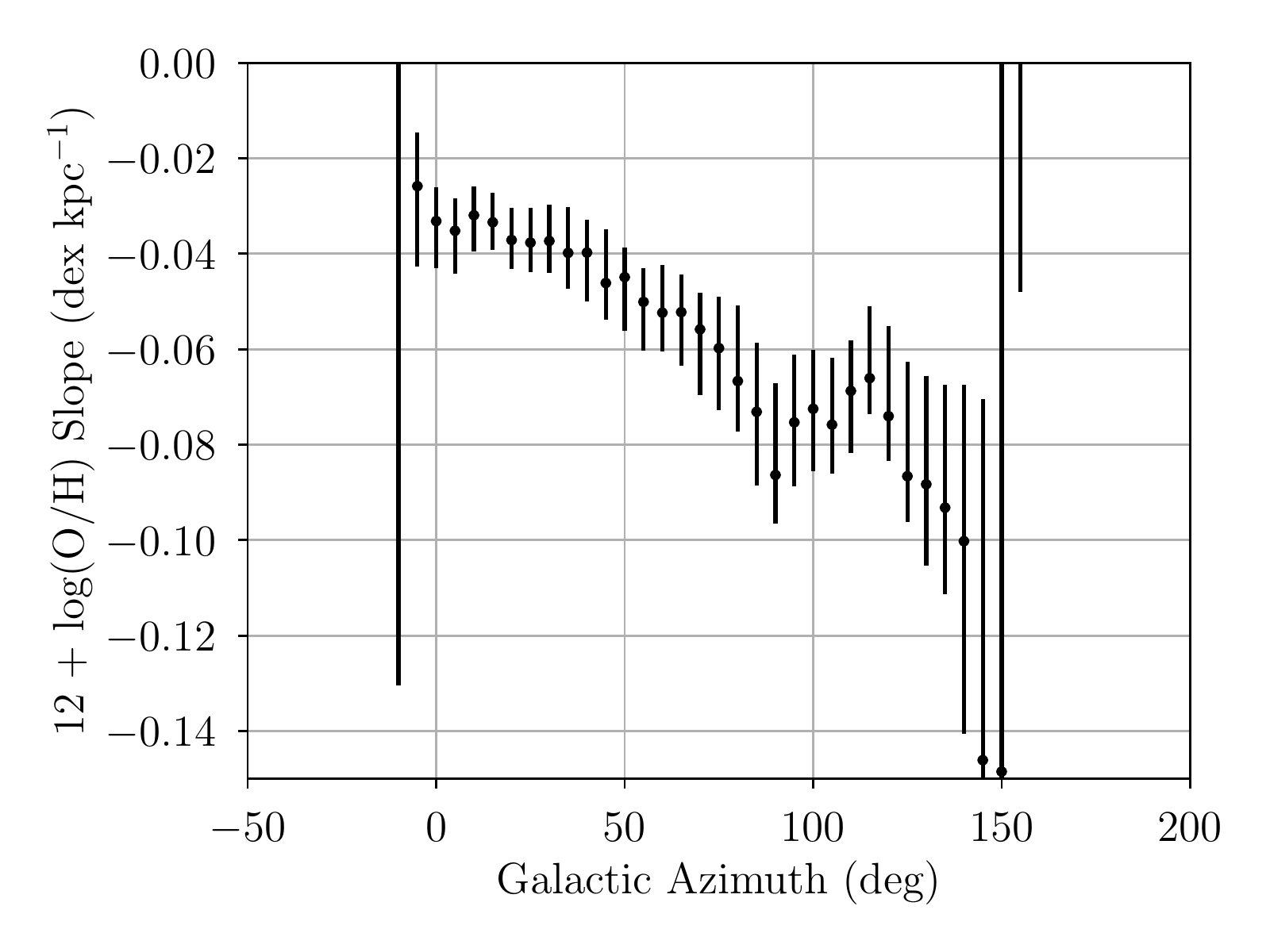} \\
  \caption{Same as Figure~\ref{fig:azimuthal_nom} for the most likely
    gradients derived from our Monte Carlo analysis. The error bars
    are the \(1\sigma\) confidence intervals on the most likely
    slopes.}
  \label{fig:azimuthal_mc}
\end{figure}

Figures~\ref{fig:te_mc} and \ref{fig:o2h_mc} show the most likely
Kriging interpolation map, most likely standard deviation map, and the
upper and lower \(1\sigma\) confidence interval bound maps for the
nebular electron temperatures and metallicities, respectively.  The
qualitative structure in the Monte Carlo Kriging interpolation maps is
similar to that in the nominal Kriging maps, though the \(1\sigma\)
confidence interval bound maps reveal where the Kriging interpolation
is ill constrained. For most of the Galactic disk, the most likely
Kriging values have \(1\sigma\) bounds \(\lesssim500\,\text{K}\) in
electron temperature and \(\lesssim0.8\,\text{dex}\) in
metallicity. These uncertainties are significantly less than the most
likely Kriging standard deviations of \({\sim}1000\,\text{K}\) and
\({\sim}0.25\,\text{dex}\), which suggests that the intrinsic scatter
in the nebular electron temperatures and metallicities exceeds the
formal uncertainties.

\section{Discussion}

The radial gradient is the most prominent feature in the metallicity
structure of the Galactic disk. Our Monte Carlo analysis of nebular
metallicities results in a most likely \hii\ region oxygen gradient of
\(-0.052\pm0.004\,\text{dex kpc\(^{-1}\)}\). \citet{molla2019a} list
the oxygen abundance gradients derived from a variety of tracers (see
their Table 2). The derived gradients range from about
\(-0.05\,\text{dex kpc\(^{-1}\)}\) for \hii\ regions and Cepheids to
about \(0\,\text{dex kpc\(^{-1}\)}\) for old stellar populations. Our
\hii\ region oxygen gradient is consistent with those found using
Cepheids \citep[e.g., \(-0.0529\pm0.0083\,\text{dex kpc\(^{-1}\)}\)
  from][]{korotin2014}, other \hii\ region samples \citep[e.g.,
  \(-0.0525\pm0.0189\,\text{dex kpc\(^{-1}\)}\)
  from][]{fernandez-martin2017}, and the \citet{molla2019a} binned
\hii\ region sample (\(-0.048 \pm 0.005\,\text{dex kpc\(^{-1}\)}\)).
  
The large variance in the measured radial metallicity gradients of
different tracers is likely due to two primary effects: (1) changes in
the metallicity gradient with time and (2) dynamical evolution of
stellar populations.  The radial gradient as traced by stars is
flatter at larger heights above the Galactic midplane
\citep{cheng2012,anders2017}. There is evidence that the stellar
metallicity gradient also flattens in the inner galaxy
\citep{hayden2015}. These stellar populations are likely older, and
thus their metallicity gradient reflects that of a younger
Galaxy. Radial migration also plays an important role in stellar
metallicity gradients \citep{sellwood2002}. The dynamical influence of
non-axisymmetric features, like spiral arms and bars, can cause stars
to migrate from their birth locations. Some studies have found that
radial migration significantly affects the observed stellar
metallicity gradients \citep[e.g.][]{minchev2013,minchev2014}, whereas
others find only an increase in the stellar metallicity dispersion at
all Galactocentric radii \citep[e.g.][]{grand2014}. These effects
should have little impact on the \hii\ region metallicity gradient,
because these nebulae are very young (\(\lesssim10\,\text{Myr}\))
compared to the dynamical timescale of the Galaxy
(\({\sim}250\,\text{Myr}\)). For example, \citet{grand2014} use a
chemodynamical simulation of a Milky Way-size galaxy to show that,
over time, the gas metallicity maintains a low dispersion at all
radii, whereas the dispersion of the stellar metallicity increases due
to radial migration.

Evidence for azimuthal variations in the radial electron temperature
and metallicity gradients has been found in the Milky Way (e.g., B15)
and other galaxies \citep[e.g.,][]{ho2017}. Here we expand upon the
B15 analysis by using a larger sample of Galactic \hii\ regions and a
more accurate kinematic distance derivation technique. Evidence for
azimuthal structure is already apparent in
Figures~\ref{fig:te_nom}--\ref{fig:o2h_mc}, and here we test the
statistical significance of these azimuthal variations.

To quantify the azimuthal structure in the nebular electron
temperature and metallicity radial gradients, we divide the Galaxy
into azimuthal bins and compute the radial gradients within each
bin. Following B15, we use bins of size \(30^\circ\) in Galactocentric
azimuth centered every \(5^\circ\) from \(-50^\circ\) to
\(200^\circ\). Using the nebulae in each bin, we make a robust least
squares linear fit to their derived electron temperatures and
metallicities as a function of their Galactocentric
radii. Figure~\ref{fig:azimuthal_nom} shows the best fit linear model
slopes as a function of Galactocentric azimuth for the nebular
electron temperature and metallicity gradients. Unlike B15, we do not
exclude bins with only a few nebulae, nor those with nebulae spanning
a small range of Galactocentric radii. The uncertainties in these bins
will be correctly determined in the subsequent Monte Carlo
analysis. In this simple least squares analysis, however, the best fit
parameters and their uncertainties are unreliable in sparsely
populated bins, such as those below \({\sim}0^\circ\) and above
\({\sim}120^\circ\). Nonetheless, we find a similar structure in the
electron temperature and metallicity gradient slopes as found by
B15. The electron temperature and metallicity slopes vary by a factor
of 2 and 3, respectively, between Galactocentric azimuths of
\({\sim}20^\circ\) and \({\sim}100^\circ\). These variations are
slightly less in magnitude than those found by B15, probably because
of our much larger sample size near \(100^\circ\) in Galactocentric
azimuth.

Multiple sources of uncertainty affect the apparent azimuthal
variations shown in Figure~\ref{fig:azimuthal_nom}. These
sources include the derived electron temperature uncertainties and the
distance uncertainties, which affect both the derived Galactocentric
radii and azimuths of the nebulae. To better quantify these sources of
uncertainty and to test the statistical significance of the apparent
azimuthal variations, we perform yet another Monte Carlo analysis. We
Monte Carlo resample the nebular electron temperatures, metallicities,
and distances to generate 1000 realizations of the data. As before,
the electron temperatures and metallacities are drawn from a Gaussian
distribution, whereas the distances are drawn from the parallax or
kinematic distance PDFs. For each realization of the data, we fit the
radial gradients in each of the several Galactocentric azimuth
bins. Finally, we fit a KDE to the linear model parameter PDFs to
estimate the most likely parameters and their confidence intervals.

Figure~\ref{fig:azimuthal_mc} shows the most likely electron
temperature and metallicity gradients from our Monte Carlo analysis.
The most obvious difference between this and the nominal gradients in
Figure~\ref{fig:azimuthal_nom} is the larger error bars. This
Monte Carlo analysis properly accounts for the uncertainties in both
the nebular electron temperatures/metallicities and distances, so
these error bars more accurately reflect the uncertainties in the
gradients within each azimuth bin. Despite the larger uncertainties,
the azimuthal variations in the radial gradients remain statistically
significant. The electron temperature gradient ranges from
\({\sim}250\,\text{K kpc\(^{-1}\)}\) at \({\sim}30^\circ\) to
\({\sim}500\,\text{K kpc\(^{-1}\)}\) at \({\sim}100^\circ\), a factor
of \({\sim}2\) increase, and the metallicity gradient ranges from
about \(-0.035\,\text{dex kpc\(^{-1}\)}\) to about \(-0.075\,\text{dex
  kpc\(^{-1}\)}\) over the same range, a factor of \({\sim}2\)
decrease.

\begin{figure}
  \centering
  \includegraphics[width=\linewidth]{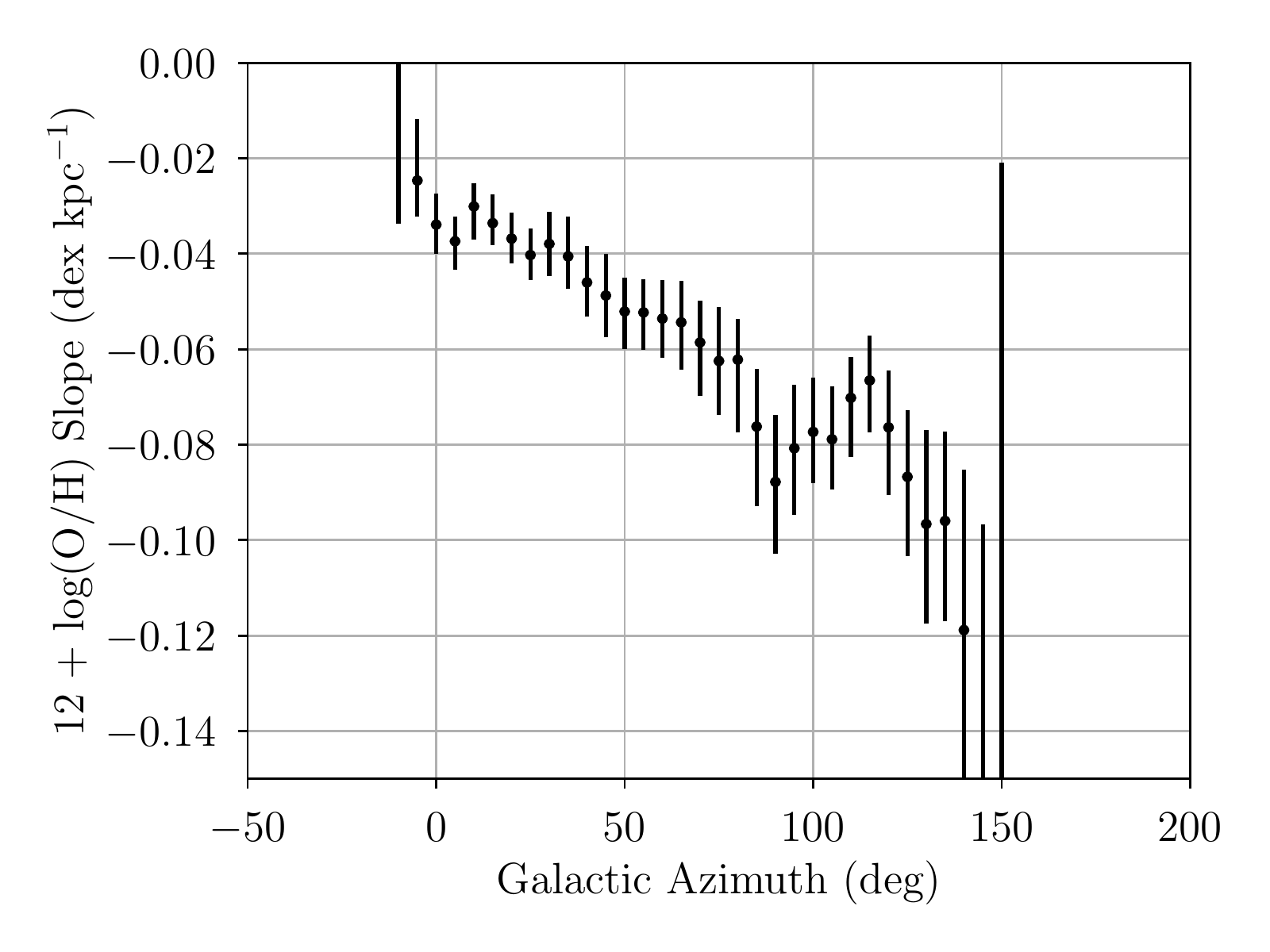}
  \includegraphics[width=\linewidth]{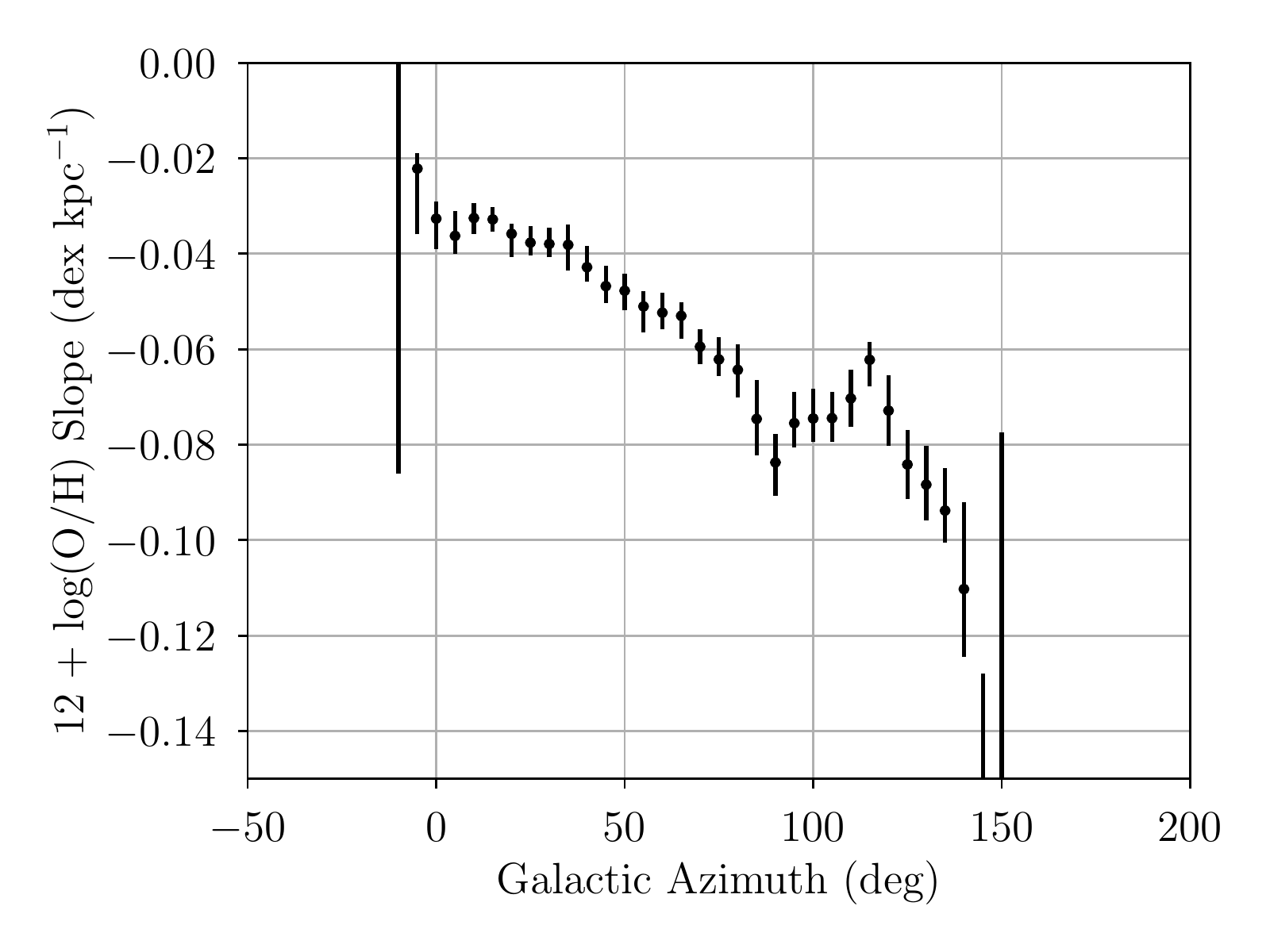} \\
  \caption{Same as the metallicity gradients in
    Figure~\ref{fig:azimuthal_mc}, except we only Monte Carlo resample
    the derived metallicities (top) or distances (bottom).}
  \label{fig:azimuthal_mctest}
\end{figure}

\begin{figure}
  \centering
  \includegraphics[width=\linewidth]{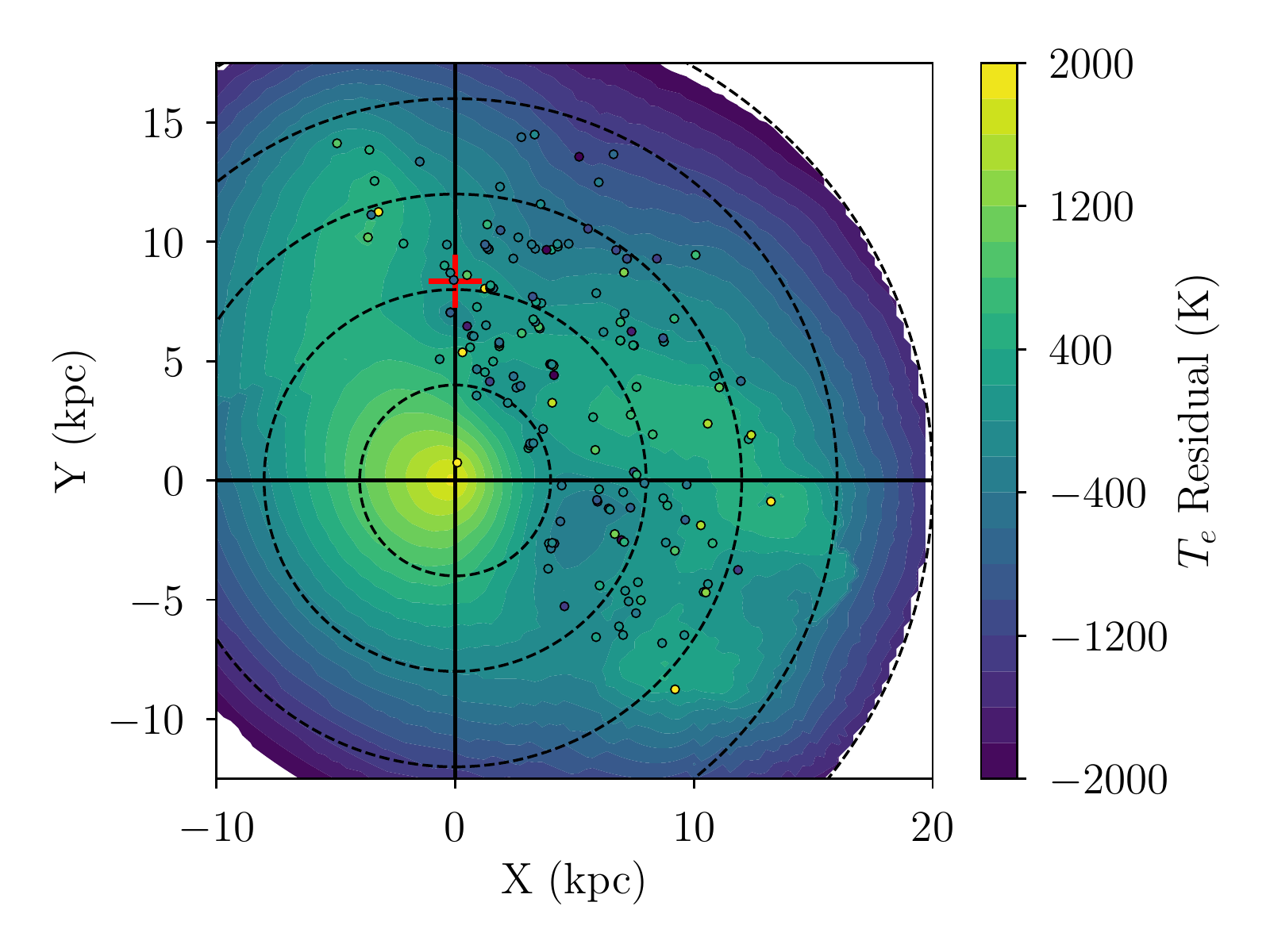}
  \includegraphics[width=\linewidth]{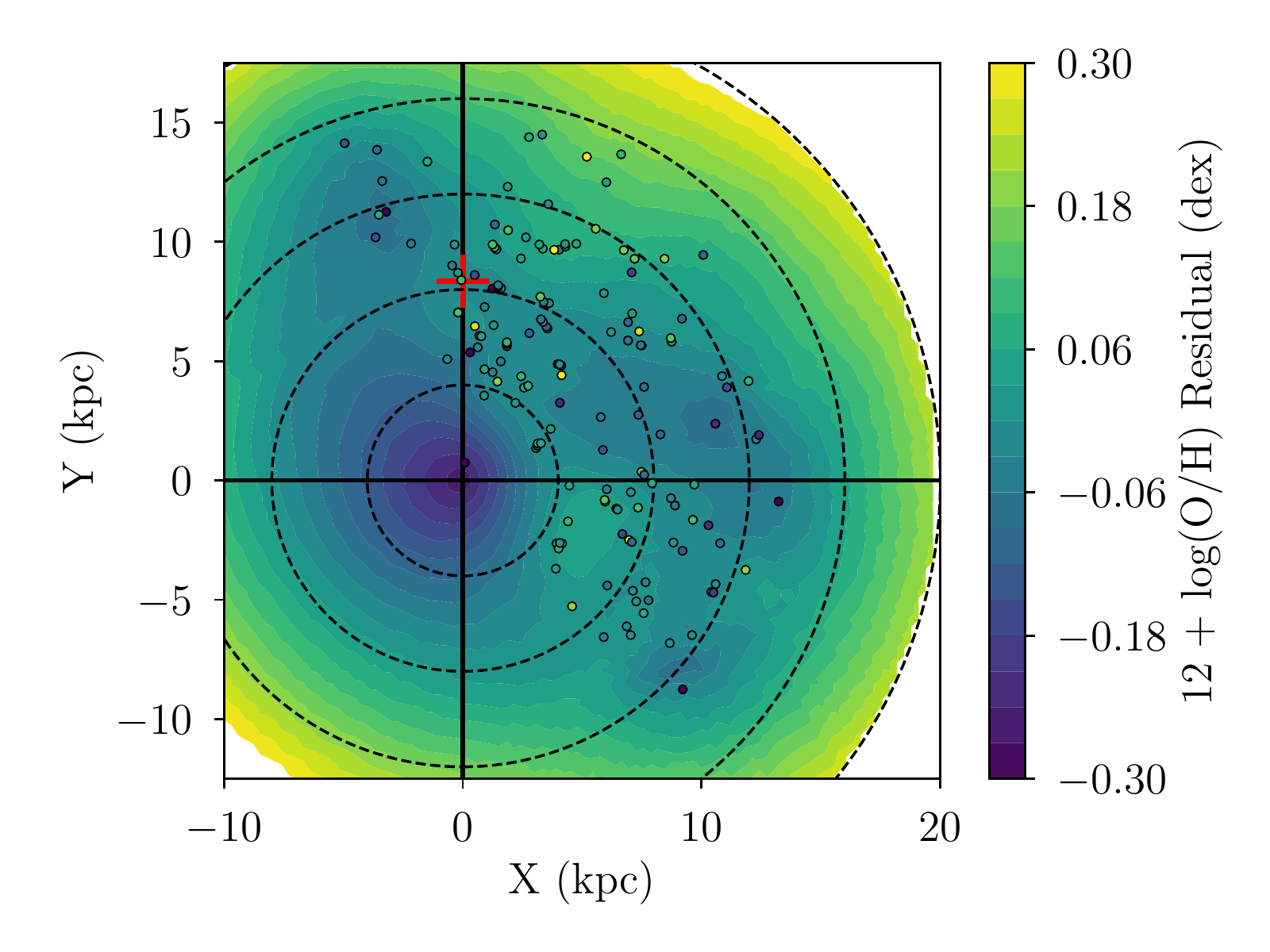}
  \caption{Most likely electron temperature (top) and metallicity
    (bottom) Kriging map residuals. The residuals are determined by
    subtracting the most likely gradient from the Monte Carlo Kriging
    maps. The features in each plot are the same as in
    Figure~\ref{fig:te_nom}.}
  \label{fig:res}
\end{figure}

The derived electron temperatures and metallicities are the largest
source of error in the radial gradient
determinations. Figure~\ref{fig:azimuthal_mctest} shows the radial
metallicity gradients in each Galactocentric azimuth bin where we
Monte Carlo resample only the metallicity (top) or only the distances
(bottom). The gradient uncertainties are a factor of \({\sim}2\)
larger when we resample only the metallicities.

The azimuthal variations in the metallicity gradient are predicted by
some simulations \citep{dimatteo2013,grand2016}. \citet{grand2016},
for example, find azimuthal metallicity structure in the young, thin
disk stellar population of a cosmological simulation of a Milky Way
analogue. The azimuthal variations are induced by the non-axisymmetric
peculiar motions near spiral arms, which drives radial migration and a
redistribution of metals. The magnitude of the azimuthal variations is
\({\sim}0.1\,\text{dex}\) in their simulation. If such stellar
azimuthal metallicity structure is persistent over long periods of
time, the enrichment of the ISM by these stars might explain the
observed azimuthal structure in the HII region metallicity
distribution. \citet{dimatteo2013} find a similar magnitude of
variation in metallicities as traced by old stars in an N-body
simulation. In Figure~\ref{fig:res} we show the residuals of the
electron temperature and metallicity Monte Carlo Kriging maps after
subtracting the most likely radial gradients. Excluding the Galactic
center and edge of the map, the magnitude of variation in the
metallicity residual map is \({\sim}0.1\,\text{dex}\), which is
consistent with the \citet{grand2016} simulation. In the first
quadrant, the residual structure between \(R{\sim}6\,\text{kpc}\) and
\({\sim}12\,\text{kpc}\) is qualitatively similar to the simulated
residuals in \citet{grand2016} and may be evidence for spiral arm
induced radial migration in the Milky Way.

Recent two-dimensional chemical evolution models also predict
azimuthal structure in the gas-phase oxygen abundance. For example,
\citet{spitoni2019} find that density fluctuations due to spiral arms
produce oxygen abundance variations on the order of
\({\sim}0.1\,\text{dex}\), with the most azimuthal structure apparent
at and beyond the corotation radius. The magnitude of these abundance
fluctuations decreases with time as the model galaxy becomes
well-mixed. This model does not consider stellar migration and
enrichment, which, according to the \citet{grand2016} simulation, are
likely important factors. \citet{molla2019b} use a 2D chemical
evolution code applied to a Milky Way analogue to conclude that spiral
arms only marginally alter the azimuthal metallicity structure. Their
model predicts present-day oxygen abundance variations of
\({\sim}0.03\,\text{dex}\) increasing to \({\sim}0.1\,\text{dex}\) in
the outer Galaxy. The oxygen abundance variations are more significant
within 1--2 Gyr after spiral arms are introduced in their model.

The nebulae in this study cover only about half of the Galactic
disk. The Southern \hii\ Region Discovery Survey
\citep[SHRDS;][]{wenger2019} is finding hundreds of new \hii\ regions
in the third and fourth Galactic quadrants, and the SHRDS
interferometric observations will allow for accurate electron
temperature and metallicity derivations. In a future work, we will
combine these northern sky nebulae with newly-discovered southern sky
\hii\ regions to create a map of \hii\ region metallicities across the
entire Galactic disk.  We will use this map to test the chemodynamical
evolution simulations by searching for evidence of metallicity
structure associated with spiral arms, the Galactic bar, and/or other
components of the Milky Way.

\section{Summary}

We use the VLA to measure the \({\sim}\)8--10 \ghz\ RRL and radio
continuum flux densities of 82 Galactic \hii\ regions. We derive the
RRL-to-continuum brightness ratio, electron temperature, and
metallicity of these nebulae. Including previous single dish
observations, the catalog of Galactic \hii\ regions with accurate
electron temperatures and distances now contains 167 nebulae spanning
Galactocentric radii \(4-16\,\text{kpc}\) and azimuths
\(-20^\circ-140^\circ\).

%% The derived VLA RRL-to-continuum brightness ratios are consistently
%% 10\% larger than the single dish values. This systematic offset is
%% partially due to the inaccurate RRL averaging techniques used by
%% previous studies. The offset may also be due to fundamental
%% differences between single dish and interferometric observations, such
%% as optical depth effects. In the future, we will test the optically
%% thin assumption be measuring the radio continuum spectral indices of
%% our nebulae. In this analysis, however, we simply scale the single
%% dish electron temperatures by \(90\%\) to offset this systematic
%% difference.

The distances to Galactic \hii\ regions are the largest source of
uncertainty in previous studies using these nebulae to trace Galactic
metallicity structure (e.g., B15). Maser parallax distances have been
determined for 46 of our nebulae. For the remainder, we use a novel
Monte Carlo kinematic distance technique to determine distances
\citep{wenger2018b}. Both the kinematic distances and distance
uncertainties to the nebulae in our sample are more accurate than in
the B15 study. In this work, the RRL-to-continuum brightness ratio
uncertainties are about twice as important as the distance
uncertainties.

Using a Monte Carlo analysis, we derive respectively the most likely
Milky Way radial electron temperature and metallicity gradients as:
\(T_e/\text{K} = 4493^{+156}_{-188} + 359^{+22}_{-18}\,R/\text{kpc}\)
and \(12 + \text{log}_{10}(\text{O/H}) = 9.130^{+0.034}_{-0.030} -
0.052^{+0.004}_{-0.004}\,R/\text{kpc}\). This metallicity gradient is
consistent with previous \hii\ region studies (e.g., B15) and young
stellar tracers, such as Cepheids \citep[e.g.,][]{korotin2014}.  We
generate maps of the electron temperature and metallicity structure of
the Galactic disk using a Monte Carlo Kriging analysis. These maps
reveal significant azimuthal variations in the Galaxy's metallicity
structure. The radial metallicity gradient varies by a factor of
\({\sim}2\) (\({\sim}0.04\,\text{dex kpc}^{-1}\)) between
Galactocentric azimuths of \({\sim}30^\circ\) and
\({\sim}100^\circ\). We find non-axisymmetric spatial metallicity
variations on the order of \({\sim}0.1\,\text{dex}\), which is
consistent with the \citet{grand2016} chemodynamical simulation. These
variations may be evidence for radial migration and metal mixing
induced by the Milky Way's spiral arms.

The Southern \hii\ Region Discovery Survey \citep{wenger2019} will add
hundreds of nebulae with electron temperature and metallicity
derivations to the third and fourth Galactic quadrants. With
\hii\ region coverage across the entire Galactic disk, we will
investigate the association between the Milky Way's metallicity
structure and the locations of spiral arms. Such structure is a test
of chemodynamical simulations and can be directly compared to
extragalactic systems.

\acknowledgments

We thank the anonymous reviewer for their constructive feedback on
this manuscript. T.V.W. is supported by the NSF through the Grote
Reber Fellowship Program administered by Associated Universities,
Inc./National Radio Astronomy Observatory, the D.N. Batten Foundation
Fellowship from the Jefferson Scholars Foundation, the Mars Foundation
Fellowship from the Achievement Rewards for College Scientists
Foundation, and the Virginia Space Grant Consortium. L.D.A. is
supported in part by NSF grant AST-1516021. T.M.B. is supported in
part by NSF grant AST-1714688.

\nraoblurb

\facility{VLA}

\software{Astropy \citep{astropy2013},
  CASA \citep{casa2007},
  KDUtils \citep{kdutils2017},
  Matplotlib \citep{matplotlib2007},
  NumPy \& SciPy \citep{numpyscipy2011},
  PyKrige \citep{murphy2014},
  Python (\url{https://www.python.org/}),
  WISP \citep{wisp2018}}

\bibliography{metallicity}

\appendix

\section{Electron Temperature Derivations}\label{sec:app_electron_temperature}

Here we derive the relationship between the nebular electron
temperature, hydrogen radio recombination line (RRL) brightness, and
radio continuum brightness of an \hii\ region. This derivation relies
on several assumptions: (1) the nebula is homogeneous, isothermal, and
in local thermodynamic equilibrium (LTE); (2) the nebula is optically
thin in both radio continuum and RRL emission; (3) the nebula is
composed solely of ionized hydrogen and singly-ionized helium; and (4)
the RRL and continuum brightness are measured with the same telescope
in the Raleigh-Jeans limit.

The free-free radio continuum absorption coefficient of an isothermal
plasma is
\begin{align}
  \frac{k_C(\nu)}{\text{pc}^{-1}} = & 3.014\times10^{-2}\left(\frac{T_e}{\rm K}\right)^{-1.5}\left(\frac{\nu}{\rm GHz}\right)^{-2} \nonumber \\
  & \times \left[\ln\left(4.955\times10^{-2}\left(\frac{\nu}{\rm GHz}\right)^{-1}\right) + 1.5\ln\left(\frac{T_e}{\rm K}\right)\right]\left(\frac{n_i\,n_e}{\text{cm\(^{-6}\)}}\right),
\end{align}
where \(T_e\) is the electron temperature, \(\nu\) is the observed
frequency, \(n_i\) is the ion number density, and \(n_e\) is the
electron number density \citep{oster1961}. \citet{altenhoff1960}
approximate the absorption coefficient as
\begin{equation}
  \frac{k_C(\nu)}{\text{pc}^{-1}} \simeq 8.235\times10^{-2}\left(\frac{T_e}{\rm K}\right)^{-1.35}\left(\frac{\nu}{\rm GHz}\right)^{-2.1}\left(\frac{n_i\,n_e}{\text{cm\(^{-6}\)}}\right), \label{eq:appD_cont_k}
\end{equation}
which is accurate within 10\% for \(100\,\mhz < \nu < 35\,\ghz\) and
\(5000\,\text{K} < T_e < 12000\,\text{K}\) \citep{mezger1967a}. The
free-free optical depth is the integral of this absorption coefficient
along the line of sight, \(\tau_C(\nu) = \int k_C(\nu)\,dl\). In a
homogeneous medium with line of sight depth \(l\), the optical depth
simplifies to \(\tau_C(\nu) = k_Cl\).

The LTE hydrogen RRL absorption coefficient for the transition from
principle quantum state \(m\) to \(n\) is
\begin{equation}
  k^*_L(\nu) = \left(\frac{h\nu_L}{kT_e}\right)\left(\frac{\pi e^2}{m_e c}\right)\left(\frac{h^2}{2\pi m_ekT_e}\right)^{3/2}n_e\,n_p\,\text{exp}\left(\frac{\chi_n}{kT_e}\right)n^2f_{nm}\phi_\nu(\nu), \label{eq:appD_rrl_start}
\end{equation}
where \(\nu_L\) is the RRL rest frequency, \(\chi_n\) is the energy
required to ionize the atom from state \(n\), \(f_{nm}\) is the
oscillator strength of the \(m\) to \(n\) transition,
\(\phi_\nu(\nu)\) is the normalized line profile with inverse
frequency units, \(n_p\) is the hydrogen number density, \(h\) is the
Planck constant, \(k\) is the Boltzmann constant, \(e\) is the
electron charge, \(m_e\) is the electron mass, and \(c\) is the speed
of light \citep{brocklehurst1972,balser1995}. The Rydberg formula
determines the transition frequency between state \(m\) and \(n\):
\begin{equation}
  \nu_L = Z^2Rc\left(n^{-2} - m^{-2}\right),
\end{equation}
where \(Z\) is the effective nuclear charge and \(R\) is the Rydberg
constant. For hydrogen, \(Z = 1\) and \(R = R_\infty(1-m_e/m_p)\),
where \(R_\infty\) is the Rydberg constant for an infinite mass and
\(m_p\) is the proton mass. If we let \(\Delta n = m - n\), then the
hydrogen transition frequencies are
\begin{equation}
  \nu_L = R_\infty c\left(1 - \frac{m_e}{m_p}\right)\left[n^{-2} - \left(n+\Delta n\right)^{-2}\right].
\end{equation}
For the low \(\Delta n\) transitions in the radio regime
(e.g. H109\(\alpha\)), \(\Delta n \ll n\) and
\begin{equation}
  \nu_L \simeq 2 R_\infty c\left(1 - \frac{m_e}{m_p}\right)\frac{\Delta n}{n^3}.
\end{equation}
Substituting these frequencies into Equation~\ref{eq:appD_rrl_start},
and assuming that we are observing in the Rayleigh-Jeans limit
\(\chi_n \ll kT_e\), the LTE RRL absorption coefficient becomes
\begin{equation}
  k^*_L(\nu) = 2R_\infty c\left(\frac{h}{kT_e}\right)\left(1 - \frac{m_e}{m_p}\right)\left(\frac{\pi e^2}{m_e c}\right)\left(\frac{h^2}{2\pi m_ekT_e}\right)^{3/2}n_e\,n_p\frac{\Delta n}{n}f_{nm}\phi_\nu(\nu). \label{eq:appD_rrl_k_full}
\end{equation}
Evaluating the constants and moving to astrophysically relevant units,
this equation becomes
\begin{equation}
  \frac{k^*_L(\nu)}{\text{pc}^{-1}} \simeq 1.070\times10^7\left(\frac{T_e}{\rm K}\right)^{-2.5}\left(\frac{n_e\,n_p}{\text{cm}^{-6}}\right)\frac{\Delta n}{n}f_{nm}\left(\frac{\phi_\nu(\nu)}{\text{Hz}^{-1}}\right). \label{eq:appD_rrl_k}
\end{equation}
The RRL optical depth is the integral of this absorption coefficient
along the line of sight, which is \(\tau^*_L(\nu) = k^*_L(\nu)l\) for
a homogeneous medium.

In an optically thin medium in LTE, the specific intensity of some
emission with optical depth \(\tau\) is \(I_\nu \simeq B_\nu(T)\tau\),
where \(B_\nu(T)\) is the Planck function at some temperature
\(T\). Assuming the RRL and continuum emission originate in the same
volume of homogeneous, isothermal gas with electron temperature,
\(T_e\), the RRL-to-continuum specific intensity ratio at \(\nu_L\) is
\begin{align}
  \frac{I_L(\nu_L)}{I_C(\nu_L)} & = \frac{\tau^*_L(\nu_L)}{\tau_C(\nu_L)} = \frac{k^*_L(\nu_L)}{k_C(\nu_L)} \nonumber \\
  \frac{I_L(\nu_L)}{I_C(\nu_L)} & = 1.300\times10^8\left(\frac{T_e}{\rm K}\right)^{-1.15}\left(\frac{\nu_L}{\rm GHz}\right)^{2.1}\frac{n_p}{n_i}\frac{\Delta n}{n}f_{nm}\left(\frac{\phi_\nu(\nu_L)}{\text{Hz}^{-1}}\right). \label{eq:appD_line2cont_deriv}
\end{align}
For a Gaussian line profile with full-width half-maximum line width
\(\Delta\nu\),
\begin{equation}
  \phi_\nu(\nu) = \frac{2}{\Delta\nu}\left(\frac{\ln 2}{\pi}\right)^{1/2}\text{exp}\left[-4\ln2\frac{(\nu-\nu_L)^2}{\Delta\nu^2}\right]
\end{equation}
and
\begin{equation}
  \phi_\nu(\nu_L) = \frac{2}{\Delta \nu}\left(\frac{\ln 2}{\pi}\right)^{1/2}. \label{eq:appD_line_profile}
\end{equation}
Using Equation~\ref{eq:appD_line_profile} in
Equation~\ref{eq:appD_line2cont_deriv}, we find
\begin{equation}
  \frac{I_L(\nu_L)}{I_C(\nu_L)} = 1.221\times10^8\frac{\Delta n}{n}\left(\frac{T_e}{\rm K}\right)^{-1.15}\left(\frac{\nu_L}{\rm GHz}\right)^{2.1}\left(\frac{\Delta \nu}{\rm Hz}\right)^{-1}\frac{n_p}{n_i}f_{nm}.
\end{equation}

If the nebulae is composed of only hydrogen and singly ionized helium,
then
\begin{equation}
  \frac{n_p}{n_i} = \frac{n_p}{n_p + n_{\text{He\(^+\)}}} = \left(1 + \frac{n_{\text{He\(^+\)}}}{n_p}\right)^{-1} = (1+y)^{-1}, \label{eq:appD_helium}
\end{equation}
where \(n_{\text{He\(^+\)}}\) is the singly ionized helium number
density and \(y \equiv n_{\text{He\(^+\)}}/n_p\) is the ratio of
singly ionized helium to hydrogen by number. We use the Doppler
equation to convert the FWHM line width from frequency to velocity
units:
\begin{equation}
  \frac{\Delta \nu}{\rm Hz} = \frac{\Delta V}{c}\left(\frac{\nu_L}{\rm Hz}\right) = 3.336\times10^3\left(\frac{\nu_L}{\rm GHz}\right)\left(\frac{\Delta V}{\text{km s}^{-1}}\right) \label{eq:appD_doppler}
\end{equation}
where \(\Delta V\) is the FWHM line width in velocity units. The
RRL-to-continuum specific intensity ratio at \(\nu_L\) is thus
\begin{equation}
  \frac{I_L(\nu_L)}{I_C(\nu_L)} = 3.661\times10^4\left(\frac{T_e}{\rm K}\right)^{-1.15}\left(\frac{\nu_L}{\rm GHz}\right)^{1.1}\left(\frac{\Delta V}{\text{km s}^{-1}}\right)^{-1}\left(1+y\right)^{-1}\frac{\Delta n}{n}f_{nm}.
\end{equation}

The expression \((f_{nm}\Delta n/n)\) is not a strong function of
\(n\) for \(\Delta n = 1\) hydrogen RRLs. For example, \((f_{nm}\Delta
n/n) = 0.19435\), \(0.19395\), and \(0.19363\) for \(\Delta n = 1\)
and \(n = 80\), \(90\), and \(100\), respectively, using the
oscillator strengths from \citet{menzel1968}. This variation is less
than \(0.3\%\) across these H\(n\alpha\) transitions, so we adopt the
H90\(\alpha\) oscillator strength to simplify the RRL-to-continuum
specific intensity equation as
\begin{equation}
  \frac{I_L(\nu_L)}{I_C(\nu_L)} = 7.100\times10^3\left(\frac{T_e}{\rm K}\right)^{-1.15}\left(\frac{\nu_L}{\rm GHz}\right)^{1.1}\left(\frac{\Delta V}{\text{km s}^{-1}}\right)^{-1}\left(1+y\right)^{-1}. \label{eq:appD_line2cont}
\end{equation}
Solving for the electron temperature, we find
\begin{equation}
   \frac{T_e}{\rm K} = \left[7.100\times10^3\left(\frac{I_C(\nu_L)}{I_L(\nu_L)}\right)\left(\frac{\nu_L}{\rm GHz}\right)^{1.1}\left(\frac{\Delta V}{\text{km s}^{-1}}\right)^{-1}\left(1+y\right)^{-1}\right]^{0.87} \label{eq:appD_te}
\end{equation}

\subsection{Single Dish Observations}

Single dish telescopes measure intensity in units of antenna
temperature, \(T_A\). In the absence of atmospheric attenuation, the
antenna temperature is related to the brightness temperature
distribution, \(T_B(\theta)\), by
\begin{equation}
  T_A = \frac{\eta_b}{\Omega_b}2\pi\int_0^\infty f(\theta)T_B(\theta)\,\sin\theta\,d\theta, \label{eq:appD_ta1}
\end{equation}
where \(\eta_b\) is the telescope beam efficiency, \(\Omega_b\) is the
telescope main beam solid angle, \(f(\theta)\) is the telescope beam
pattern, and the integral is the convolution of the source brightness
distribution with the telescope beam \citep{mezger1967a}. For a
Gaussian beam with half-power beam width (HPBW) \(\theta_b\), the beam
pattern is \(f(\theta) = \text{exp}[-4\ln(2)\,\theta^2/\theta_b^2]\)
and the beam solid angle is \(\Omega_b = 2\pi\int_0^\infty f(\theta)
\theta\,d\theta = \pi\theta_b^2/(4\ln 2)\).  Similarly, if the source
brightness distribution is Gaussian with amplitude \(T_B\) and
half-power width \(\theta_s\), then the source brightness temperature
distribution is \(T_B(\theta) =
T_B\text{exp}[-4\ln(2)\,\theta^2/\theta_s^2]\). In astronomy,
\(\theta\) is typically very small, \(\sin\theta \simeq \theta\), and
the integral in Equation~\ref{eq:appD_ta1} is
\begin{align}
  \int_0^\infty f(\theta)T_B(\theta)\,\sin\theta\,d\theta & \simeq \int_0^\infty f(\theta)T_B(\theta)\,\theta\,d\theta \nonumber \\
  & = T_B\int_0^\infty \text{exp}\left[-4\ln(2)\theta^2\left(\frac{\theta_s^2 + \theta_b^2}{\theta_s^2\theta_b^2}\right)\right]\,\theta\,d\theta \nonumber \\
  & = \frac{T_B}{8\ln2}\left(\frac{\theta_s^2\theta_b^2}{\theta_s^2 + \theta_b^2}\right),
\end{align}
where we use \(\int_0^\infty x\,\text{exp}(-ax^2)\,dx = 1/(2a)\). The
antenna temperature is thus
\begin{equation}
  T_A = \eta_bT_B\left(\frac{\theta_s^2}{\theta_s^2 + \theta_b^2}\right). \label{eq:appD_ta}
\end{equation}
For a resolved source, \(\theta_s \gg \theta_b\) and \(T_A \simeq
\eta_bT_B\). For an unresolved source, \(\theta_s \ll \theta_b\) and
\(T_A \simeq \eta_b T_B (\theta_s^2/\theta_b^2)\).

Brightness temperature is defined as
\begin{equation}
  T_B \equiv \frac{c^2}{2k\nu^2}I_\nu,
\end{equation}
where \(I_\nu\) is the specific intensity. For an optically thin
medium, \(I_\nu = B_\nu(T)\tau\), where \(T\) is the blackbody
temperature of the emission and \(\tau\) is the optical depth.
In the Rayleigh-Jeans limit, the brightness temperature is
simply
\begin{equation}
  T_B = \frac{c^2}{2k\nu^2}B_\nu(T)\tau \simeq T\tau. \label{eq:appD_tb}
\end{equation}
Substituting Equation~\ref{eq:appD_tb} into Equation~\ref{eq:appD_ta},
we find
\begin{equation}
  T_A = \eta_bT\tau\left(\frac{\theta_s^2}{\theta_s^2 + \theta_b^2}\right).
\end{equation}

If the RRL and continuum antenna temperatures are measured with the
same telescope and at the same frequency, and if both sources of
emission originate from the same volume of homogeneous and isothermal
gas with electron temperature \(T_e\),
Equation~\ref{eq:appD_line2cont} is trivially
\begin{equation}
  \frac{T_L(\nu_L)}{T_C(\nu_L)} = 7.100\times10^3\left(\frac{T_e}{\rm K}\right)^{-1.15}\left(\frac{\nu_L}{\rm GHz}\right)^{1.1}\left(\frac{\Delta V}{\text{km s}^{-1}}\right)^{-1}\left(1+y\right)^{-1}, \label{eq:appD_line2cont_sd}
\end{equation}
where \(T_C(\nu_L)\) and \(T_L(\nu_L)\) are the continuum and RRL
antenna temperatures measured at the RRL frequency \(\nu_L\),
respectively. Equation~\ref{eq:appD_te} becomes
\begin{equation}
  \frac{T_e}{\rm K} \simeq \left[7.100\times10^3\left(\frac{T_C(\nu_L)}{T_L(\nu_L)}\right)\left(\frac{\nu_L}{\rm GHz}\right)^{1.1}\left(\frac{\Delta V}{\text{km s}^{-1}}\right)^{-1}\left(1+y\right)^{-1}\right]^{0.87}. \label{eq:appD_te_sd}
\end{equation}

\subsection{Averaging Single Dish RRLs}

In Galactic \hii\ region surveys, we average multiple RRL transitions
to increase the RRL signal-to-noise ratio. Each RRL transition is an
independent measurement of the nebular electron temperature, so the
electron temperature derived from many RRL-to-continuum antenna
temperature measurements is
\begin{equation}
  \frac{T_e}{\rm K} = \left[7.100\times10^3\left(\frac{\Delta V}{\text{km s}^{-1}}\right)^{-1}\left(1+y\right)^{-1}\right]^{0.87}\left<\left[\left(\frac{T_C(\nu_L)}{T_L(\nu_L)}\right)\left(\frac{\nu_L}{\rm GHz}\right)^{1.1}\right]^{0.87}\right>, \label{eq:appD_avg_te}
\end{equation}
assuming that adjacent RRL transitions have similar FWHM line widths
in velocity units \citep[e.g.,][]{balser2011}. Previous single dish
RRL studies have used different strategies for averaging RRL
transitions. \citet{balser2011} and \citet{balser2015}, for example,
scale each RRL antenna temperature to account for the variations in
telescope beam size with frequency, then average the re-scaled RRL
spectra. They measure the continuum antenna temperature at one
frequency within the RRL frequency range, then take the ratio of the
average RRL antenna temperatures to this continuum temperature. This
strategy is an approximation to Equation~\ref{eq:appD_avg_te}.

Here we compute the difference between the true electron temperature
and the \citet{balser2011} and \citet{balser2015} approximation using
multiple RRL transitions. From Equation~\ref{eq:appD_avg_te}, the
factor we need to derive is
\begin{align}
  X_{\rm true} & = \left<\left[\left(\frac{T_C(\nu_L)}{T_L(\nu_L)}\right)\left(\frac{\nu_L}{\rm GHz}\right)^{1.1}\right]^{0.87}\right> \nonumber \\
  & = \left(\frac{T_e}{\rm K}\right)\left[7.100\times10^3\left(\frac{\Delta V}{\text{km s}^{-1}}\right)^{-1}(1+y)^{-1}\right]^{-0.87},
\end{align}
where \(T_e \propto X_{\rm true}\) and \(X_{\rm true}\) is the only variable in
Equation~\ref{eq:appD_avg_te} that depends on the RRL
transition. \citet{balser2011} and \citet{balser2015} approximate this
factor as
\begin{equation}
  X = \left[\left(\frac{T_C(\nu_C)}{\langle T^*_L(\nu_L)\rangle}\right)\left(\frac{\langle \nu_L\rangle}{\rm GHz}\right)^{1.1}\right]^{0.87},
\end{equation}
where \(\nu_C\) is the observed continuum frequency and
\(T^*_L(\nu_L)\) is the RRL antenna temperature corrected for the
variation of telescope beam size with frequency. They re-scale the
observed RRL antenna temperature, \(T_L(\nu_L)\), using
\begin{equation}
  T^*_L(\nu_L) = T_L(\nu_L)\left(\frac{\theta_s^2 + \theta_b^2}{\theta_s^2 + (\theta^*_b)^2}\right),
\end{equation}
where \(\theta_b\) is the HPBW at \(\nu_L\), and \(\theta^*_b\) is the
HPBW at \(\nu_C\), and \(\theta_s\) is the half-power width of the
source. The observed source brightness distribution is the convolution
of the actual source brightness and the telescope beam. With the
assumption that the telescope beam and source brightness distribution
are Gaussian, the convolution is also a Gaussian with half-power width
\(\theta_o^2 = \theta_s^2 + \theta_b^2\). \citet{balser2011} and
\citet{balser2015} measure the source half-power width at \(\nu_C\),
which is \((\theta_o^*)^2 = \theta_s^2 + (\theta^*_b)^2\). The true,
deconvolved source size is \(\theta_s^2 = (\theta_o^*)^2 -
(\theta^*_b)^2\), and the re-scaled antenna temperature in terms of
observables is
\begin{equation}
  T^*_L(\nu_L) = T_L(\nu_L)\left(\frac{(\theta_o^*)^2 - (\theta^*_b)^2 + \theta_b^2}{(\theta^*_o)^2}\right).
\end{equation}
For a point source, \((\theta_o^*)^2 \simeq (\theta^*_b)^2\) and
\(T^*_L(\nu_L) \simeq T_L(\nu_L)[\theta_b^2/(\theta^*_b)^2]\), whereas
if the source is very resolved, \((\theta_o^*)^2 \gg (\theta^*_b)^2\)
and \(T^*_L(\nu_L) \simeq T_L(\nu_L)\).

Using Equations~\ref{eq:appD_rrl_k}, \ref{eq:appD_line_profile}, and
\ref{eq:appD_doppler}, the line center LTE optical depth of the
\(i\)th RRL transition is
\begin{equation}
  \tau^*_{L,i}(\nu_{L,i}) = 584.47\left(\frac{T_e}{\rm K}\right)^{-2.5}\left(\frac{n_e\,n_p}{\text{cm}^{-6}}\right)\left(\frac{\nu_{L,i}}{\rm GHz}\right)^{-1}\left(\frac{\Delta V}{\text{km s}^{-1}}\right)^{-1}\left(\frac{l}{\rm pc}\right),
\end{equation}
where we have assumed \(\tau^*_{L,i} = \int k^*_{L,i}\,dl =
k^*_{L,i}l\) for a homogeneous medium with an LTE absorption
coefficient \(k^*_{L,i}\) and a line of sight depth \(l\). The
antenna temperature of this transition is
\begin{align}
  T_{L,i}(\nu_{L,i}) & = T_e\tau^*_{L,i}(\nu_{L,i})\eta_b\left(\frac{\theta^2_s}{\theta^2_s + \theta^2_b}\right) \nonumber \\
  & = 584.47\eta_b\left(\frac{T_e}{\rm K}\right)^{-1.5}\left(\frac{n_e\,n_p}{\text{cm}^{-6}}\right)\left(\frac{\nu_{L,i}}{\rm GHz}\right)^{-1}\left(\frac{\Delta V}{\text{km s}^{-1}}\right)^{-1}\left(\frac{l}{\rm pc}\right) \nonumber \\
  & \times \left(\frac{\theta^2_s}{\theta^2_s + \theta^2_b}\right) \label{eq:appD_tl}
\end{align}
and the re-scaled RRL antenna temperature is
\begin{align}
  T^*_{L,i}(\nu_{L,i}) & = 584.47\eta_b\left(\frac{T_e}{\rm K}\right)^{-1.5}\left(\frac{n_e\,n_p}{\text{cm}^{-6}}\right)\left(\frac{\nu_{L,i}}{\rm GHz}\right)^{-1}\left(\frac{\Delta V}{\text{km s}^{-1}}\right)^{-1}\left(\frac{l}{\rm pc}\right) \nonumber \\
   & \times \left(\frac{\theta^2_s}{\theta^2_s + (\theta^*_b)^2}\right).
\end{align}
The average re-scaled RRL antenna temperature of several RRL
transitions is
\begin{align}
  \langle T^*_{L}(\nu_L)\rangle & = 584.47\eta_b \left(\frac{T_e}{\rm K}\right)^{-1.5} \left(\frac{n_e\,n_p}{\text{cm}^{-6}}\right)\left(\frac{\Delta V}{\text{km s}^{-1}}\right)^{-1}\left(\frac{l}{\rm pc}\right) \nonumber \\
  & \times \left<\left(\frac{\nu_{L}}{\rm GHz}\right)^{-1}\left(\frac{\theta^2_s}{\theta^2_s + (\theta^*_b)^2}\right)\right> \label{eq:appD_balser_tl}
\end{align}
assuming that the RRLs have similar FWHM line widths in velocity
units.

From Equation~\ref{eq:appD_cont_k}, the continuum optical depth at
frequency \(\nu_C\) is
\begin{equation}
  \tau_{C}(\nu_C) = 8.235\times10^{-2}\left(\frac{T_e}{\rm K}\right)^{-1.35}\left(\frac{\nu_C}{\rm GHz}\right)^{-2.1}\left(\frac{n_i\,n_e}{\text{cm\(^{-6}\)}}\right)\left(\frac{l}{\rm pc}\right),
\end{equation}
where, again, we have assumed that the medium is homogeneous. The
continuum antenna temperature is
\begin{align}
  T_C(\nu_C) & = T_e\tau_C(\nu_C)\eta_b\left(\frac{\theta^2_s}{\theta^2_s + (\theta_b^*)^2}\right) \nonumber \\
  & = 8.235\times10^{-2}\eta_b\left(\frac{T_e}{\rm K}\right)^{-0.35}\left(\frac{\nu_C}{\rm GHz}\right)^{-2.1}\left(\frac{n_i\,n_e}{\text{cm\(^{-6}\)}}\right)\left(\frac{l}{\rm pc}\right) \nonumber \\
  & \times \left(\frac{\theta_s}{\theta_s + (\theta_b^*)^2}\right). \label{eq:appD_balser_tc}
\end{align}

Substituting Equations~\ref{eq:appD_balser_tl} and
\ref{eq:appD_balser_tc} into the \citet{balser2011} and
\citet{balser2015} approximation of \(X\), we find
\begin{align}
  X = & \left(\frac{T_e}{\rm K}\right)\left[1.409\times10^{-4} \left(\frac{\nu_C}{\rm GHz}\right)^{-2.1}\left(\frac{\Delta V}{\text{km s}^{-1}}\right)(1+y)\left(\frac{\langle \nu_L\rangle}{\rm GHz}\right)^{1.1}\right]^{0.87} \nonumber \\
      & \times \left<\left(\frac{\nu_L}{\rm GHz}\right)^{-1}\right>^{-0.87}.
\end{align}
The ratio of the \(X\) approximation and \(X_{\rm true}\) is
\begin{equation}
  \frac{X}{X_{\rm true}} = \left(\frac{\nu_C}{\rm GHz}\right)^{-1.827}\left(\frac{\langle \nu_L\rangle}{\rm GHz}\right)^{0.957}\left<\left(\frac{\nu_L}{\rm GHz}\right)^{-1}\right>^{-0.87}.
\end{equation}
As a sanity check on this expression, if the RRL and continuum antenna
temperatures are measured at only one RRL frequency, then \(\nu_C =
\nu_L = \langle \nu_L\rangle\) and this ratio is
unity. \citet{balser2015} measured the continuum antenna temperature
at \(\nu_C = 8.556\ghz\) and the RRL antenna temperature for 6
H\(n\alpha\) transitions (H87\(\alpha\) to H93\(\alpha\), excluding
H90\(\alpha\)). The average RRL frequency is \(\langle \nu_L\rangle =
8.903\ghz\), but \citet{balser2015} use \(\langle \nu_L\rangle =
9\ghz\). With these values, the \(X\) ratio is
\begin{equation}
  \frac{X}{X_{\rm true}} = 1.057.
\end{equation}
Therefore, \citet{balser2015} and other studies that average the same
RRL transitions and observe the same continuum frequency will
overestimate the derived electron temperatures by \({\sim}5.7\%\).

\citet{quireza2006a} and \citet{quireza2006b} use different RRL
transitions and calibration strategies to derive electron
temperatures.  In their \cii\ survey, they observe H91\(\alpha\) and
H92\(\alpha\).  They assume both transitions have the same antenna
temperature in the bright \hii\ region W3, and they use this
assumption to calibrate H92\(\alpha\) relative to H91\(\alpha\). From
Equation~\ref{eq:appD_tl}, this calibration factor is
\begin{equation}
  \frac{T_{L,\,\text{H91}\alpha}}{T_{L,\,\text{H92}\alpha}} = \left(\frac{\nu_{\text{H92}\alpha}}{\nu_{\text{H91}\alpha}}\right)\left(\frac{\theta^2_s + \theta^2_{b,\,\text{H92}\alpha}}{\theta^2_s + \theta^2_{b,\,\text{H91}\alpha}}\right).
\end{equation}
W3 is unresolved in their survey, so \(\theta^2_s + \theta^2_b \simeq
\theta^2_b\). Using \(\theta^2_b \propto \nu^{-2}\) for a Gaussian
beam, this factor becomes
\begin{equation}
  \frac{T_{L,\,\text{H91}\alpha}}{T_{L,\,\text{H92}\alpha}} \simeq \left(\frac{\nu_{\text{H91}\alpha}}{\nu_{\text{H92}\alpha}}\right) = 1.033.
\end{equation}
Therefore, the average RRL antenna temperature in their surveys is
\begin{align}
  \langle T_L(\nu_L) \rangle & = \frac{1}{2}\left(T_{L,\,\text{H91}\alpha} + 1.033T_{L,\,\text{H92}\alpha}\right) \nonumber \\
  \langle T_L(\nu_L) \rangle & = 292.235\eta_b\left(\frac{T_e}{\rm K}\right)^{-1.5}\left(\frac{n_e\,n_p}{\rm cm^{-6}}\right)\left(\frac{\Delta V}{\text{km s}^{-1}}\right)^{-1}l \nonumber \\
  & \times \left[\left(\frac{\nu_{\text{H91}\alpha}}{\rm GHz}\right)^{-1}\left(\frac{\theta_s^2}{\theta_s^2+\theta^2_{b,\,\text{H91}\alpha}}\right) + 1.033\left(\frac{\nu_{\text{H92}\alpha}}{\rm GHz}\right)^{-1}\left(\frac{\theta_s^2}{\theta_s^2+\theta^2_{b,\,\text{H92}\alpha}}\right)\right].
\end{align}
The continuum antenna temperature at \(\nu_C\) is given by
Equation~\ref{eq:appD_balser_tc}, and the
\citet{quireza2006a,quireza2006b} \(X\) is
\begin{align}
  X & = \left(\frac{T_e}{\rm K}\right)\left[2.818\times10^{-4}\left(\frac{\nu_C}{\rm GHz}\right)^{-2.1}\left(\frac{\Delta V}{\text{km s}^{-1}}\right)(1+y)\left(\frac{\langle \nu_L\rangle}{\rm GHz}\right)^{1.1}\right]^{0.87} \nonumber \\
  & \times \left[\left(\frac{\nu_{\text{H91}\alpha}}{\rm GHz}\right)^{-1}\left(\frac{\theta_s^2+(\theta_b^*)^2}{\theta_s^2+\theta^2_{b,\,\text{H91}\alpha}}\right) + 1.033\left(\frac{\nu_{\text{H92}\alpha}}{\rm GHz}\right)^{-1}\left(\frac{\theta_s^2+(\theta_b^*)^2}{\theta_s^2+\theta^2_{b,\,\text{H92}\alpha}}\right)\right]^{-0.87}.
\end{align}
The ratio of this \(X\) approximation to \(X_{\rm true}\) is
\begin{align}
  \frac{X}{X_{\rm true}} & = 1.829\left(\frac{\nu_C}{\rm GHz}\right)^{-1.827}\left(\frac{\langle \nu_L\rangle}{\rm GHz}\right)^{0.957} \nonumber \\
  & \times \left[\left(\frac{\nu_{\text{H91}\alpha}}{\rm GHz}\right)^{-1}\left(\frac{\theta_s^2+(\theta_b^*)^2}{\theta_s^2+\theta^2_{b,\,\text{H91}\alpha}}\right) + 1.033\left(\frac{\nu_{\text{H92}\alpha}}{\rm GHz}\right)^{-1}\left(\frac{\theta_s^2+(\theta_b^*)^2}{\theta_s^2+\theta^2_{b,\,\text{H92}\alpha}}\right)\right]^{-0.87}.
\end{align}

\citet{quireza2006a} and \citet{quireza2006b} did not account for the
variation in telescope beam size in their analysis. Therefore, their
ratio \(X/X_{\rm true}\) has a dependence on the source size. In the
limit that the source is unresolved at all frequencies, \(\theta_s^2
\ll \theta^2_b\). For Gaussian beams with \(\theta^2 \propto
\nu^{-2}\), the ratio in this limit becomes
\begin{align}
  \lim_{\theta_s^2 \ll \theta^2_b} \frac{X}{X_{\rm true}} & = 1.829\left(\frac{\nu_C}{\rm GHz}\right)^{-1.827}\left(\frac{\langle \nu_L\rangle}{\rm GHz}\right)^{0.957} \nonumber \\
   & \times \left[\left(\frac{\nu_{\text{H91}\alpha}}{\rm GHz}\right)^{-1}\left(\frac{(\theta_b^*)^2}{\theta^2_{b,\,\text{H91}\alpha}}\right) + 1.033\left(\frac{\nu_{\text{H92}\alpha}}{\rm GHz}\right)^{-1}\left(\frac{(\theta_b^*)^2}{\theta^2_{b,\,\text{H92}\alpha}}\right)\right]^{-0.87} \nonumber \\
  \lim_{\theta_s^2 \ll \theta^2_b} \frac{X}{X_{\rm true}} & = 1.829\left(\frac{\nu_C}{\rm GHz}\right)^{-1.827}\left(\frac{\langle \nu_L\rangle}{\rm GHz}\right)^{0.957} \nonumber \\
   & \times \left[\left(\frac{\nu_{\text{H91}\alpha}}{\rm GHz}\right)^{-1}\left(\frac{\nu_{\text{H91}\alpha}^2}{\nu_C^2}\right) + 1.033\left(\frac{\nu_{\text{H92}\alpha}}{\rm GHz}\right)^{-1}\left(\frac{\nu_{\text{H92}\alpha}^2}{\nu_C^2}\right)\right]^{-0.87} \nonumber \\
  \lim_{\theta_s^2 \ll \theta^2_b} \frac{X}{X_{\rm true}} & = 1.829\left(\frac{\nu_C}{\rm GHz}\right)^{-0.087}\left(\frac{\langle \nu_L\rangle}{\rm GHz}\right)^{0.957}\left[\left(\frac{\nu_{\text{H91}\alpha}}{\rm GHz}\right) + 1.033\left(\frac{\nu_{\text{H92}\alpha}}{\rm GHz}\right)\right]^{-0.87}.
\end{align}
Using the RRL frequencies, \(\nu_C = 8.665\ghz\), and \(\langle \nu_L\rangle = \nu_{\text{H91}\alpha}\),
which is what \citet{quireza2006a,quireza2006b} use, we find
\begin{equation}
  \lim_{\theta_s^2 \ll \theta^2_b} \frac{X}{X_{\rm true}} = 1.0.
\end{equation}
For unresolved sources, \citet{quireza2006a,quireza2006b} correctly
calculate the electron temperatures in their \cii\ survey. In the
resolved case, \(\theta_s^2 \gg \theta^2_b\) and
\begin{align}
  \lim_{\theta_s^2 \gg \theta^2_b} \frac{X}{X_{\rm true}} & = 1.829\left(\frac{\nu_C}{\rm GHz}\right)^{-1.827}\left(\frac{\langle \nu_L\rangle}{\rm GHz}\right)^{0.957}\left[\left(\frac{\nu_{\text{H91}\alpha}}{\rm GHz}\right)^{-1} + 1.033\left(\frac{\nu_{\text{H92}\alpha}}{\rm GHz}\right)^{-1}\right]^{-0.87} \nonumber \\
  \lim_{\theta_s^2 \gg \theta^2_b} \frac{X}{X_{\rm true}} & = 0.956.
\end{align}
Therefore, the \citet{quireza2006b} electron temperatures for the
\cii\ survey nebulae are underestimated by up to \(5\%\) depending on
the source morphology.

In their \(^3\)He survey, \citet{quireza2006a,quireza2006b} only
observe the H91\(\alpha\) transition. The \(X\) ratio for this survey
is simpler:
\begin{equation}
  \frac{X}{X_{\rm true}} = \left(\frac{\nu_C}{\rm GHz}\right)^{-1.827}\left(\frac{\nu_{\text{H91}\alpha}}{\rm GHz}\right)^{1.827}\left(\frac{\theta_s^2+(\theta_b^*)^2}{\theta_s^2+\theta^2_{b,\,\text{H91}\alpha}}\right)^{-0.87},
\end{equation}
with limits
\begin{align}
  \lim_{\theta_s^2 \ll \theta^2_b} \frac{X}{X_{\rm true}} & = \left(\frac{\nu_C}{\rm GHz}\right)^{-1.827}\left(\frac{\nu_{\text{H91}\alpha}}{\rm GHz}\right)^{1.827}\left(\frac{(\theta_b^*)^2}{\theta^2_{b,\,\text{H91}\alpha}}\right)^{-0.87} \nonumber \\
  & = \left(\frac{\nu_C}{\rm GHz}\right)^{-0.087}\left(\frac{\nu_{\text{H91}\alpha}}{\rm GHz}\right)^{0.087} \nonumber \\
  \lim_{\theta_s^2 \ll \theta^2_b} \frac{X}{X_{\rm true}} & = 1.0
\end{align}
and
\begin{align}
  \lim_{\theta_s^2 \gg \theta^2_b} \frac{X}{X_{\rm true}} & = \left(\frac{\nu_C}{\rm GHz}\right)^{-1.827}\left(\frac{\nu_{\text{H91}\alpha}}{\rm GHz}\right)^{1.827} \nonumber \\
  \lim_{\theta_s^2 \gg \theta^2_b} \frac{X}{X_{\rm true}} & = 0.983.
\end{align}
\citet{quireza2006b} underestimate their electron temperatures by up
to \(2\%\) in their \(^3\)He survey.

\begin{deluxetable}{lcccc}
%\tabletypesize{\normalsize}
\tablecaption{Single Dish Electron Temperature Corrections\label{tab:appD_sd_corrections}}
\tablehead{
  \colhead{Author} & \colhead{RRLs} & \colhead{\(\nu_C\)} & \colhead{\(<\nu_L>\)\tablenotemark{a}} & \colhead{\(X/X_{\rm true}\)} \\
  \colhead{}       & \colhead{}     & \colhead{\ghz}      & \colhead{\ghz}                        & \colhead{}
}
\startdata
  \citet{quireza2006a,quireza2006b} \cii\ Survey & H91\(\alpha\);H92\(\alpha\) & 8.665 & 8.585 & \(0.956\) to \(1.0\) \\
  \citet{quireza2006a,quireza2006b} \(^3\)He Survey & H91\(\alpha\) & 8.665 & 8.585 & \(0.983\) to \(1.0\) \\
  \citet{balser2011,balser2015} & H87\(\alpha\) to H93\(\alpha\) & 8.665 & 9.0 & 1.057 \\
\enddata
\tablenotetext{a}{Average RRL frequency used by the author, which is not the actual average RRL frequency}
\end{deluxetable}

In Table~\ref{tab:appD_sd_corrections} we list the \(X/X_{\rm true}\)
factors for the \citet{quireza2006a}, \citet{quireza2006b},
\citet{balser2011}, and B15 single dish studies. For each survey, we
list the author; the observed RRL transitions; the observed continuum
frequency; the average RRL frequency they used in the electron
temperature equation; and the \(X/X_{\rm true}\) factor.

\subsection{Interferometer Observations}

Interferometers measure intensity in units of flux density per
synthesized beam, \(S\), which is related to brightness temperature,
\(T_B\), by the Rayleigh-Jeans law:
\begin{equation}
  S = \frac{2kc^2}{\nu^2}T_B. \label{eq:appD_rj_law}
\end{equation}
If the RRL and continuum flux densities are measured at the same
frequency, with the same telescope, and with the same synthesized beam
size, the RRL-to-continuum flux density ratio and electron temperature
are given by Equations~\ref{eq:appD_line2cont} and \ref{eq:appD_te},
respectively, where \(I_C\) is the continuum flux density and \(I_L\)
is the RRL flux density.

\subsection{Averaging Interferometer RRLs}

An important difference between interferometric observations and
single dish observations is that interferometers measure the RRL and
continuum emission simultaneously.  At each RRL frequency, we measure
the RRL flux density and continuum flux density with the same
synthesized beam. If the source is homogeneous and isothermal, we can
ignore all effects of the varying beam size.

In our VLA survey analysis, we extract spectra from our data cubes in
two ways: from the pixel of brightest continuum emission, such that
the spectrum has units of flux density per beam, or from the sum of
all pixels within a region, such that the spectrum has units of flux
density. We average these spectra weighted by the continuum brightness
and rms noise in the line-free regions, so our interferometric \(X\)
factor is
\begin{equation}
  X = \left[\left(\frac{\langle S_C(\nu_L) \rangle^*}{\langle S_L(\nu_L) \rangle^*}\right)\left(\frac{\langle \nu_L\rangle^*}{\rm GHz}\right)^{1.1}\right]^{0.87},
\end{equation}
where \(S_C\) and \(S_L\) are the continuum and RRL brightness or flux
density, respectively, and \(\langle\rangle^*\) indicates a weighted
average. If we assume that the spectral rms noise is the same in each
RRL transition, then the weighted average values are simply
\begin{align}
  \langle S_C(\nu_L) \rangle^* & = \frac{\sum_i S_C^2(\nu_{L,i})}{\sum_i S_C(\nu_{L,i})} \\
  \langle S_L(\nu_L) \rangle^* & = \frac{\sum_i S_L(\nu_{L,i})S_C(\nu_{L,i})}{\sum_i S_C(\nu_{L,i})} \\
  \langle \nu_L\rangle^* & = \frac{\sum_i \nu_{L,i}S_C(\nu_{L,i})}{\sum_i S_C(\nu_{L,i})}.
\end{align}

From Equations~\ref{eq:appD_cont_k} and \ref{eq:appD_rj_law}, the
continuum brightness at the \(i\)th RRL frequency is
\begin{align}
  S_C(\nu_{L,i}) & = \frac{2k\nu_{L,i}^2}{c^2}T_e\tau_C(\nu_{L,i}) \nonumber \\
  \frac{S_C(\nu_{L,i})}{\text{Jy sr}^{-1}} & = 2.530\times10^3\left(\frac{T_e}{\rm K}\right)^{-0.35}\left(\frac{\nu_{L,i}}{\rm GHz}\right)^{-0.1}\left(\frac{n_i\,n_e}{\text{cm}^{-6}}\right)\left(\frac{l}{\rm pc}\right)
\end{align}
for a homogeneous nebula with depth \(l\). The RRL frequency is the
only factor that depends on RRL transition, so
\begin{equation}
  \sum_i \frac{S_C(\nu_{L,i})}{\text{Jy sr}^{-1}} = 2.530\times10^3\left(\frac{T_e}{\rm K}\right)^{-0.35}\left(\frac{n_i\,n_e}{\text{cm}^{-6}}\right)\left(\frac{l}{\rm pc}\right)\sum_i\left(\frac{\nu_{L,i}}{\rm GHz}\right)^{-0.1}
\end{equation}
and
\begin{align}
  \frac{\langle S_C(\nu_L) \rangle^*}{\text{Jy sr}^{-1}} & = 2.530\times10^3\left(\frac{T_e}{\rm K}\right)^{-0.35}\left(\frac{n_i\,n_e}{\text{cm}^{-6}}\right)\left(\frac{l}{\rm pc}\right) \nonumber \\
  & \times \left[\sum_i\left(\frac{\nu_{L,i}}{\rm GHz}\right)^{-0.2}\right]\left[\sum_i\left(\frac{\nu_{L,i}}{\rm GHz}\right)^{-0.1}\right]^{-1}.
\end{align}

Using Equations~\ref{eq:appD_rrl_k} and \ref{eq:appD_rj_law}, the
LTE brightness of the \(i\)th RRL is
\begin{align}
  S_L(\nu_{L,i}) & = \frac{2k\nu_{L,i}^2}{c^2}T_e\tau_L^*(\nu_{L,i}) \nonumber \\
  \frac{S_L(\nu_{L,i})}{\text{Jy sr}^{-1}} & = 1.796\times10^7\left(\frac{T_e}{\rm K}\right)^{-1.5}\left(\frac{\nu_{L,i}}{\rm GHz}\right)\left(\frac{n_pn_e}{\text{cm}^{-6}}\right)\left(\frac{\Delta V}{\text{km s}^{-1}}\right)^{-1}\left(\frac{l}{\rm pc}\right)
\end{align}
for a homogeneous medium with depth \(l\). The average RRL brightness
is
\begin{align}
  \frac{\langle S_L(\nu_L) \rangle^*}{\text{Jy sr}^{-1}} & = 1.796\times10^7\left(\frac{T_e}{\rm K}\right)^{-0.5}\left(\frac{n_pn_e}{\text{cm}^{-6}}\right)\left(\frac{\Delta V}{\text{km s}^{-1}}\right)^{-1}\left(\frac{l}{\rm pc}\right) \nonumber \\
  & \times \left[\sum_i\left(\frac{\nu_{L,i}}{\rm GHz}\right)^{0.9}\right]\left[\sum_i\left(\frac{\nu_{L,i}}{\rm GHz}\right)^{-0.1}\right]^{-1}.
\end{align}

The average RRL frequency is
\begin{equation}
  \frac{\langle \nu_L\rangle^*}{\rm GHz} = \left[\sum_i\left(\frac{\nu_{L,i}}{\rm GHz}\right)^{0.9}\right]\left[\sum_i\left(\frac{\nu_{L,i}}{\rm GHz}\right)^{-0.1}\right]^{-1},
\end{equation}
so the interferometric \(X\) approximation simplifies to
\begin{equation}
  X = \left(\frac{T_e}{\rm K}\right)\left[1.409\times10^{-4}\left(\frac{\Delta V}{\text{km s}^{-1}}\right)(1+y)\left(\frac{\sum_i \nu_{L,i}^{-0.2}}{\sum_i \nu_{L,i}^{0.9}}\right)\left(\frac{\sum_i \nu_{L,i}^{0.9}}{\sum_i \nu_{L,i}^{-0.1}}\right)^{1.1}\right]^{0.87}.
\end{equation}
The ratio of this approximation to \(X_{\rm true}\) is
\begin{equation}
  \frac{X}{X_{\rm true}} = \left[\left(\frac{\sum_i \nu_{L,i}^{-0.2}}{\sum_i \nu_{L,i}^{0.9}}\right)\left(\frac{\sum_i \nu_{L,i}^{0.9}}{\sum_i \nu_{L,i}^{-0.1}}\right)^{1.1}\right]^{0.87}
\end{equation}
We observe the seven RRL transitions from H87\(\alpha\) to
H93\(\alpha\). Our \(X\) factor ratio is thus
\begin{equation}
  \frac{X}{X_{\rm true}} = 1.0
\end{equation}
Our strategy for averaging multiple RRL transitions to compute
electron temperatures is accurate.

\end{document}